%% file: ms.tex
\documentclass[a4paper,fleqn,usenatbib,useAMS,times]{mnras}
\usepackage{graphicx}	
\usepackage{amsmath}	
\usepackage{amssymb}	
\usepackage{multicol}   
\usepackage{bm}		
\usepackage{pdflscape}	
\usepackage[T1]{fontenc}
\usepackage{ae,aecompl}
\usepackage{newtxtext,newtxmath} 
\usepackage{fancyhdr}  
\usepackage{supertabular}
\usepackage{longtable}
\usepackage{url}
\usepackage{dcolumn}
\usepackage{morefloats}	 
\usepackage{rotating,caption}  	 
\usepackage{tikz}    	 
\usepackage{adjustbox}
\usepackage{blindtext} 
\usepackage{natbib} 
\usepackage{threeparttable} 
\title[A survey of low-$z$ weak absorbers]{   
\vskip-0.5cm 
{\LARGE COS-Weak: Probing the CGM using analogs of weak Mg~{\sc ii} absorbers at $z<0.3$}}    
\author[S. Muzahid et al.]
{
\parbox{\textwidth}{\vskip-0.5cm  
S. Muzahid$^{1,2}$\thanks{E-mail: sowgat@strw.leidenuniv.nl},     
G. Fonseca$^{1,3}$, 
A. Roberts$^{1,4} $,  
B. Rosenwasser$^{1,5}$,   
P. Richter$^{6,7}$,           
A. Narayanan$^{8}$, 
C. Churchill$^{9}$, and  
J. Charlton$^{1}$,  
} 
\vspace*{4pt}\\   
$^{1}$ Department of Astronomy \& Astrophysics, The Pennsylvania State University, 525 Davey Lab, University Park, State College, PA 16802, USA \\ 
$^{2}$ Leiden Observatory, Leiden University, PO Box 9513, NL-2300 RA Leiden, the Netherlands  \\  
$^{3}$ Department of Physics, University of Connecticut, Storrs, CT 06269, USA  \\ 
$^{4}$ Departamento de Astronom\'{i}a, Universidad de Chile, Camino del Observatorio 1515, Las Condes, Santiago, Chile \\ 
$^{5}$ Department of Astronomy, University of Wisconsin-Madison, 475 N. Charter Street, Madison, WI 53706, USA \\ 
$^{6}$ Institut f$\ddot{u}$r Physik und Astronomie, Universit$\ddot{a}$t Potsdam, Haus 28, Karl-Liebknecht-Str. 24/25, 14476 Golm (Potsdam), Germany \\ 
$^{7}$ Leibniz-Institute for Astrophysics Potsdam (AIP), An der Sternwarte 16, D-14482 Potsdam, Germany \\ 
$^{8}$ Indian Institute of Space Science and Technology, Thiruvananthapuram 695547, Kerala, India \\ 
$^{9}$ Department of Astronomy, New Mexico State University, Las Cruces, NM 88003, USA \\  
}   
\date{\vskip-1cm Accepted. Received; in original form} 
\pubyear{2016}
\begin{document}
\label{firstpage}
\pagerange{\pageref{firstpage}--\pageref{lastpage}}   
\maketitle
\newcommand{\sqcm}{cm$^{-2}$}  
\newcommand{\lya}{Ly$\alpha$}
\newcommand{\lyb}{Ly$\beta$}
\newcommand{\lyg}{Ly$\gamma$}
\newcommand{\lyd}{Ly$\delta$}
\newcommand{\HeI}{\mbox{He\,{\sc i}}}
\newcommand{\HeII}{\mbox{He\,{\sc ii}}}
\newcommand{\HI}{\mbox{H\,{\sc i}}}
\newcommand{\HII}{\mbox{H\,{\sc ii}}}
\newcommand{\HH}{\mbox{H{$_2$}}}
\newcommand{\hh}{\mbox{\tiny H{$_2$}}} 
\newcommand{\hi}{\mbox{\tiny H{\sc i}}} 
\newcommand{\OI}{\mbox{O\,{\sc i}}}
\newcommand{\OII}{\mbox{O\,{\sc ii}}}
\newcommand{\OIII}{\mbox{O\,{\sc iii}}}
\newcommand{\OIV}{\mbox{O\,{\sc iv}}}
\newcommand{\OV}{\mbox{O\,{\sc v}}}
\newcommand{\OVI}{\mbox{O\,{\sc vi}}}
\newcommand{\OVII}{\mbox{O\,{\sc vii}}}
\newcommand{\OVIII}{\mbox{O\,{\sc viii}}}
\newcommand{\CaVIII}{\mbox{Ca\,{\sc viii}}}
\newcommand{\CaVII}{\mbox{Ca\,{\sc vii}}}
\newcommand{\CaVI}{\mbox{Ca\,{\sc vi}}}
\newcommand{\CaV}{\mbox{Ca\,{\sc v}}}
\newcommand{\CIV}{\mbox{C\,{\sc iv}}}
\newcommand{\CV}{\mbox{C\,{\sc v}}}
\newcommand{\CVI}{\mbox{C\,{\sc vi}}}
\newcommand{\CII}{\mbox{C\,{\sc ii}}}
\newcommand{\CI}{\mbox{C\,{\sc i}}}
\newcommand{\CIIs}{\mbox{C\,{\sc ii}}$^\ast$}
\newcommand{\CIII}{\mbox{C\,{\sc iii}}}
\newcommand{\SiI}{\mbox{Si\,{\sc i}}}
\newcommand{\SiII}{\mbox{Si\,{\sc ii}}}
\newcommand{\SiIII}{\mbox{Si\,{\sc iii}}}
\newcommand{\SiIV}{\mbox{Si\,{\sc iv}}}
\newcommand{\SiXII}{\mbox{Si\,{\sc xii}}}
\newcommand{\SV}{\mbox{S\,{\sc v}}}
\newcommand{\SIV}{\mbox{S\,{\sc iv}}}
\newcommand{\SIII}{\mbox{S\,{\sc iii}}}
\newcommand{\SII}{\mbox{S\,{\sc ii}}}
\newcommand{\SI}{\mbox{S\,{\sc i}}}
\newcommand{\ClI}{\mbox{Cl\,{\sc i}}}
\newcommand{\ArI}{\mbox{Ar\,{\sc i}}}
\newcommand{\NI}{\mbox{N\,{\sc i}}}
\newcommand{\NII}{\mbox{N\,{\sc ii}}}
\newcommand{\NIII}{\mbox{N\,{\sc iii}}}
\newcommand{\NIV}{\mbox{N\,{\sc iv}}}
\newcommand{\NV}{\mbox{N\,{\sc v}}}
\newcommand{\PV}{\mbox{P\,{\sc v}}}
\newcommand{\PII}{\mbox{P\,{\sc ii}}}
\newcommand{\PIII}{\mbox{P\,{\sc iii}}}
\newcommand{\PIV}{\mbox{P\,{\sc iv}}}
\newcommand{\NeVIII}{\mbox{Ne\,{\sc viii}}}
\newcommand{\ArVIII}{\mbox{Ar\,{\sc viii}}}
\newcommand{\NeV}{\mbox{Ne\,{\sc v}}}
\newcommand{\NeVI}{\mbox{Ne\,{\sc vi}}}
\newcommand{\NeX}{\mbox{Ne\,{\sc x}}} 
\newcommand{\NaIX}{\mbox{Na\,{\sc ix}}} 
\newcommand{\MgII}{\mbox{Mg\,{\sc ii}}}
\newcommand{\FeII}{\mbox{Fe\,{\sc ii}}}
\newcommand{\MgX}{\mbox{Mg\,{\sc x}}}
\newcommand{\AlXI}{\mbox{Al\,{\sc xi}}}
\newcommand{\FeIII}{\mbox{Fe\,{\sc iii}}}
\newcommand{\NaI}{\mbox{Na\,{\sc i}}}
\newcommand{\CaII}{\mbox{Ca\,{\sc ii}}}
\newcommand{\zabs}{$z_{\rm abs}$}
\newcommand{\zmin}{$z_{\rm min}$}
\newcommand{\zmax}{$z_{\rm max}$}
\newcommand{\zqso}{$z_{\rm qso}$}
\newcommand{\subHe}{_{\it HeII}}
\newcommand{\subH}{_{\it HI}}
\newcommand{\subHLy}{_{\it H Ly}}
\newcommand{\degree}{\ensuremath{^\circ}}
\newcommand{\lapp}{\mbox{\raisebox{-0.3em}{$\stackrel{\textstyle <}{\sim}$}}}
\newcommand{\gapp}{\mbox{\raisebox{-0.3em}{$\stackrel{\textstyle >}{\sim}$}}}
\newcommand{\be}{\begin{equation}}
\newcommand{\en}{\end{equation}}
\newcommand{\di}{\displaystyle}
\def\tworule{\noalign{\medskip\hrule\smallskip\hrule\medskip}} 
\def\onerule{\noalign{\medskip\hrule\medskip}} 
\def\bl{\par\vskip 12pt\noindent}
\def\bll{\par\vskip 24pt\noindent}
\def\blll{\par\vskip 36pt\noindent}
\def\rot{\mathop{\rm rot}\nolimits}
\def\alf{$\alpha$}
\def\refff{\leftskip20pt\parindent-20pt\parskip4pt}
\def\zabs{$z_{\rm abs}$}
\def\zqso{$z_{\rm qso}$}
\def\zem{$z_{\rm em}$}
\def\mgii{Mg\,{\sc ii}~} 
\def\feiia{Fe\,{\sc ii}$\lambda$2600}
\def\mgia{Mg\,{\sc i}$\lambda$2852}
\def\mgiia{Mg\,{\sc ii}$\lambda$2796}
\def\mgiib{Mg\,{\sc ii}$\lambda$2803}
\def\mgiiab{Mg\,{\sc ii}$\lambda\lambda$2796,2803}
\def\wobs{$w_{\rm obs}$}
\def\kms{km~s$^{-1}$}
\def\bnt{$b_{\rm nt}$}
\def\fosc{$f_{\rm osc}$}
\def\chisq{$\chi^{2}$}
\def\dtype{$\delta_{\rm type}$}  
\def\Msun{$M_{\odot}$}  
\begin {abstract} 
\noindent 
We present a sample of 34 weak metal line absorbers at $z<0.3$ selected by the simultaneous $>3\sigma$ detections of the \SiII$\lambda1260$ and \CII$\lambda1334$ absorption lines, with $W_{r}(\SiII)<0.2$~\AA\ and $W_{r}(\CII)<0.3$~\AA, in archival $HST/$COS spectra. Our sample increases the number of known low-$z$ ``weak absorbers'' by a factor of $>5$. The column densities of \HI\ and low-ionization metal lines obtained from Voigt profile fitting are used to build simple photoionization models. The inferred densities and line of sight thicknesses of the absorbers are in the ranges of $-3.3 < \log n_{\rm H}/{\rm cm^{-3}} < -2.4$ and $\sim$1 pc -- 50 kpc (median  $\approx$~500~pc), respectively. Most importantly, 85\% (50\%) of these absorbers show a metallicity of $\rm [Si/H] > -1.0~(0.0)$. The fraction of systems showing near-/super-solar metallicity in our sample is significantly higher than in the \HI-selected sample \citep[][]{Wotta16} and the galaxy-selected sample \citep[][]{Prochaska17} of absorbers probing the circum-galactic medium at similar redshift. A search for galaxies has revealed a significant galaxy-overdensity around these weak absorbers compared to random positions with a median impact parameter of 166~kpc from the nearest galaxy. Moreover, we find the presence of multiple galaxies in $\approx80$\% of the cases, suggesting group environments. The observed $d\mathcal{N}/dz$ of $0.8\pm0.2$ indicates that such metal-enriched, compact, dense structures are ubiquitous in the halos of low-$z$ group galaxies. We suggest that these are transient structures that are related to galactic outflows and/or stripping of metal-rich gas from galaxies.     
\end {abstract}
\vskip-1cm 
\section{Introduction} 
\label{sec:intro}  

Metal absorption lines detected in quasar spectra are thought to arise from the circum-galactic medium (CGM) of intervening galaxies close to the lines of sight \citep[e.g.,][]{Bergeron86,Bergeron91,Steidel92,Adelberger05,Kacprzak08,Kacprzak15,Chen09,Steidel10,Stocke13,Werk13,Bordoloi14b,Turner14,Johnson15,Keeney17}. The absorption strength of a metal line depends primarily on the metallicity and on the ionization parameter of the absorbing gas. The metallicity of an absorber is an important diagnostic of the physical processes at play. For example, high metallicity gas is likely to observe in galactic outflows \citep[]{Tripp11,Muzahid14,Muzahid15b}, whereas the gas infalling from the inter-galactic medium (IGM) on to a galaxy halo is expected to be metal-poor \citep[]{Ribaudo11,Churchill12,Kacprzak12b}. The ionization parameter, usually determined from the ratio of two consecutive ionic states of the same element, provides the density of the absorbing gas, provided the intensity of the incident radiation is known. It is generally assumed that the extra-galactic ultraviolet (UV) background radiation \citep[UVB;][]{Haardt96,Haardt12,Khaire15a} prevails in the CGM of a $\sim L_*$ galaxy at impact parameters greater than several tens of kpc \citep[]{Werk14,Narayanan10b}. The relative abundances of heavy elements in an absorption system provide important clues regarding the star formation history and the initial mass function of the host-galaxy \cite[]{Pettini99,Pettini02,Becker12,Zahedy17}. Metal lines are, thus, essential for determining the physical conditions of the otherwise invisible, diffuse gas in the CGM and for inferring the physical processes that determine the evolution of a galaxy.   

The rest-frame wavelengths of the \MgII~$\lambda\lambda$2796,2803 doublet transitions are such that it is easily observable using ground based optical telescopes for a wide redshift range ($0.3\lesssim z \lesssim 2.5$). Consequently, \MgII\ absorbers have been studied for the last couple of decades by a significant number of authors \citep[e.g.,][and references therein]{Lanzetta87,Petitjean90,Steidel92,Charlton98,Churchill99a,Churchill03,Rigby02,Nestor05,Kacprzak08,Chen10b,Zhu13,Nielsen13a,Nielsen13b,Joshi17,Dutta17}. In addition, recent infrared surveys have identified \MgII\ absorbers all the way up to $z\approx7$ \citep[]{Matejek12,ChenSF16,Bosman17,Codoreanu17}. In all these studies, \MgII\ absorbers are found to probe a wide variety of astrophysical environments. Based on the rest-frame equivalent width of the $\lambda$2796 transition ($W_r^{2796}$) \MgII\ absorbers are historically classified into weak ($W_r^{2796}<$0.3~\AA) and strong ($W_r^{2796}\ge$0.3~\AA) groups. The strong absorbers are sometimes categorized into very strong ($W_r^{2796}\ge$1~\AA) and ultrastrong ($W_r^{2796}\ge$3~\AA) systems \citep[]{Nestor05,Nestor07}. The very strong/ultrastrong \MgII\ absorbers are generally found to be associated with luminous galaxies ($L\gtrsim0.1L_*$). They often show velocities spread over several hundreds of \kms\ and hence are thought to be related to galactic superwind driven by supernova and/or starburst \citep[e.g.,][but see \cite{Gauthier13} for an alternate scenario]{Bond01,Prochter06,Zibetti07}.           

In a detailed study of the \MgII\ absorbing CGM, \citet{Nielsen13b} have found a covering fraction ($C_f$) of 60--100\% (15--40\%) for systems with $W_r^{2796}>$0.1~\AA\ ($W_r^{2796}>$1.0~\AA) out to 100~kpc. The $C_f$ does not change significantly with the host-galaxy's color and/or luminosity \citep[see also][]{Werk13}. However, a bimodality in the azimuthal angle distribution of \MgII\ absorbing galaxies has been reported in the literature, such that the gas is preferentially detected near the projected major and minor axes of star-forming galaxies \citep[][]{Kacprzak12a,Bouche12}. Galaxies with little star-formation do not show any such preference. The observed bimodality is thought to be driven by gas accreted along the major axis and outflowing along the minor axis. 

The weak absorbers with $W_r^{2796}<0.3$~\AA, that are distinct from the strong population in several aspects \citep[e.g.][]{Rigby02}, are of particular interest in this paper. The first systematic survey of weak \MgII\ absorbers was conducted by \citet{Churchill99a} in the redshift range $0.4 < z < 1.4$ using HIRES/Keck spectra. They found that the number of weak absorbers per unit redshift of $d\mathcal{N}/dz\approx1.74\pm0.10$ is $\approx$4 times higher than that of the Lyman Limit Systems (LLSs) which have neutral hydrogen column density $N(\HI)>10^{17.2}$~\sqcm. Therefore it is suggested that a vast majority of the weak \MgII\ absorbers must arise in sub-LLS environments. The evolution of the $d\mathcal{N}/dz$ for the weak absorbers shows a peak at $z=1.2$ \citep[]{Narayanan05,Narayanan07} which was thought to be connected with the evolution of the cosmic star-formation rate density of dwarf galaxies \citep[see also Fig.~9 of][]{Mathes17}. However, in a recent survey, \citet{Codoreanu17} have found that the $d\mathcal{N}/dz$ of $1.35\pm0.58$ for the weak absorbers with $0.1 < W_r < 0.3$~\AA\ at $z\approx2.3$, increases to $2.58\pm0.67$ by $z\approx4.8$ \cite[see also][]{Bosman17}. The number of weak absorbers exceeds the number expected from an exponential fit to strong systems with $W_r^{2796}>0.3$~\AA. The $d\mathcal{N}/dz$ at these high redshifts, however, is consistent with cosmological evolution of the population, suggesting that the processes responsible for weak absorbers are already in place within the first 1~Gyr of cosmic history. One of the most intriguing properties of the weak absorbers is that, in most cases, they exhibit near-solar to super-solar metallicities \citep[]{Rigby02,Lynch07,Misawa08,Narayanan08} in spite of the fact that luminous galaxies are rarely found within a $\sim$50~kpc impact parameter \citep[]{Churchill05,Milutinovic06}. Studying weak absorbers at low-$z$ ($z\approx0.1$) is advantageous since it is relatively easy to search for the host-galaxies.   

Because of the atmospheric cutoff of optical light, a direct search for the \MgII\ doublet at $z<0.3$ is not viable. Instead, \citet{Narayanan05}, for the first time, used weak \SiII~$\lambda$1260 ($W_r^{1260}<0.2$~\AA) and \CII~$\lambda1334$ ($W_r^{1334}<0.3$~\AA) lines as proxies for the weak \MgII\ absorbers. Both $\rm Si_{14}^{28}$ and $\rm Mg_{12}^{24}$ are $\alpha$-process elements. The creation and destruction ionization potentials of \SiII\ (i.e., 8.1 and 16.3 eV, respectively) are very similar to \MgII\ (i.e., 7.6 and 15.0 eV, respectively). The abundance of Si in the  solar neighborhood (i.e., $\rm [Si/H]=-4.49$) is also very similar to that of Mg, i.e., $\rm [Mg/H]=-4.40$ \citep[]{Asplund09}. All these facts indicate that \SiII\ and \MgII\ arise from the same gas phase and \SiII\ is a good proxy for \MgII\ \citep[see also][]{Narayanan05,Herenz13}. In Appendix~\ref{MgIIconnection} we have shown that the weak metal line absorbers studied here are indeed analogous to the known weak \MgII\ absorbers based on our model predicted \MgII\ column densities.            

The only known previous systematic survey for weak absorbers at low-$z$ was conducted by \citet{Narayanan05}. They searched for weak \SiII~$\lambda$1260 and \CII~$\lambda1334$ absorbers in high resolution $HST$ STIS Echelle spectra of 25 quasars. They found only six weak \MgII\ analog absorbers over a redshift pathlength of $\Delta z\approx5.3$. Their estimated $d\mathcal{N}/dz$ of $\approx1.0$ is consistent with cosmological evolution of the population. By considering the effect of the change in the UVB on an otherwise static absorbing gas, their photoionization models suggested that the low-$z$ weak \MgII\ population is likely composed of both kiloparsec-scale, low density structures that only gave rise to \CIV\ absorption at $z\approx1$ and parsec-scale, higher density structures that produced weak \MgII\ absorption at $z\approx1$. Clearly the constancy of $d\mathcal{N}/dz$ need not necessarily suggest the same physical origin for the population at different redshifts, and it warrants a detailed ionization modelling of the absorbers.     

In this paper we have studied weak \MgII\ absorber analogs detected in $HST/$COS spectra via the \SiII~$\lambda$1260 and \CII~$\lambda1334$ transitions. With 34 absorbers in total, here we present the first-ever statistically significant sample of weak absorbers at low-$z$. The paper is organized as follows: In Section~\ref{sec:data} we provide the details of the observations, data reduction, absorber search techniques, and absorption line measurements. In Section~\ref{sec:ana} we present our analysis which includes estimating $d\mathcal{N}/dz$, measuring \HI\ and ionic column densities, and building photoionization models. The main results based on the ionization models are presented in Section~\ref{sec:result}. In Section~\ref{sec:diss} we discuss our main results, followed by a summary in Section~\ref{sec:summary}. Throughout this study we adopt a flat $\Lambda$CDM cosmology with $H_0=71$~\kms~$\rm Mpc^{-1}$, $\Omega_{\rm M}=0.3$, $\Omega_{\Lambda}=0.7$, and $\Omega_{b}h^2=0.02$. Abundances of heavy elements are given in the notation $\rm [X/Y] = \log (X/Y) - \log (X/Y)_{\odot}$ with solar relative abundances taken from \cite{Asplund09}. All the distances given are in physical units.

\section{DATA}      
\label{sec:data}

\subsection{Observations and data reduction}     
\label{subsec:observations}

We have searched for weak \MgII\ analogs in 363 medium resolution, far-UV (FUV) spectra of active galactic nuclei (AGN)/quasars (QSOs) observed with the {\it Hubble Space Telescope}$/$Cosmic Origins Spectrograph ($HST/$COS). These spectra were available in the public $HST$ archive before February, 2016. Note that all these spectra were obtained under programs prior to the $HST$ Cycle-22. The properties of COS and its in-flight operations are discussed in \cite{Osterman11} and \cite{Green12}. We have used only spectra that were obtained with the medium resolution ($R\approx20000$) FUV COS gratings (i.e. G130M and/or G160M). The data were retrieved from the $HST$ archive and reduced using {\sc calcos} pipeline software. The pipeline reduced data (`$x1d$' files) were flux calibrated. The individual exposures were aligned and co-added using the {\sc idl} code (`$coadd\_{x1d}$'), developed by \citet{Danforth10} and subsequently improved by \cite{Keeney12} and \cite{Danforth16}, in order to increase the spectral signal-to-noise ratio ($\rm S/N$). The final co-added spectra typically have spectral coverage of $\approx$~1150--1450~\AA\ for the G130M grating and $\approx$~1450--1800 \AA\ for the G160M grating. Since these archival spectra come from various different observing programs, they show a wide range in $\rm S/N$ (i.e., 2--60 per resolution element). In our analysis, we do not include 67 spectra that show $\rm S/N<5$ per resolution element over most of it spectral coverage. The details of the remaining 296 spectra used in this study are given in Table~\ref{tab:COSlos}. There are 174 spectra with data from both the G130M and G160M gratings, 114 spectra with only G130M grating data, and 8 spectra with only G160M grating data. The median $S/N$ of the 288 G130M grating spectra is 9 per resolution element. The median $S/N$ of the 182 G160M grating spectra is 7 per resolution element. In Table~\ref{tab:nCOSlos} we list the 67 spectra that are available in the archive but are not used in this study due to poor $\rm S/N$ (i.e. $<5$ per resolution element). The COS FUV spectra are highly oversampled with 6 raw pixels per resolution element. We, thus, binned the spectra by three pixels. All our measurements and analyses are performed on the binned data. We note that while the binning improves $\rm S/N$ per pixel by a factor of $\sqrt3$, it does not affect the absorption line measurements (i.e. $W_r$ and $N$). For each spectrum continuum normalization was done by fitting the line-free regions with smooth low-order polynomials.

\input{tables/systems-ordered.tex}

\begin{figure} 
\includegraphics[width=0.50\textwidth]{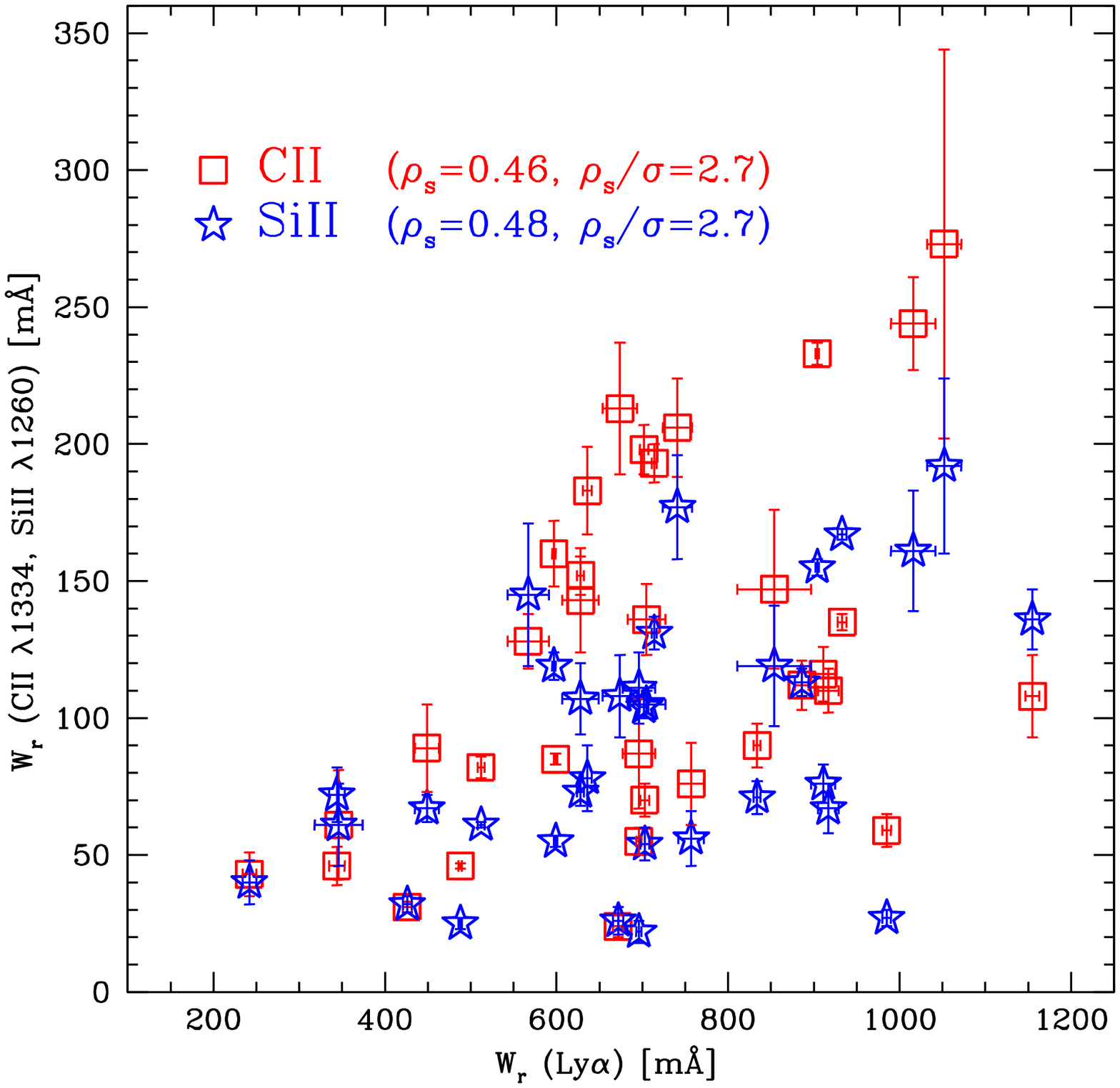}     
\vskip-0.2cm  
\caption{The rest-frame equivalent widths of the low-ionization metal lines (\CII\ and \SiII) against $W_r$(\lya). All the ``weak'' absorbers have $W_r(\SiII~\lambda1260)<200$~m\AA\ and $W_r(\CII~\lambda1334)<300$~m\AA\ by design. A mild $2.7\sigma$ correlation between $W_r(\CII, \SiII)$ and $W_r$(\lya) is present. An upper envelope with an apparent lack of data points in the upper-left corner (i.e. the region with $W_r$(\lya)~$<600$~m\AA\ and $W_r(\CII, \SiII)>100$~m\AA) is evident.}  
\label{fig:ew}  
\end{figure}

\begin{figure} 
\includegraphics[width=0.50\textwidth]{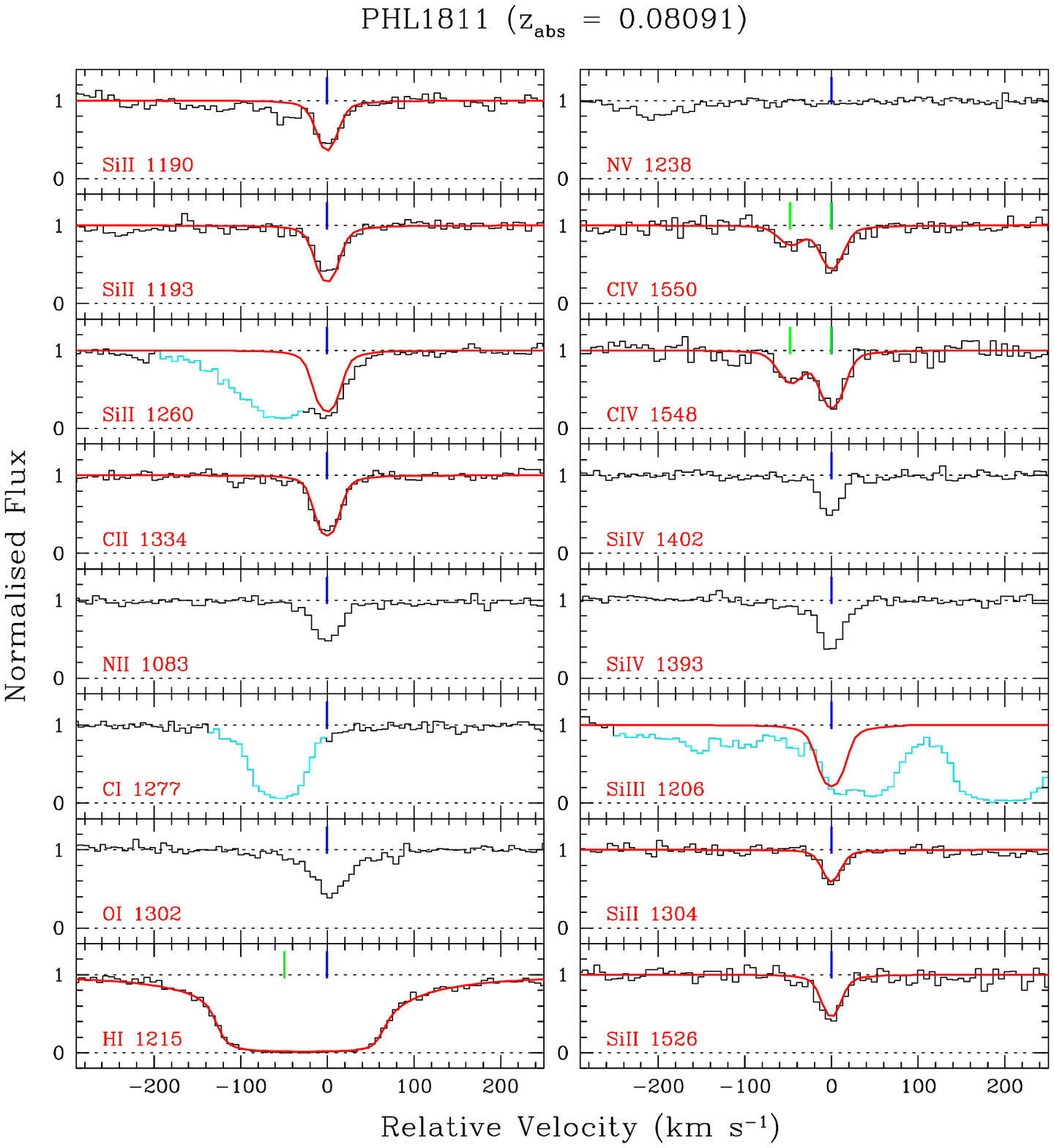}   
\vskip-0.2cm  
\caption{Velocity plot of a weak absorber at \zabs~$=$~0.08091 detected towards PHL1811. The smooth solid red curves over-plotted on top of data (black histogram) are the best fitting model profiles. Unrelated absorption lines are shown in cyan. Since the \SiIII\ line is blended with the Galactic \SiII$\lambda1304$ line, we have generated a synthetic profile corresponding to the maximum allowed column density. Single component \CII\ and \SiII\ lines are detected at $\sim0$~\kms. The \CIV\ absorption shows another component at $\sim-50$~\kms. Both the \CIV\ and \lya\ lines are fitted with two Voigt profile components. The line centroids of the individual components are marked by the vertical ticks in each panel. Velocity plots, Voigt profile fits, and comments about each absorbers can be found in Appendix~\ref{appendix:vplots} as online material.}  
\label{fig:vplot-exam}  
\end{figure}

\subsection{Search techniques and the sample}       
\label{subsec:search}  

Our search for the weak absorbers relies on the simultaneous presence of the \SiII$\lambda1260$ and \CII$\lambda1334$ lines. Note that the presence of \CII$\lambda1334$ restricts our survey to a maximum redshift of $z_{\rm max}\approx0.35$. First, we assume every detected absorption in a spectrum above $\lambda_{\rm obs}>1260$~\AA\ is due to the \SiII$\lambda1260$ line from redshift \zabs~$=(\lambda_{\rm obs}/\lambda_{\rm rest}) -1$, where $\lambda_{\rm rest}$ is the rest-frame wavelength of the \SiII$\lambda1260$ transition \footnote{In this work, $\lambda_{\rm rest}$ and oscillator strengths $f_{\rm osc}$ of different lines are taken from \cite{Morton03}}. Next we check for the presence of the corresponding \CII$\lambda1334$ line using the \zabs. We build a ``primary'' list of absorbers using each of the identified coincidence. We then check for the presence of other common transitions ( e.g., \lya, \SiII$\lambda1193$, \CII$\lambda1036$, \SiIII). Since some amount of neutral hydrogen is expected to be associated with these weak low-ionization absorbers, we have excluded any ``primary'' system that does not exhibit any detectable \lya\ absorption. The presence of other metal lines would depend on the phase structure, ionization parameter, and metallicity of the absorber. We, therefore, do not impose the detection of any other metal line as a necessary criterion to confirm a weak absorber. Next, we investigate if the identified \SiII$\lambda1260$ and$/$or \CII$\lambda1334$ lines can have any other identity (i.e. other lines from other redshifts). We include a system in our ``secondary'' list when both the lines are free from significant contamination.  

For each absorber in the ``secondary'' list we measure rest-frame equivalent widths ($W_r$) of \SiII$\lambda1260$ and \CII$\lambda1334$ lines. We consider only systems in which both the lines are detected with $\geqslant 3\sigma$ significance. Next we impose the $W_r$ threshold criteria so that each system in our final sample, listed in Table~\ref{tab:sample}, has $W_r(\SiII\lambda1260)<200$ m\AA\ and $W_r(\CII\lambda1334)<300$ m\AA. The equivalent widths of the \CII\ and \SiII\ lines of our final sample are plotted against the $W_r$(\lya) in Fig.~\ref{fig:ew}. A mild correlation between the equivalent widths of low-ionization metal lines and \lya\ is apparent. A Spearman rank correlation test suggests a correlation coefficient, $\rho_s$, of $\approx0.47$ with a $2.7\sigma$ significance. We notice an upper envelope in the figure which is likely due to a metallicity effect. In particular, the systems with $W_r$(\lya)~$\lesssim600$~m\AA\ would require unreasonably high metallicities in order to produce \SiII\ and/or \CII\ absorption with $W_r \gtrsim$~100~m\AA. The systems with $W_r$(\lya)~$>600$~m\AA, on the other hand, show a wide range in \SiII\ and \CII\ equivalent widths. This could be because of poor small scale metal mixing so that the observed \HI\ need not entirely be associated with the low-ionization metal line \citep[e.g.,][]{Schaye07}. This is certainly the case for systems for which the \HI\ is not centered in velocity around the low ionization absorption (see e.g., Fig.~\ref{fig:vplot-exam}).    

The \SiII/\CII\ systems that show $W_r$ higher than our thresholds, i.e. the strong absorbers, are listed in Table~\ref{tab:strong} for completeness. The systems with \zabs\ within 5000~\kms\ of the emission redshift (\zem) of the background AGN$/$QSOs, as listed in Table~\ref{tab:intrinsic}, are excluded for this study since they might have different origins \citep[see e.g.,][]{Muzahid13} than the intervening absorbers we are interested in. Additionally, the systems with $cz < 1000$~\kms\ are not considered here, since they are likely to be related to our own Galaxy and$/$or local group galaxies. We refer the reader to \citet{Richter16} for a detailed analysis of such absorbers. There are a total of 34 weak \MgII\ absorber analogs in our sample satisfying all of the above requirements.            

\begin{figure} 
\includegraphics[width=0.49\textwidth]{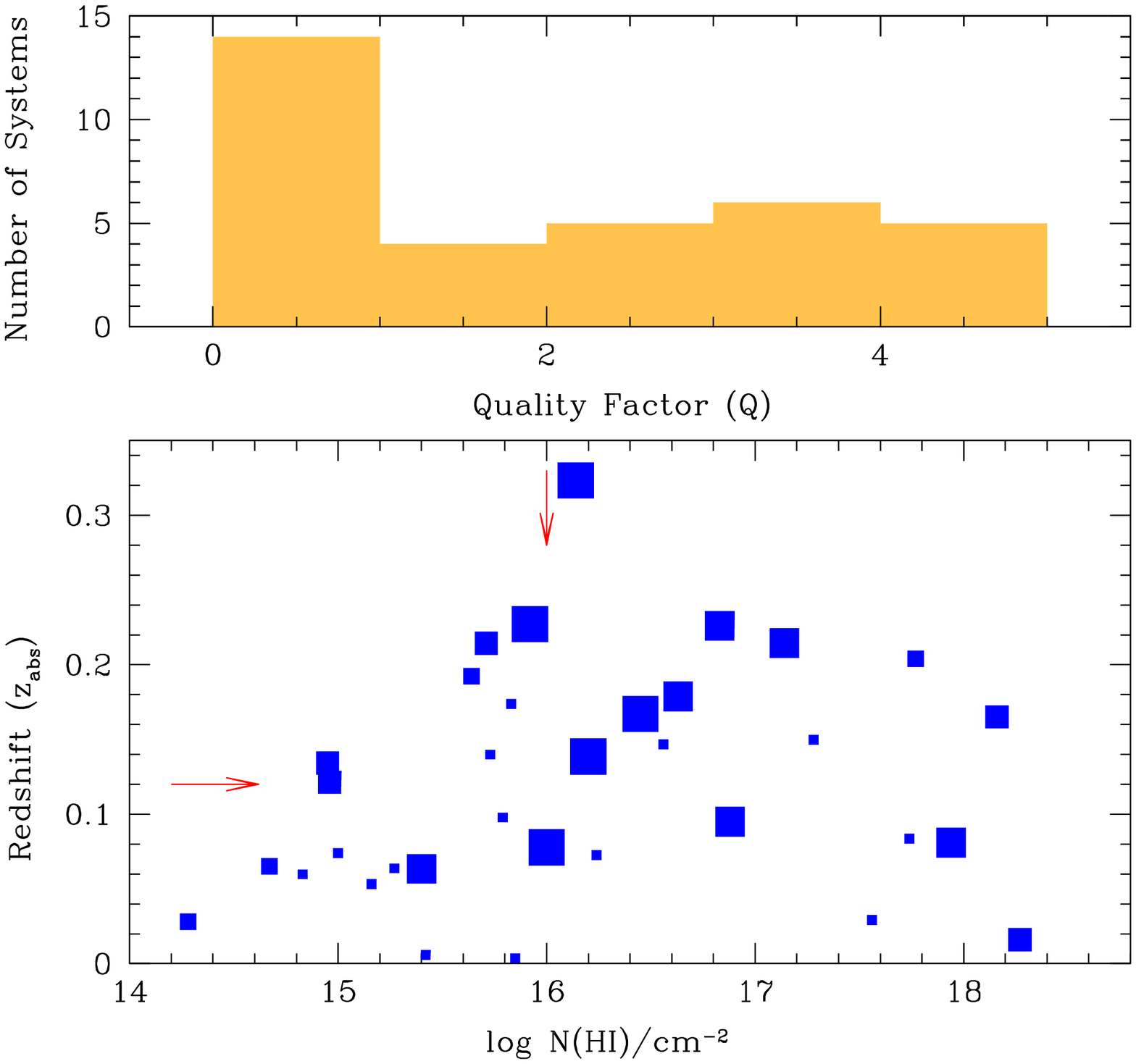}   
\vskip-0.2cm  
\caption{The absorption redshifts plotted against the $N(\HI)$ values estimated for individual systems. The median values of \zabs\ and $\log N(\HI)$ are indicated by the horizontal and vertical arrows, respectively. Symbol sizes are scaled by the corresponding quality factors ($Q$). The $Q$ distribution is shown in the top panel. In total there are 16 systems with $Q\ge3$ for which $N(\HI)$ estimates are reliable.}
\label{fig:NHIdist}  
\end{figure}

\subsection{Absorption Line Measurements}           
\label{subsec:measurement}  
   
Besides measuring equivalent widths, we used {\sc vpfit} \footnote{http://www.ast.cam.ac.uk/$\sim$rfc/vpfit.html} to obtain column densities of \HI, \SiII, and \CII. The column densities of \SiIII, \CIV, and \OVI\ are also measured whenever available. Since the line spread function (LSF) of the COS spectrograph is not a Gaussian, we use the LSF given by \cite{Kriss11}. The LSF was obtained by interpolating the LSF tables at the observed central wavelength for each absorption line and was convolved with the model Voigt profile while fitting absorption lines or generating synthetic profiles. For fitting an absorption line, we have used the minimum number of components required to achieve a reduced $\chi^2 \approx1$. However, owing to the limited resolution of the COS spectrograph, we may be missing the ``true'' component structure. As a consequence, the Doppler parameters generally turn out to be larger compared to the high-$z$ weak absorbers detected in high-resolution $VLT/$UVES spectra \citep[e.g.,][]{Narayanan08}. For a strong line this would lead to an underestimation of the corresponding column density. Nevertheless, the column density can be accurately measured for a weak line that falls on the linear part of the curve-of-growth (COG). COS wavelength calibration is known to have uncertainties at the level of $\approx5-15$~\kms\ \citep[]{Savage11b,Muzahid15a}. We do see velocity misalignments of the same order between different absorption lines from the same system. An example of a weak absorber with the model profiles is shown in Fig.~\ref{fig:vplot-exam}. System plots with Voigt profiles fits of the full sample are available in Appendix~\ref{appendix:vplots}.

\section{Analysis}   
\label{sec:ana}

\begin{figure} 
\includegraphics[width=0.5\textwidth]{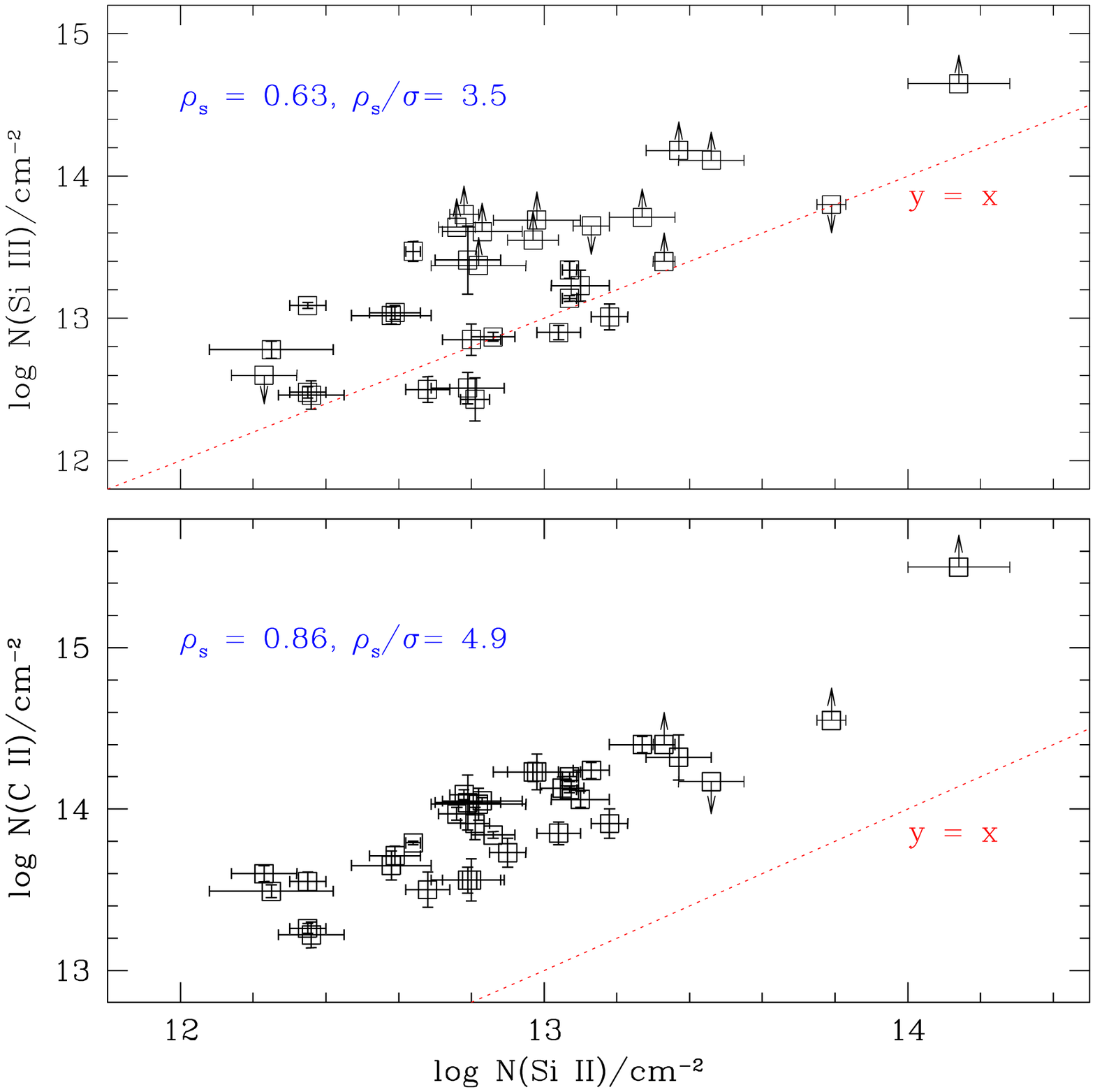}   
\vskip-0.2cm  
\caption{$\log N(\CII)$ versus $\log N(\SiII)$ (bottom) and $\log N(\SiIII)$ versus $\log N(\SiII)$ (top) for each absorber. Note that $N(\SiII)$ is more tightly correlated with $N(\CII)$ as compared to $N(\SiIII)$ in our  sample. The $y=x$ (dotted) line is shown just for comparing the relative strengths of the low-ionization metal lines. The tight correlation between $N(\CII)$ and $N(\SiII)$ indicates that they trace the same phase of the absorbing gas. The somewhat larger scatter in the top panel possibly suggests that \SiIII\ might have an additional contribution from the high-ionization gas phase for a fraction of the absorbers.}           
\label{fig:metal-metal}        
\end{figure}

\subsection{Absorber's Frequency, $d\mathcal{N}/dz$}   
\label{sec:dndz} 
 
We have read through the titles and the abstracts of all the relevant program IDs, under which the 296 COS spectra were obtained, in order to inspect whether the spectra were taken to probe the CGM of known foreground galaxies. The spectra obtained from dedicated CGM programs (e.g. PID: 11598, 12248) were assigned to ``flag = 0'', meaning that they are not part of our statistical sample. The remaining spectra with ``flag = 1'' constitute our statistical sample. There are 178 (118) ``flag = 1'' (``flag = 0'') spectra in our sample. The redshift path-length covered for simultaneously detecting \SiII$\lambda1260$ ($\approx$0.2~\AA) and \CII~$\lambda1334$ ($\approx$0.3~\AA) in the 178 spectra in our statistical sample is $\Delta z=27.8$. There are 22 bona fide weak absorbers identified in these spectra (see Table~\ref{tab:sample}). This yields the number density of weak absorbers, $d\mathcal{N}/dz = 0.8\pm0.2$ at $z<0.3$. This is consistent with the value ($1.00\pm0.20$) obtained by \citet{Narayanan05} using a much smaller sample size. We note that the $d\mathcal{N}/dz$ of \HI\ absorbers \citep[]{Danforth16} with $N(\HI)\approx10^{16}$~\sqcm\ at $z\approx0.1$ (the median redshift of our sample) matches with that of the weak absorbers. Interestingly, the median $N(\HI)$ in our sample is $10^{16}$~\sqcm, as we describe in the next section. This re-confirms the finding of \cite{Churchill99a} that the majority of the weak absorbers arise in a sub-LLS environment.  

Recently \citet{Richter16b} have presented a sample of \SiIII\ absorbers at similar redshift using archival COS spectra. They have found $d\mathcal{N}/dz$ of $2.5\pm0.4$ for \SiIII\ down to a $N(\SiIII)$ of $10^{12.2}$~\sqcm. Comparing this to the frequency of weak absorbers, we suggest that roughly half of the \SiIII\ population should show associated weak \SiII\ absorption. From the middle panel of Fig.~6 of \citet{Richter16b}, there are 24 \SiII\ absorbers with $W_r(\SiII\lambda1260)<200$~m\AA. This leads to a  $d\mathcal{N}/dz$ of $1.0\pm0.2$ (using $\Delta z =24$ for their sample) which is consistent with our estimate.

In Table~\ref{tab:sample}, there are 12 absorbers that are detected in the ``flag = 0'' spectra. We did not use them for the $d\mathcal{N}/dz$ calculation in order to avoid any possible observational bias. However, those systems are considered in all other analyses related to the chemical/physical conditions of the absorbing gas.

\subsection{Neutral Hydrogen Column Density, $N(\HI)$} 

Owing to the low redshifts of the weak absorbers ($z<0.3$), higher order Lyman series lines are not covered by the COS spectra for most cases. 17/34 systems have only \lya\ and 7/34 systems have both \lya\ and \lyb\ coverages. The remaining 10 systems have \lya, \lyb, and \lyg\ (or higher order lines) coverages. Estimating reliable $N(\HI)$ values in the absence of unsaturated higher order Lyman series lines is challenging. We, thus, assign a quality factor ``$Q$'' for each of our $N(\HI)$ measurements based on the absorption strength (level of saturation) and the presence (or absence) of the higher order lines. The quality factor can take values from 1--5, with 5 being the best/secure measurements. $N(\HI)$ values with $Q=1$ and $2$ are not well constrained. For each absorber listed in Table~\ref{tab:sample}, we checked the literature to see if a $N(\HI)$ measurement is available from {\sc fuse} data. For 7 systems we have adopted $N(\HI)$ values from the literature that were well constrained from higher-order Lyman series lines covered by the {\sc fuse} spectra. These systems are \zabs~$=0.22597$ towards HE0153-4520 \citep[]{Savage11b}, \zabs~$=0.16710$ towards PKS0405-123 \citep[]{Prochaska04b}, \zabs~$=0.13850$ towards PG1116+215 \citep[]{Sembach04}, \zabs~$=0.06350$ towards 3C263 \citep[]{Savage12}, \zabs~$=0.09495$ towards PKS1302-102 \citep[]{Cooksey08}, \zabs~$=0.07774$ towards PHL1811 \citep[]{Lacki10}, and \zabs~$=0.08091$ towards PHL1811 \citep[]{Jenkins05}. There are 3 additional systems for which we have adopted $N(\HI)$ values from the literature. These systems are \zabs~$=0.22722$ towards Q0107--025A \citep[]{Muzahid14}, \zabs~$=0.21451$ towards SDSSJ1322+4645, and \zabs~$=0.17885$ towards SDSSJ1419+4207. The $N(\HI)$ estimates for the latter two systems are obtained from the Lyman limit breaks seen in the G140L, G130/1222 grating observations using COS \citep[see][]{Prochaska17}. For all of these 10 systems we assign $Q\ge3$. The absorption redshifts and $N(\HI)$ estimates are shown in Fig.~\ref{fig:NHIdist}. The median $N(\HI)$ is $10^{16.0}$~\sqcm\ for the full sample. The median value becomes $10^{15.9}$~\sqcm\ for the systems with $Q\ge3$. The median value of redshifts changes from $0.12$ to $0.22$ when we consider the systems with $Q\ge3$ as opposed to the full sample. This follows from the fact that the most of the lower redshift systems do not have higher order Lyman series lines covered.

\begin{figure} 
\includegraphics[width=0.45\textwidth]{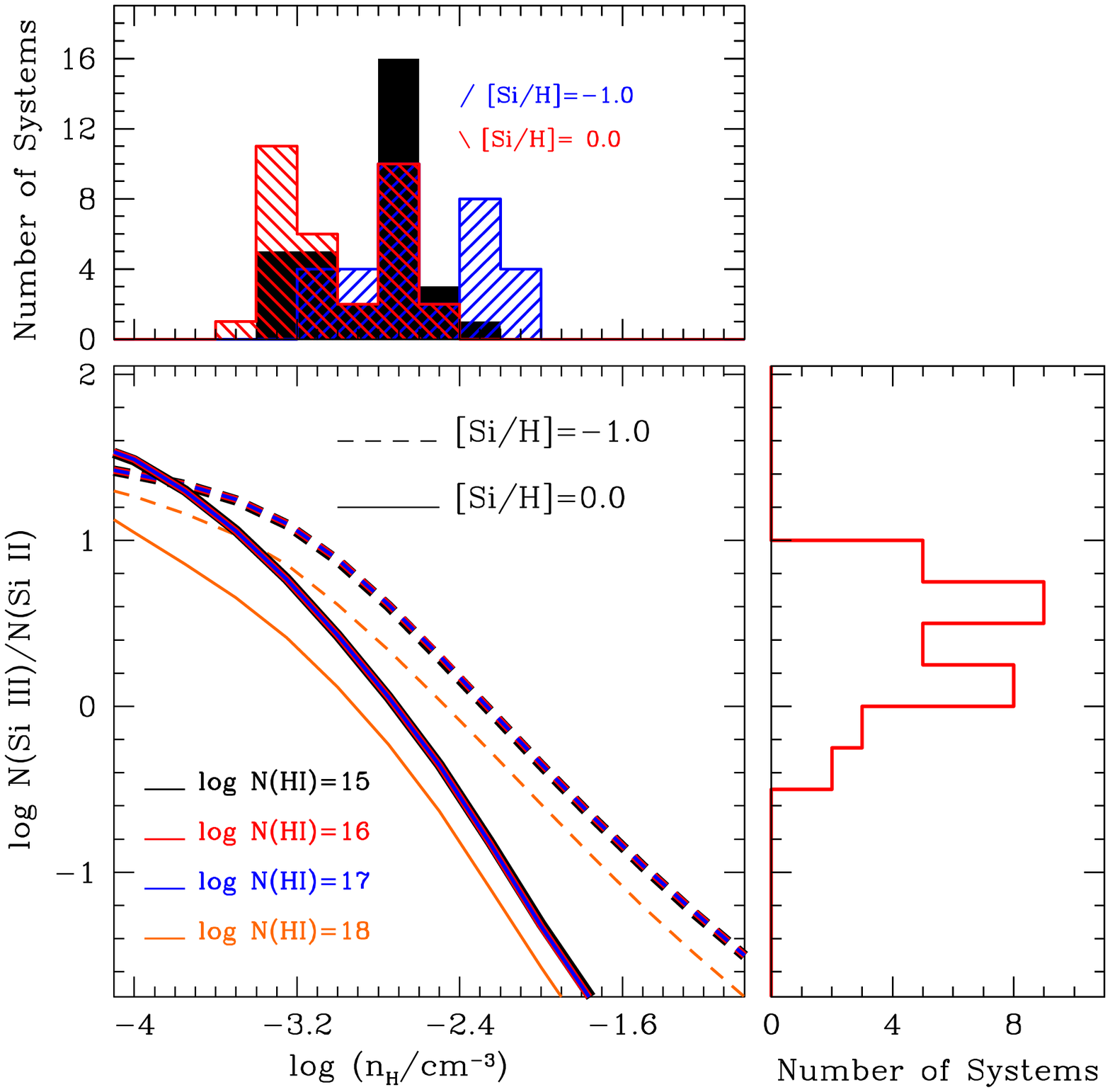}   
\vskip-0.2cm  
\caption{Bottom: {\sc cloudy} predicted \SiIII\ to \SiII\ ratios against the gas density for two different metallicities as indicated by the two different line styles. For each metallicity, four different models are run corresponding to four different $N(\HI)$ values as shown by different colors. Clearly, the ratio is degenerate with respect to the assumed model metallicity. The distribution of observed $N(\SiIII)/N(\SiII)$ ratio is shown in the right panel. Top: The density distribution corresponding to the observed $N(\SiIII)/N(\SiII)$ ratio distribution for the two different model metallicities. The solid black histogram indicates the final density distribution after correcting for the metallicity degeneracies (see text).}           
\label{fig:SiIIIbySiII}        
\end{figure}

\begin{figure*} 
\includegraphics[width=0.33\textwidth]{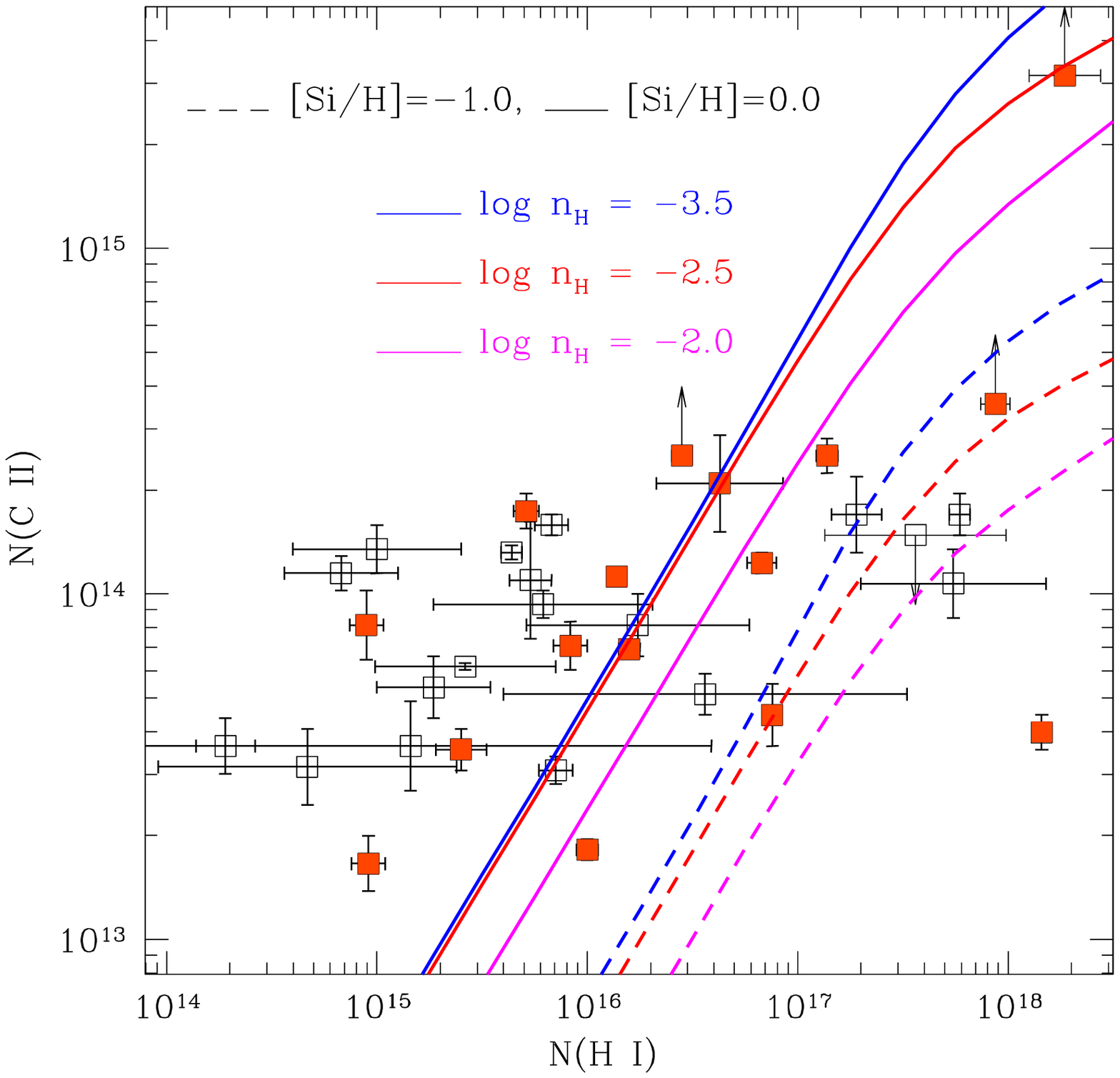}   
\includegraphics[width=0.33\textwidth]{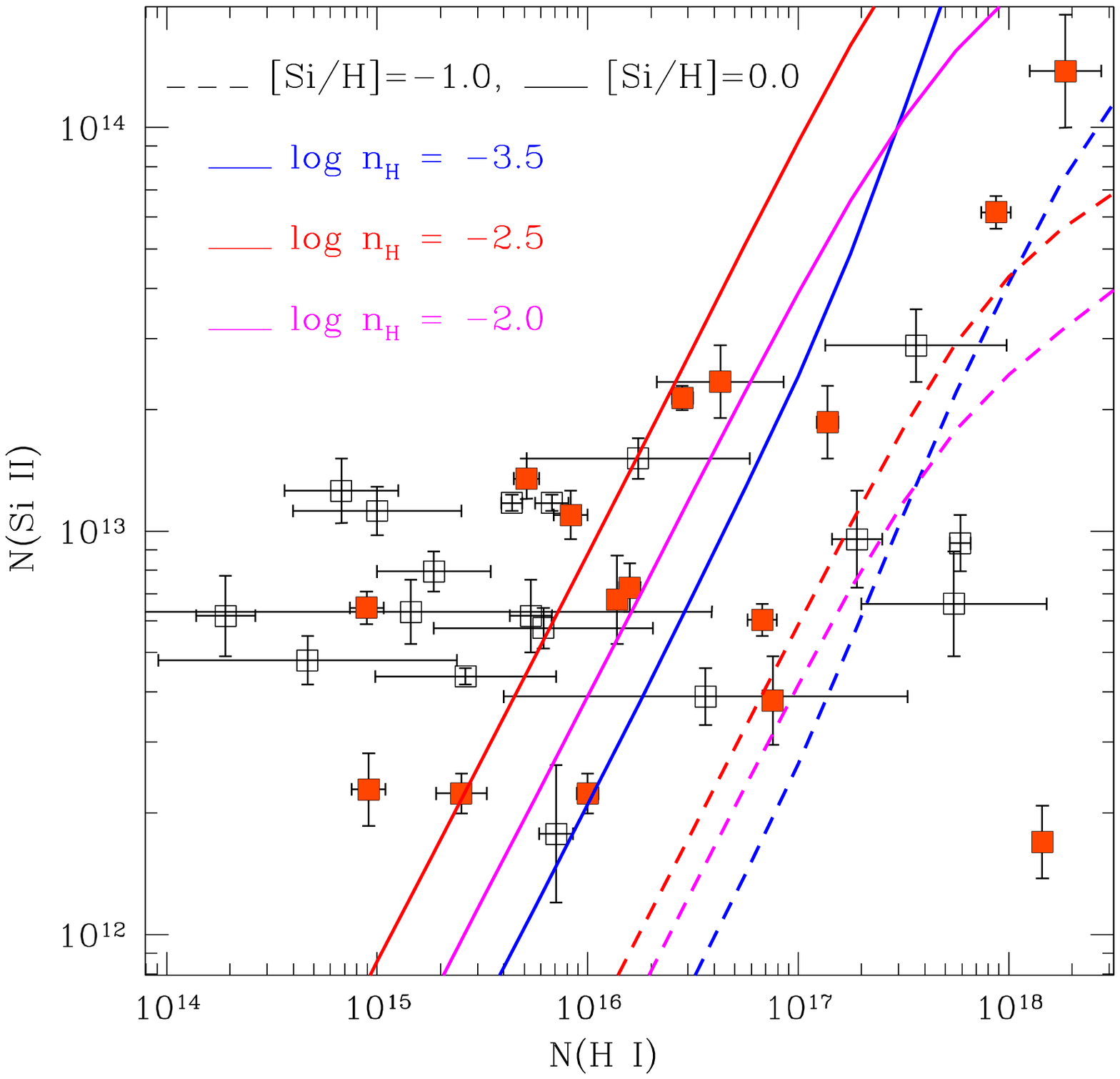}   
\includegraphics[width=0.33\textwidth]{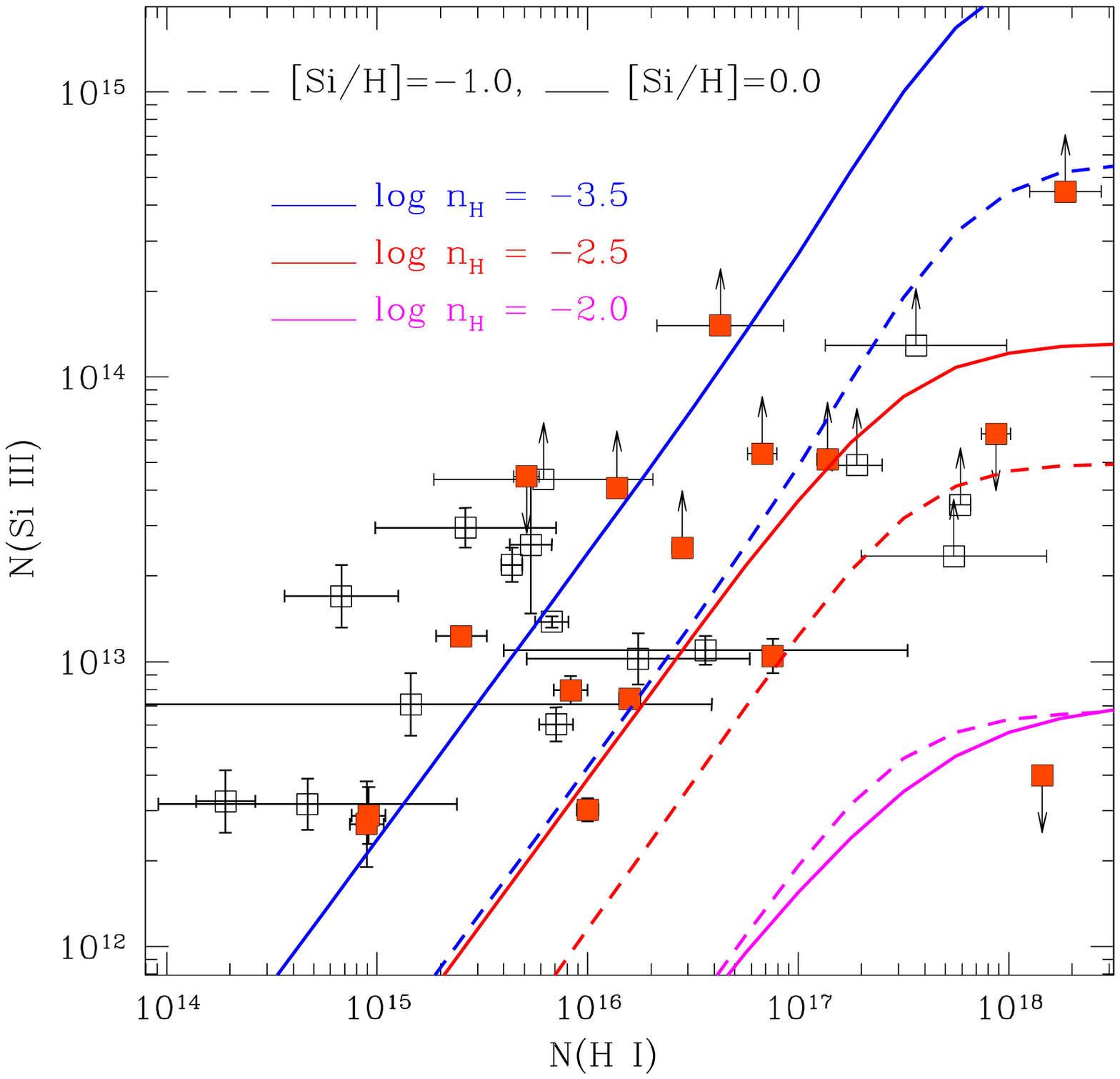}   
\vskip-0.2cm  
\caption{The \CII\ (left), \SiII\ (middle), and \SiIII\ (right) column densities against $N(\HI)$, as measured in different systems. The filled squares represent systems with $Q\ge3$. The PI model curves corresponding to two different metallicities are shown in two different line-styles. For each model, we show three curves in three different colors corresponding to three representative densities (i.e., $\log n_{\rm H} = -2.0, -2.5, \& -3.5$).}   
\label{fig:Nmetals-NHI}      
\end{figure*}

\subsection{Metal Column Densities}  

Besides \HI, we have measured the column densities of \CII, \SiII, \SiIII, \CIV\ and \OVI\ when available. All the measured total column densities (i.e. the sum of the component column densities) are listed in Table~\ref{tab:summary}. The systems in our sample are weak by design and therefore it is expected that the \CII\ and \SiII\ lines are not saturated in most cases. In fact, for 29/34 cases we could measure the $N(\CII)$ adequately. The minimum and maximum values of $\log N(\CII)/\rm cm^{-2}$ for these systems are 13.22 and 14.40, respectively, with a median value of 13.9. The median value increases only by $\approx0.1$ dex for the full sample. For all 34 systems we could measure $N(\SiII)$ reliably. The minimum and maximum values of $\log N(\SiII)/\rm cm^{-2}$ are 12.23 and 14.14, respectively, with a median value of 12.9. The median value of $N(\SiII)$ is an order of magnitude lower than that of $N(\CII)$.  

Owing to the large $f_{\rm osc}\lambda_{\rm rest}$ value (i.e. 2027) of the \SiIII$\lambda1206$ transition, the \SiIII\ absorption lines in our sample are saturated for nearly half (15/32) of the systems. We note that for two systems (i.e. \zabs~$=0.06376$ towards UKS-0242-724 and \zabs~$=0.07407$ towards SDSSJ1214+0825) \SiIII\ lines fall in the spectral gap and hence no $N(\SiIII)$ information is available. The minimum and maximum values of $\log N(\SiIII)/\rm cm^{-2}$ are, respectively, 12.43 and 13.47 for the 18 systems in which we could measure $N(\SiIII)$ adequately. The median value of the $N(\SiIII)$ distribution is $10^{13.0}$~\sqcm, which is close to that of the $N(\SiII)$ distribution but a factor of $\approx10$ smaller than the $N(\CII)$ distribution. The median value of $\log N(\SiIII)/\rm cm^{-2}$ increases to $\approx13.2$ for the full sample.     

In the bottom panel of Fig.~\ref{fig:metal-metal}, the \CII\ column densities are plotted against $N(\SiII)$. Note that when multiple components are present, we have summed the component column densities. The $N(\CII)$ values are tightly correlated with $N(\SiII)$. A Spearman rank correlation test gives $\rho_s=0.86$ with a $4.9\sigma$ significance. If we consider only systems in which both $N(\CII)$ and $N(\SiII)$ are measured adequately (i.e. excluding the limits) we obtain $\rho_s=0.81$ with a $4.4\sigma$ significance. The strong correlation coefficient ensures that the \CII\ and \SiII\ arise from the same phase of the absorbing gas.     

In the top panel of Fig.~\ref{fig:metal-metal}, the \SiIII\ column densities are plotted against $N(\SiII)$. A statistically significant correlation between $N(\SiIII)$ and $N(\SiII)$, with $\rho_s = 0.63$ ($3.5\sigma$), is apparent from the figure. Nevertheless, there is more scatter in this plot compared to the bottom panel ($N(\CII)$ vs $N(\SiII)$). It is likely due to the fact that some amount of \SiIII\ might be contributed by the high-ionization gas phase giving rise to \CIV/\OVI\ absorption for at least some fraction of the systems. We will discuss this issue in more detail in the following section.

\subsection{Ionization Modeling}

We use the photoionization (PI) simulation code {\sc cloudy} \citep[v13.03, last described by][]{Ferland13} to infer the overall physical and chemical properties of the weak absorbers in our sample. In our constant density PI models, we have assumed that (a) the absorbing gas has a plane parallel geometry, (b) the gas is subject to the extragalactic UV background (UVB) radiation at $z=0.1$ as computed by \citet[KS15 hereafter]{Khaire15a}\footnote{Ideally one should use different UVBs corresponding to the \zabs\ of the different absorbers. Since the majority of the systems in our sample have \zabs\ in the range 0.0--0.2 (only one system at \zabs~$>0.3$) with a median \zabs\ of $\approx0.1$, we have used the UVB at $z=0.1$ for all the absorbers. We have found that the use of $z =0.1$ UVB for all the absorbers could lead to a maximum uncertainty in the ionization parameter, $\log U$, of $\approx0.2$ dex, where $\log U= \log n_{\gamma}-\log n_{\rm H}$, is the ratio of number density of \HI\ ionizing photons to the number density of protons.}, (c) the relative abundances of heavy elements in the absorbing gas are similar to the solar values as measured by \citet{Asplund09}, and (d) the gas is dust free. Models are run for two different gas-phase metallicities, i.e., $\rm[X/H] =-1.0$ and $\rm [X/H] = 0.0$. The KS15 UVB calculations make use of an updated QSO emissivity and star-formation rate density \citep[see][for details]{Khaire15b}. Their models with escape fraction (of \HI\ ionizing photons from galaxies, $f_{\rm esc}$) of $0\%$ provide \HI\ photoionization rates ($\Gamma_{\HI}$) at $z<0.5$ that are consistent with the measurements of \citet[]{Shull15} and \citet[]{Gaikwad16}. We therefore use their model corresponding to $f_{\rm esc}=0\%$. 

PI models for all the absorbers are done in a uniform fashion. First, we assume that the \SiII\ and \SiIII\ absorption lines originate in the same phase of the absorbing gas and hence the ratio of $N(\SiIII)/N(\SiII)$ uniquely fixes the $\log U$ (or $\log n_{\rm H}$) of the absorber. We, however, note that the \SiIII\ need not entirely stem from the low-ionization gas phase that produces \SiII. If some amount of \SiIII\ arise from high-ionization gas phase, the \SiIII\ to \SiII\ ratio would provide an upper limit on the ionization parameter (a lower limit on density). Next, we use the ionization corrections of \HI\ and \SiII\ ($f_{\HI}$ and $f_{\SiII}$, respectively) at the derived $n_{\rm H}$ and calculate the Si abundance, [Si/H] $=  \log N(\SiII)/N(\HI) - \log f_{\SiII}/f_{\HI} - \log {\rm (Si/H)_{\odot}}$, of the absorber using the observed $N(\HI)$ and $N(\SiII)$. The $f_{\SiII}/f_{\HI}$ ratio shows a relatively flat peak for a large range in density of $\rm 10^{-3.4}-10^{-2.4} cm^{-3}$, and falls off rapidly at both below and above this density range. Thus, even if the measured $N(\SiIII)$ has some contribution from a high ionization gas phase, our inferred metallicity will essentially be unaltered (see also Section~\ref{sec:dissmetal}).       

\section{Results based on the PI models}    
\label{sec:result}  

\subsection{The density (ionization parameter) distribution}     

In the bottom panel of Fig.~\ref{fig:SiIIIbySiII} we show the variation of the model predicted $N(\SiIII)/N(\SiII)$ ratio with density for two different model metallicities, i.e., solar and 1/10th of solar values. We note that the $N(\SiIII)/N(\SiII)$ ratio depends on the metallicity assumed in the {\sc cloudy} model. This is likely due to the increased electron density in high metallicity gas increasing the recombination/cooling rate for a fixed density and temperature. For each model metallicity we have used four different $\log N(\HI)/\rm cm^{-2}$ values (i.e., 15, 16, 17, and 18) as the stopping conditions for the simulations. Note that, for a given metallicity, the ratio is independent of the $N(\HI)$ values as long as the gas is optically thin to the hydrogen ionizing photons (i.e., $N(\HI)\lesssim10^{17.2}$~\sqcm). We, however, used the observed $N(\HI)$ values as stopping conditions for {\sc cloudy} for each of the absorbers in order to derive individual constraints.     

\newpage 
\input{tables/summaryall.tex}    
\clearpage 

The red (135$^{\degree}$) hashed histogram in the top panel of Fig.~\ref{fig:SiIIIbySiII} represents the $\log n_{\rm H}$ distribution corresponding to the observed $N(\SiIII)/N(\SiII)$ ratios, shown in the bottom-right panel, for an assumed metallicity of $\rm [X/H] =0.0$. The blue (45$^{\degree}$) hashed histogram, on the other hand, represents the same for an assumed metallicity of $\rm [X/H] = -1.0$. Clearly, an assumption of any particular model metallicity for all the absorbers in a sample leads to an incorrect density distribution\footnote{We, however, note that the effect is less important for the derived metal abundances.}. In order to overcome this problem we have performed our PI modelling in an iterative way. First, we used the density solution corresponding to the $\rm [X/H] = -1.0$ model and calculate the Si abundances for each absorber. We then select the systems with $\rm [Si/H] > -0.3$ (i.e. half solar) and redo the density calculation using the $\rm [X/H] = 0.0$ model. The distribution of our final density solutions is shown by the black histogram in the top panel of Fig.~\ref{fig:SiIIIbySiII}. The $\log n_{\rm H}/\rm cm^{-3}$ distribution has a median value of $-2.8$ with minimum and maximum values of $-3.3$ and $-2.4$, respectively. For the adopted UVB, this leads to a range in ionization parameter, $\log U$, of $-2.7$ to $-3.6$ with a median value of $-3.2$.    

\subsection{The column density trends}     

In the different panels of Fig.~\ref{fig:Nmetals-NHI}, column densities of low-ionization metal lines are plotted against the corresponding $N(\HI)$. A mild trend of increasing low-ionization metal line column densities with increasing $N(\HI)$ is noticeable. A similar trend has also been reported for the COS-Halos sample \citep[i.e.,][]{Werk13,Prochaska17}. A Spearman rank correlation analysis, assuming limits are detections, also confirms the trends with a $\rho_s\gtrsim0.3$ when the full sample is considered (see Table~\ref{tab:spear}). It is important to note that when we consider the sub-sample with $N(\HI)>10^{16}$~\sqcm\ (median value), the correlation coefficients remain high (i.e. $\rho_s\gtrsim0.2$). The sub-sample with $N(\HI)\le10^{16}$~\sqcm, on the contrary, show $\rho_s$ consistent with 0.0. Though the sample size is small, it seems that the mild trends seen in the full sample are mainly driven by the systems with $N(\HI)>10^{16}$~\sqcm.            

In each panel in Fig.~\ref{fig:Nmetals-NHI}, six different PI model curves, corresponding to two different metallicities and three different densities, are overplotted. Note that the three representative densities span the density range we obtained from the $N(\SiIII)/N(\SiII)$ ratios as discussed above. It is apparent from the three panels that the majority of the systems with $\log N(\HI)/\rm cm^{-2}<16.0$ are not reproduced by the models with $1/10$th of solar metallicity. Moreover, even the solar-metallicity-models fail to reproduce the systems with $\log N(\HI)/\rm cm^{-2}<15.7$. It is, therefore, clear that the vast majority of the weak absorbers studied here have very high metallicities.  

\subsection{Trends in the model parameters}

The PI model predicted densities ($n_{\rm H}$), Si-abundances ($\rm [Si/H]$), total hydrogen column densities ($N_{\rm H}$), and line of sight thicknesses ($L$) are listed in columns 8, 9, 10, and 11 of Table~\ref{tab:summary}, respectively. As mentioned before, the densities of the absorbers show a range of $10^{-3.3}$--$10^{-2.4}$ $\rm cm^{-3}$ with a median value of $10^{-2.8}$~$\rm cm^{-3}$. The density range and the median value do not change when only $Q>2$ systems are considered. The inferred $\rm [Si/H]$ values show a range of $-2.5$ to $+1.6$ with a median value of $0.0$ (i.e. solar abundance). However, when we considered only $Q>2$ systems the maximum value drops to $\rm [Si/H] = 0.9$ (i.e. $\approx8$ times solar) and the median value becomes $-0.1$. The total hydrogen column densities for the weak absorbers show a range of 4 orders of magnitude, i.e., $N_{\rm H}=10^{16.0-20.3}$~\sqcm, with a median value of $10^{18.1}$~\sqcm. The line of sight thicknesses, $L$, also exhibit a wide range of 1~pc -- 53~kpc with a median value of $\approx500$~pc. The median value changes to $\approx780$~pc for the systems with $Q>2$.  

\begin{figure*} 
\vspace*{+0.5cm}
\hspace*{+0.5cm}
\includegraphics[width=0.70\textwidth,angle=90]{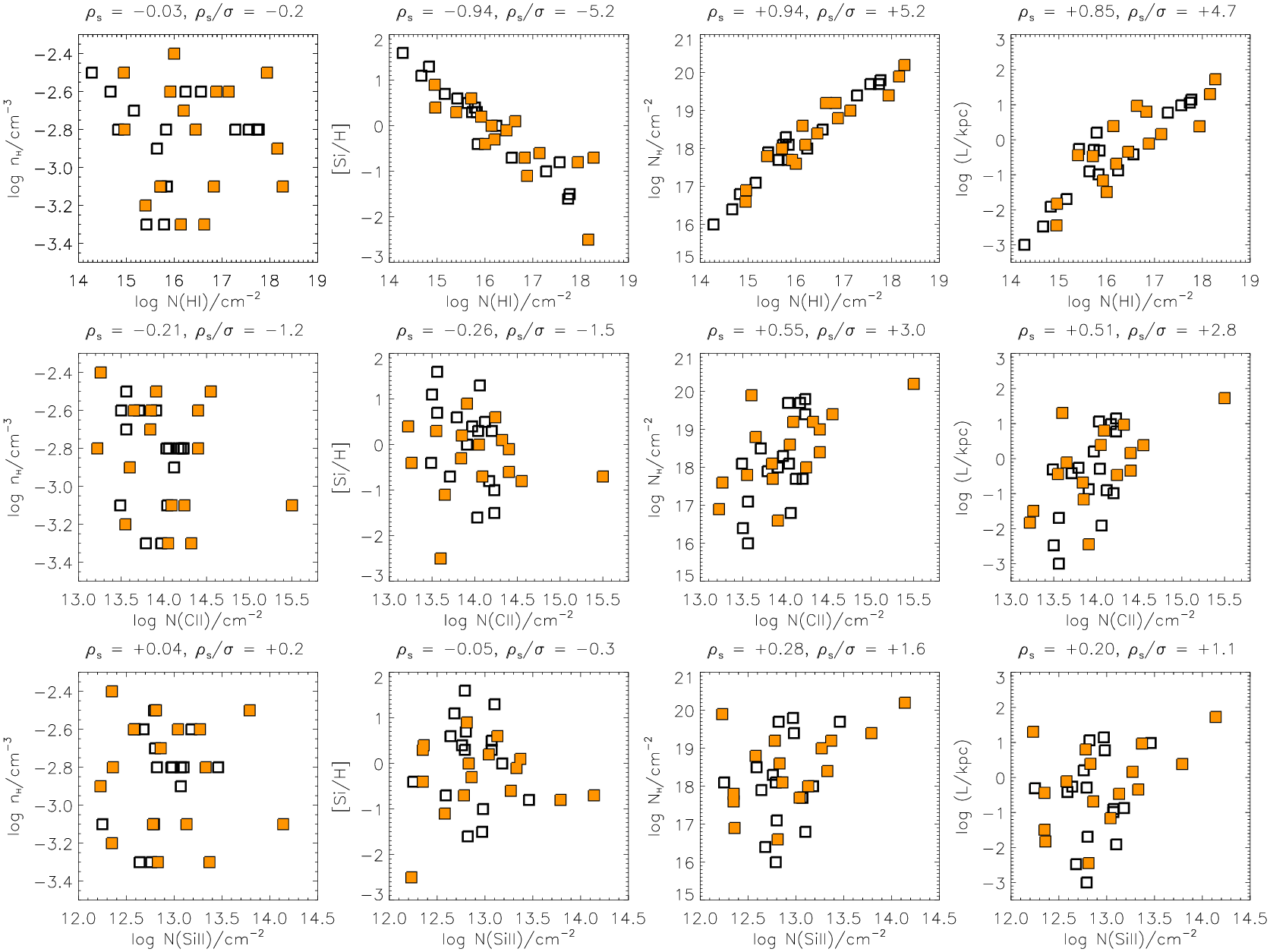}   
\vskip-0.1cm  
\caption{Scatter plots of PI model parameters ($\log n_{\rm H}$, 1st column; $\rm [Si/H]$, 2nd column, $\log N_{\rm H}$, 3rd column; $\log L$, 4th column) and the observed column densities ($N(\HI)$, top row; $N(\CII)$, middle row; $N(\SiII)$, bottom row). The filled squares represent the systems with $Q>2$. The results of the Spearman rank correlation analysis are mentioned on the top of each panel. Possible origins of the observed correlations are discussed in the text. In particular, we argue that the $5.2\sigma$ anti-correlation seen between $\rm [Si/H]$ and $\log N(\HI)$ {\em cannot} be entirely attributed to selection bias.}            
\label{fig:model-param}      
\end{figure*}

The PI model parameters are plotted against the observed column densities of \HI\ (top row), \CII\ (middle row), and \SiII\ (bottom row) in Fig.~\ref{fig:model-param}. The inferred densities do not show any trend with the neutral hydrogen and/or low-ionization metal line column densities. No significant correlation is seen between $\rm [Si/H]$ and $N(\CII)$ or $N(\SiII)$. A strong anti-correlation ($\rho_s=-0.94$, $\rho_s/\sigma=-5.2$) between $\rm [Si/H]$ and $N(\HI)$, however, is apparent from the figure. The trend remains statistically significant ($\rho_s=-0.89$, $\rho_s/\sigma=-3.5$) when only systems with $Q>2$ are considered. A similar trend has been recently reported by \citet[]{Prochaska17} in the COS-Halos sample. Such a trend is generally expected in a metal line selected sample of absorbers, because, in practice, there are many low-$N(\HI)$ systems without any detectable low-ionization metal lines that are not considered in our study. In order to investigate the dependence of the correlation on $N(\HI)$, we have performed the Spearman rank correlation analysis for two different sub-samples, one with $N(\HI)>10^{16}$~\sqcm\ and the other with $N(\HI)\le10^{16}$~\sqcm, separately. We obtain $\rho_s = -0.75$ and $\rho_s/\sigma =-2.9$ ($\rho_s = -0.89$ and $\rho_s/\sigma = -3.5$) for the upper-$N(\HI)$ (lower-$N(\HI)$) sub-samples, suggesting a statistically significant anti-correlation between $\rm [Si/H]$ and $N(\HI)$ for the systems with $N(\HI)>10^{16}$~\sqcm. Note that, $>$80\% of the low-$z$ \HI\ absorbers with $N(\HI)>10^{16}$~\sqcm\ exhibit low-ionization metal lines \citep[see Fig. 8 in][]{Danforth16}. Thus, the strong anti-correlation seen in the full sample {\em cannot} be fully attributed to sample selection bias \citep[see also the discussion of][]{Misawa08}. However, we do point out that our survey is limited to weak absorbers only. Inclusion of strong absorbers would introduce more scatter in the relation, particularly at the higher $N(\HI)$ end.            

The strong $5.2\sigma$ correlation between $N_{\rm H}$ and $N(\HI)$ is not surprising since the expected variation in $f_{\HI}$ over the inferred narrow density range of $10^{-3.3}-10^{-2.4}$~cm$^{-3}$ is small and $N_{\rm H}$ ($\equiv N(\HI)/f_{\HI}$) is directly proportional to $N(\HI)$. The $4.7\sigma$ correlation between $L~(\equiv N_{\rm H}/n_{\rm H})$ and $N(\HI)$ follows from the fact that $N_{\rm H}$ is strongly correlated with $N(\HI)$ whereas $n_{\rm H}$ is not.     
The $N_{\rm H}$ also show a significant (mild) correlation with $N(\CII)$ ($N(\SiII)$). If the column density of the (i+1)-th ionization state of an element $\rm X$, with ionization fraction of $f_{{\rm X_i}}$, is $N({\rm X_i})$, then     
\begin{eqnarray}  
N_{\rm H} & = & \frac{N({\rm X_i})}{f_{{\rm X_i}} 10^{\rm [X/H]} {\rm (X/H)_{\odot}}}~, {\rm and} \\  
L & = & \frac{N({\rm X_i})}{f_{{\rm X_i}} 10^{\rm [X/H]} {\rm (X/H)_{\odot}}  n_{\rm H}}~. 
\end{eqnarray}    
Thus, a correlation between $N_{\rm H}$ (and/or $L$) and metal line column density is generally anticipated. Any anti-correlation between the metallicity ($\rm [X/H]$) and the ionic column density, as seen for \CII, would further strengthen the correlation. The apparent lack of any correlation between $\rm [Si/H]$ and $N(\SiII)$ has been manifested by the mild $1.6\sigma$ correlation seen between $N_{\rm H}$ and $N(\SiII)$ in contrast to the strong $3.0\sigma$ correlation seen between $N_{\rm H}$ and $N({\CII})$.

\begin{table}  
\begin{threeparttable}[b] 
\caption{The details of Spearman rank correlation analysis} 
\begin{tabular}{cccrr} 
\hline \hline 
Sample & Data1 & Data2 & $\rho_s$  & $\rho_s/\sigma$ \\ 
\hline  
Full 	        	 & $N(\HI)$  &  $N(\CII)$   & 0.48 & 2.8  \\ 
Full     		 & $N(\HI)$  &  $N(\SiII)$  & 0.28 & 1.6  \\ 
\hline 
$N(\HI)>10^{16}$~\sqcm\   & $N(\HI)$  &  $N(\CII)$   &  0.30 & 1.2  \\  
$N(\HI)>10^{16}$~\sqcm\   & $N(\HI)$  &  $N(\SiII)$  &  0.20 & 0.8  \\  
\hline 
$N(\HI)\le10^{16}$~\sqcm\  & $N(\HI)$  &  $N(\CII)$   &  0.10   & 0.4 \\  
$N(\HI)\le10^{16}$~\sqcm\  & $N(\HI)$  &  $N(\SiII)$  & $-0.12$ & $-0.5$ \\  
\hline 
\end{tabular} 
\label{tab:spear}  
\end{threeparttable}  
\end{table} 

\subsection{High-ionization metal lines}  
\label{sec:highions} 

Although the main focus of this paper is the weak absorbers detected via the low-ionization metal lines, in all but a few cases high-ionization metal lines (e.g. \CIV\ and/or \OVI) are also covered. For example, $N(\CIV)$ information is available for a total of 23 systems (15 measurements and 8 upper limits, see Table~\ref{tab:summary}). All the upper limits on $N(\CIV)$ are consistent with our PI model predictions. For the 15 systems with positive detection of \CIV, the column densities are found to be in the range $10^{13.2-15.0}$~\sqcm\ with a median value of $10^{14.1}$~\sqcm. Our $N(\CIV)$ measurements are at least 0.5~dex higher than the model predicted values for 12/15 systems, suggesting a separate gas phase for \CIV\ absorption in those absorbers. For the \zabs~$=0.00369$ system towards IRAS-F04250-5718, the difference is only 0.2~dex, which is well within our PI model uncertainties. For the two other systems (\zabs~$=0.00575$ towards RXJ1230.8+0115 and \zabs~$=0.09784$ towards SDSSJ1357+1704) the model predicted \CIV\ column densities are higher than the observed ones. In both these systems the \SiIV\ line is detected. It is likely that the majority of the \SiIII\ absorption in these systems arises from the gas phase giving rise to \SiIV\ (and \CIV) absorption. In that situation the ionization parameter of the low-ionization phase can be a lot lower which, in turn, will reduce the predicted $N(\CIV)$. In the \zabs~$=0.09784$ system towards SDSSJ1357+1704, the stronger \SiIII\ component at $\approx0$~\kms, which dominates the total $N(\SiIII)$, is not seen in the low-ionization lines but is seen in the \SiIV\ line (see Appendix~\ref{appendix:vplots}). We further note that the majority of systems for which the predicted $N(\CIV)$ values are considerable (i.e. $>10^{14.0}$~\sqcm) show strong \SiIV\ absorption. These facts demonstrate the need for component-by-component, multi-phase photoionization models \citep[see e.g.,][]{Charlton03,Milutinovic06,Misawa08}, which is beyond the scope of this paper.  

The \OVI\ doublet is covered for 18/34 of the weak absorbers, of which 12 show detectable absorption. The \OVI\ column densities in these absorbers vary between $10^{13.8-14.7}$~\sqcm\ with a median value of $10^{14.4}$~\sqcm. The maximum predicted $N(\OVI)$ by our PI models (i.e., $10^{13.2}$~\sqcm) is 0.6~dex lower than the lowest measured $N(\OVI)$ value. It clearly indicates that in all these systems \OVI\ must originate from a separate gas phase. 
In 10/34 of the weak absorbers both the \CIV\ and the \OVI\ doublets are covered. In 6/10 cases both \CIV\ and \OVI\ are detected, and it is likely that they arise in the same gas phase. In one case (\zabs~$=$~0.09784 towards SDSSJ1357+1704 discussed in the paragraph above) only \CIV\ is detected. Finally, in 3/10 cases neither \CIV\ or \OVI\ are detected. We conclude that about two thirds of the weak absorbers have a separate, higher ionization or hotter phase.

\section{Discussion}    
\label{sec:diss}

\subsection{Redshift Evolution} 
\label{sec:dissred}

Our sample of 34 weak \MgII\ absorber analogs at $z<0.3$ represents a substantial increase over the sample of 6 absorbers found by \citet{Narayanan05} at low redshift.  However, the value of $dN/dz =0.8\pm0.2$ that we have derived is consistent with that earlier study, and a factor of a few smaller than what would be expected if static populations of weak \MgII\ and \CIV\ absorbers were simply evolved to the present era subject to the decreasing UVB. \citet{Narayanan05} noted that both pc-scale structures that produce weak \MgII\ absorption at $z\approx1$ and kpc-scale structures that are detected in \CIV\ absorption, but not in low ionization transitions, would evolve from $z\approx1$ to produce weak low ionization absorption at $z\approx0.1$. To some extent, both populations must be less abundant at low redshift than they were at $z\approx1$. The present sample of 34 absorbers is large enough that we can attempt to determine how the low redshift population compares to the $z\approx1$ population, however \citet{Narayanan05} warned that ``hidden phases'' would in some cases make it impossible to extract accurate physical properties (see their Figs. 10--12).    

We begin by noting that 7/34 of our weak absorbers are very small ($<32$~pc) structures with solar or super-solar metallicity and with derived densities $\log n_{\rm H}/\rm cm^{-3} \approx -2.6$ in the upper half of our sample. Although these overlap with the densities derived by \citet{Narayanan08} for weak {\MgII} absorbers (using coverage of {\FeII} to constrain density), they are at the low end of the range. All but one of these seven small absorbers has only one low ionization component, and \CIV\ is either not detected at all (in 4/7) or it could be in the same phase with, and centered on the low ionization absorption. Even these absorbers, for which the metallicity is inferred to be quite high, could have two phases that contribute roughly equally to the \SiII/\CII\ absorption, as in the simulated system in Fig. 11 of \citet{Narayanan05}.  However, it is still clear that there is a smaller, higher density phase which has a surprisingly high metallicity.  

Furthermore, many of the remaining 27/34 absorbers, though they have larger derived line of sight thicknesses based on our conservative estimates, are likely to have a hidden phase. The inferred density of that phase would be higher than we estimate if some of the \SiIII\ is in fact in a lower density phase that produced \CIV\ absorption. The inferred metallicity of that phase would be higher if some of the \HI\ is associated with the higher ionization transitions, \CIV\ and/or \OVI.  

Thus, despite our larger sample, the situation remains unclear. We can, however, affirm that there are fewer weak, low ionization absorbers at present than we would expect if we took the absorber population at $z\approx1$ and simply evolved the same structures to the present day subject to a lower UVB. However, as we discuss below, it seems more likely that the weak absorbers are transient, and that the processes that create these small, high metallicity structures faraway from galaxies are less active at present than they were in the past.

\subsection{Cosmological Importance}   
\label{sec:cosmo}    

If the comoving number density of the weak absorbers at $z\approx0.1$ is $n_{cl}$ and the proper cross-section is typically $\pi R_{cl}^2$ then $d\mathcal{N}/dz$ can be expressed as: 

\begin{equation} 
\frac{d\mathcal{N}}{dz} = n_{cl} \pi R_{cl}^2 \frac{c}{H_0} \frac{(1+z)^2}{\sqrt{(1+z)^3\Omega_{\rm M}+\Omega_\Lambda}}~.  
\end{equation}  
If we assume a spherical geometry for the weak absorbers with a characteristic radius of $R_{cl}\equiv L/2$, where $L\approx500$~pc is the median line of sight thickness we obtained from PI model, then the comoving number density becomes $n_{cl} \sim 8\times10^{2}$ Mpc$^{-3}$ for the observed $d\mathcal{N}/dz$ of $\approx0.8$. Integrating the $r$-band luminosity function of $z=0.1$ galaxies in SDSS \citep[]{Blanton03}, the number density of galaxies down to $0.01L_*$ is $n_{\rm gal}\sim2\times10^{-2}h^3$ Mpc$^{-3}$. Therefore, the population of weak absorber clouds must have been huge, and outnumbered bright galaxies by tens of thousands to one. Similar conclusions have been drawn for the weak absorber population at high-$z$ \citep[]{Rigby02}, for metal-rich \CIV\ absorbers at $z\approx2.3$ \citep[]{Schaye07}, and for \NeVIII\ absorbers at $z\approx0.7$ \citep[]{Meiring13}.  

Under the assumption of spherical geometry, the cloud mass can be expressed as: 
\begin{equation} 
M_{cl} = \frac{4\pi}{3}~ R_{cl}^3~ n_{\rm H}~ \mu m_{p}~,   
\end{equation}                    
where, $\mu\approx1.4$ is the mean atomic weight. Using the median $L$ and $n_{\rm H}$ values we derived from the PI models, we obtain $<M_{cl}> \sim10^{3}$~\Msun. The cosmic mass density of the weak absorbers at $z\approx0.1$ is then $\Omega \equiv n_{cl}M_{cl}/\rho_{cr} \sim10^{-6}$, where $\rho_{cr}\approx2\times10^{11}$ \Msun~Mpc$^{-3}$ is the critical density of the universe. Clearly, the weak absorbers at $z\approx0.1$ carry a negligible fraction of cosmic baryons ($\Omega_b\approx0.04$).           

If the weak absorber clouds are associated with the CGM of a galaxy population with comoving number density $n_{\rm gal}$ at $z\approx0.1$, then the derived $d\mathcal{N}/dz$ can, alternatively, be used to estimate the characteristic radius of the halo using the following relation:  
\begin{equation} 
R_h \approx  {\rm 130~kpc} \left(\frac{d\mathcal{N}/dz}{0.8}\right)^{1/2} \left(\frac{n_{\rm gal}}{10^{-2}\rm Mpc^{-3}}\right)^{-1/2} C_f^{-1/2}~.   
\label{eqn:halo}
\end{equation}  
Here $C_f$ is the covering fraction of the weak absorbers. Note that the halo radius can be much larger if the covering fraction of the weak absorbers is significantly lower than unity. Following the prescription detailed in Section~5 of \cite{Richter16b}, Eqn.~\ref{eqn:halo} can inversely be used to obtain $C_f$. Assuming that the absorber's population extends out to the virial radii of galaxies with $L/L_*>0.001$, they have derived a relation: $d\mathcal{N}/dz=3.6 C_f$. The observed $d\mathcal{N}/dz$ for weak absorbers, thus, corresponds to a covering fraction of $\approx0.3$. This is roughly half of the covering fraction derived for the \SiIII\ population studied in \cite{Richter16b}. This is expected since they have noted that about half of the \SiIII\ absorber population with $\log n_{\rm H}/\rm cm^{-3}\gtrsim-3.0$ is associated with \SiII\ absorption.       

\begin{figure} 
\includegraphics[width=0.50\textwidth]{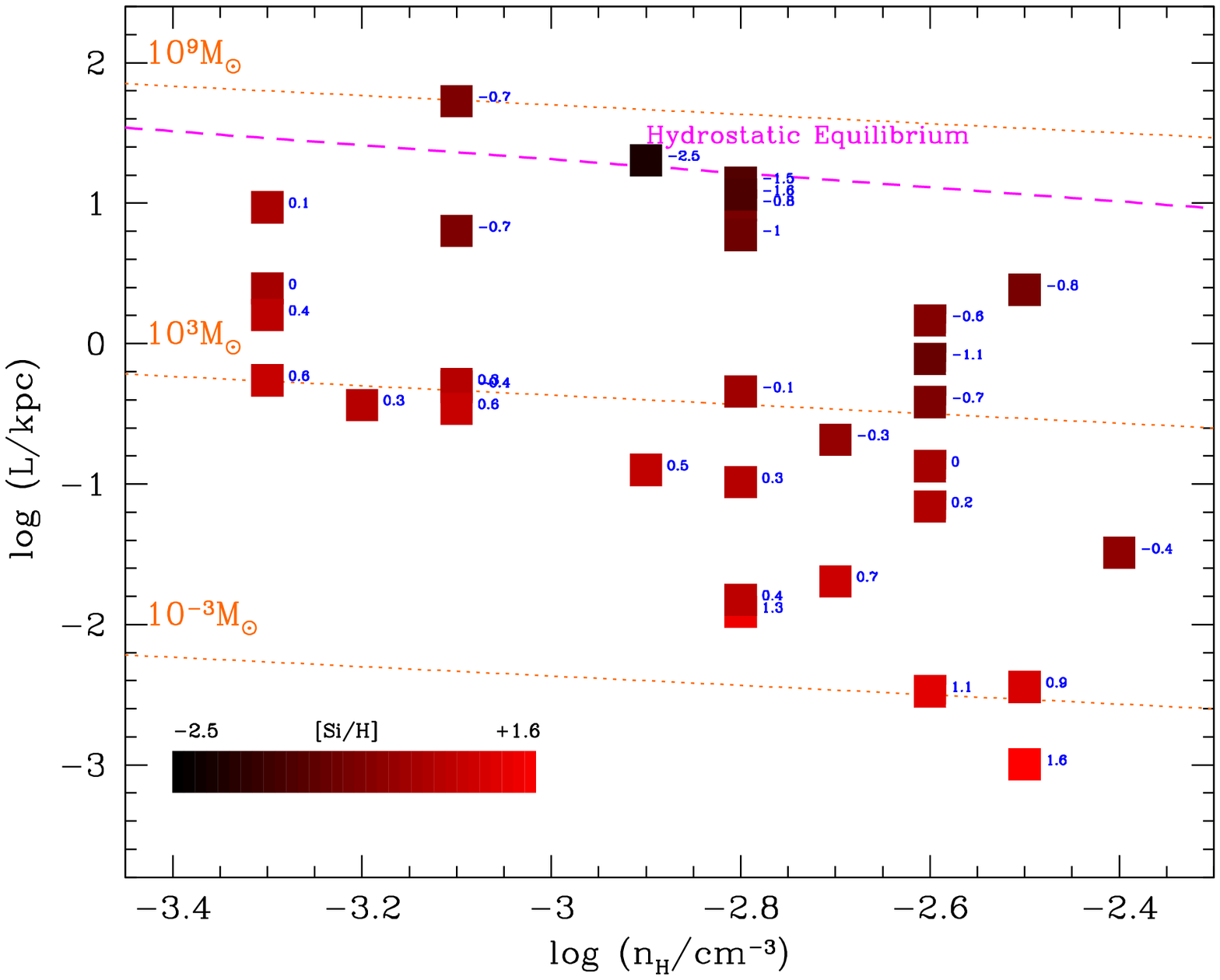}   
\vskip-0.2cm  
\caption{The line of sight thickness versus gas density as obtained from the PI models. Each data point is color-coded by the corresponding metallicity. The dashed line marks the relationship under local hydrostatic equilibrium \citep[]{Schaye01}. High metallicity systems are too small to be self-gravitating. The dotted lines indicate the lines of constant masses. The majority of the systems contain gas masses of $\lesssim10^{5}M_{\odot}$.}                
\label{fig:hydrostatic}      
\end{figure}

\begin{figure} 
\includegraphics[width=0.50\textwidth]{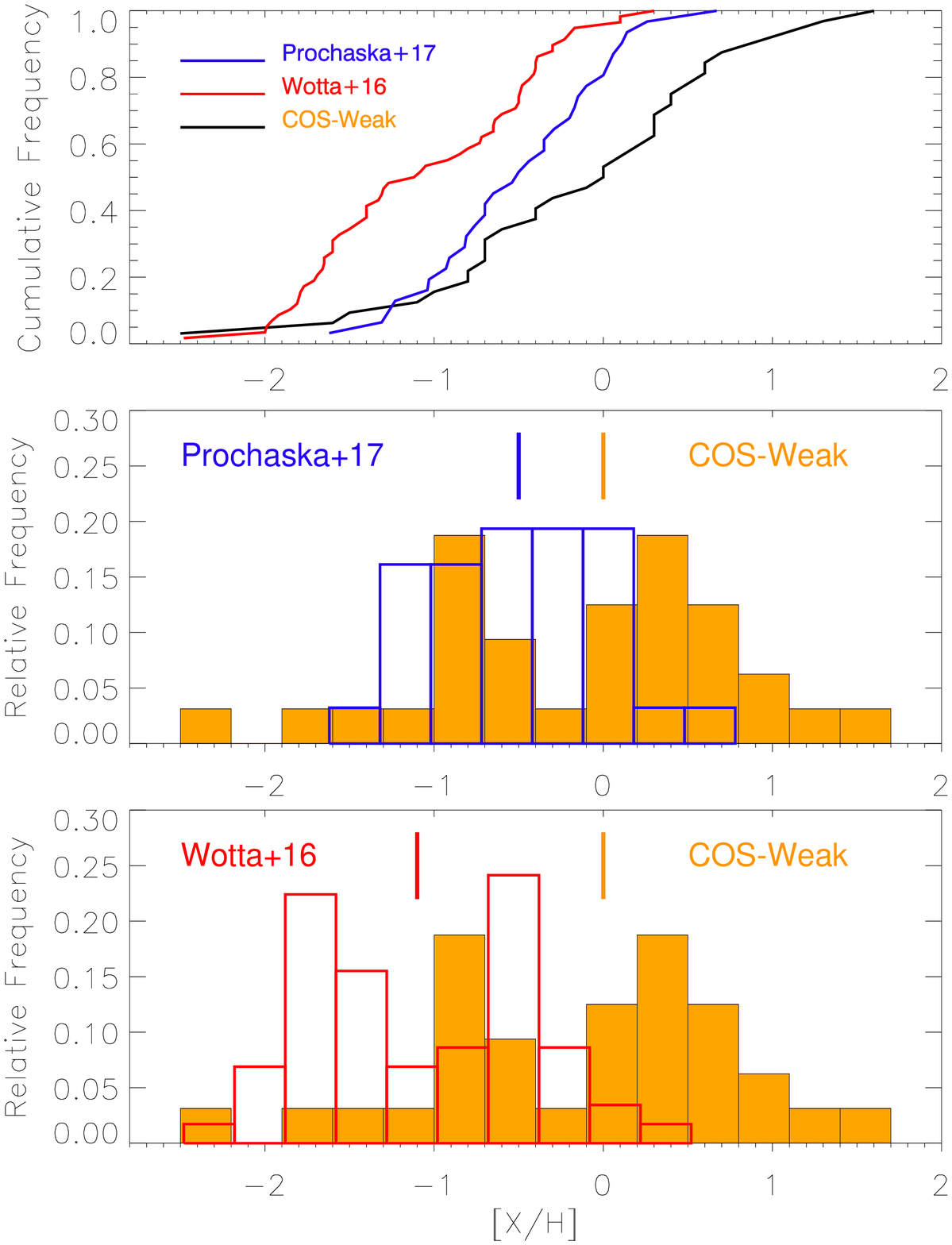}   
\vskip-0.2cm  
\caption{Bottom: Metallicity distribution of the weak absorbers in our sample (black histogram with orange shade). The red histogram in the bottom panel shows the bimodal metallicity distribution of \HI-selected low-$z$ Lyman limit systems studied by \citet{Wotta16}. The blue histogram in the middle panel shows the metallicity distribution of the COS-Halos (galaxy-selected) sample from \citet{Prochaska17}. The top panel shows the cumulative distributions of all three sample with respective colors. Both the galaxy-selected and \HI-selected absorbers show very different metallicity distributions when compared to the weak absorbers. Our sample shows the highest fraction ($\approx50$\%) of solar/super-solar metallicity absorbers.           
}           
\label{fig:logZdist}       
\end{figure}

Following \citet{Stocke13}, the number of metal-rich clouds inside the characteristic radius is $N_{cl} = C_f S (\frac{R_h}{R_{cl}})^2 \approx 3\times10^{5}$, assuming the ``shadowing factor'' ($S$) and $C_f$ to be of order unity. The volume filling factor of these clouds is then $N_{cl} (R_{cl}/R_h)^{3}\approx0.002$ (i.e. only $\approx0.2$\%). The total gas mass ($M_{\rm gas} = N_{cl}<M_{cl}>$) associated with these absorbers is $\sim 3\times10^{8}$~\Msun. This mass estimate is significantly lower compared to the COS-Halos sample \cite[i.e., $>6\times10^{10}$~\Msun;][]{Werk14} but roughly matches with the calculation of \cite{Stocke13} for their sample of dwarf galaxies. The mass in Silicon ($N_{cl}<M_{cl}>\rm <[Si/H]> (Si/H)_{\odot}$) is $\sim10^{4}$~\Msun. It corresponds to an Oxygen mass of $M_{O}\sim10^{5}$~\Msun, assuming solar relative abundance. This Oxygen mass estimate is $\approx2$ orders of magnitude lower than the estimate of \cite{Tumlinson11} for the \OVI\ absorbing gas in the COS-Halos sample. Here we note that, all our mass estimates are lower limits in the sense that we do not take the possible contribution from the high-ionization gas phase into account. 

From the discussion above, it is clear that the weak absorbers may stem from extended gaseous halos that are filled with large numbers of pc--kpc-scale clouds. Given the high metallicities of the majority of the weak absorbers, it is most natural to think that they arise in large scale galactic outflows. An important question in this context is the stability of the absorbing clouds. Self-gravity and external pressure due to an ambient medium are generally thought to be the two main channels that can confine the small, metal-rich clouds \cite[see e.g.][]{Schaye07}. The typical size for an optically thin, self-gravitating, purely gaseous cloud with density $\sim10^{-2.8\pm0.5}\rm cm^{-3}$ is $\sim$10--30~kpc \citep[see Eqn.~7 in][]{Schaye07}, whereas, the median line of sight thickness of our sample is \linebreak $\approx500$~pc. As demonstrated in Fig.~\ref{fig:hydrostatic}, a vast majority of the weak absorbers in our sample are too tiny to be supported by self-gravity unless they are significantly dark matter dominated structures.      

The presence of another gas phase, giving rise to the high-ionization metal lines, particularly the \OVI, is almost certain (see Section~\ref{sec:highions}). This phase, presumably with lower density and higher temperature, can, in principle, confine the low-ionization gas phase. However, in cases where detailed ionization modelling has been done the two phases are not found to be in pressure equilibrium \citep[e.g.][]{Meiring13,Muzahid14,Muzahid15b,Hussain15}. But the high-ionization gas phase traced by \OVI\ could arise from the mixing layers of cool and an ambient medium too hot for UV line diagnostics. This hotter unobserved gas, e.g., hot halo created via accretion shocks and/or shocks due to large-scale galactic outflows, might well be the confining medium \citep[e.g.,][]{Mulchaey09,Narayanan11}. Indeed, X-ray observations have revealed the presence of such a medium with $T>10^{6}$~K in the halo of Milky Way \citep[]{Gupta12}. In the absence of any confining agent, the clouds will expand freely until they reach pressure equilibrium or will eventually be evaporated. For a $\sim10^{4}$~K gas cloud with $R_{cl}\sim10^{2}$~pc, the free expansion time scale is $R_{cl}/c_{s} \sim10^{7}$~yr, much shorter than the Hubble time. This suggests that the high metallicity weak absorbers could be transient in nature.       

\begin{figure*} 
\vbox{\hbox{
\includegraphics[width=0.50\textwidth]{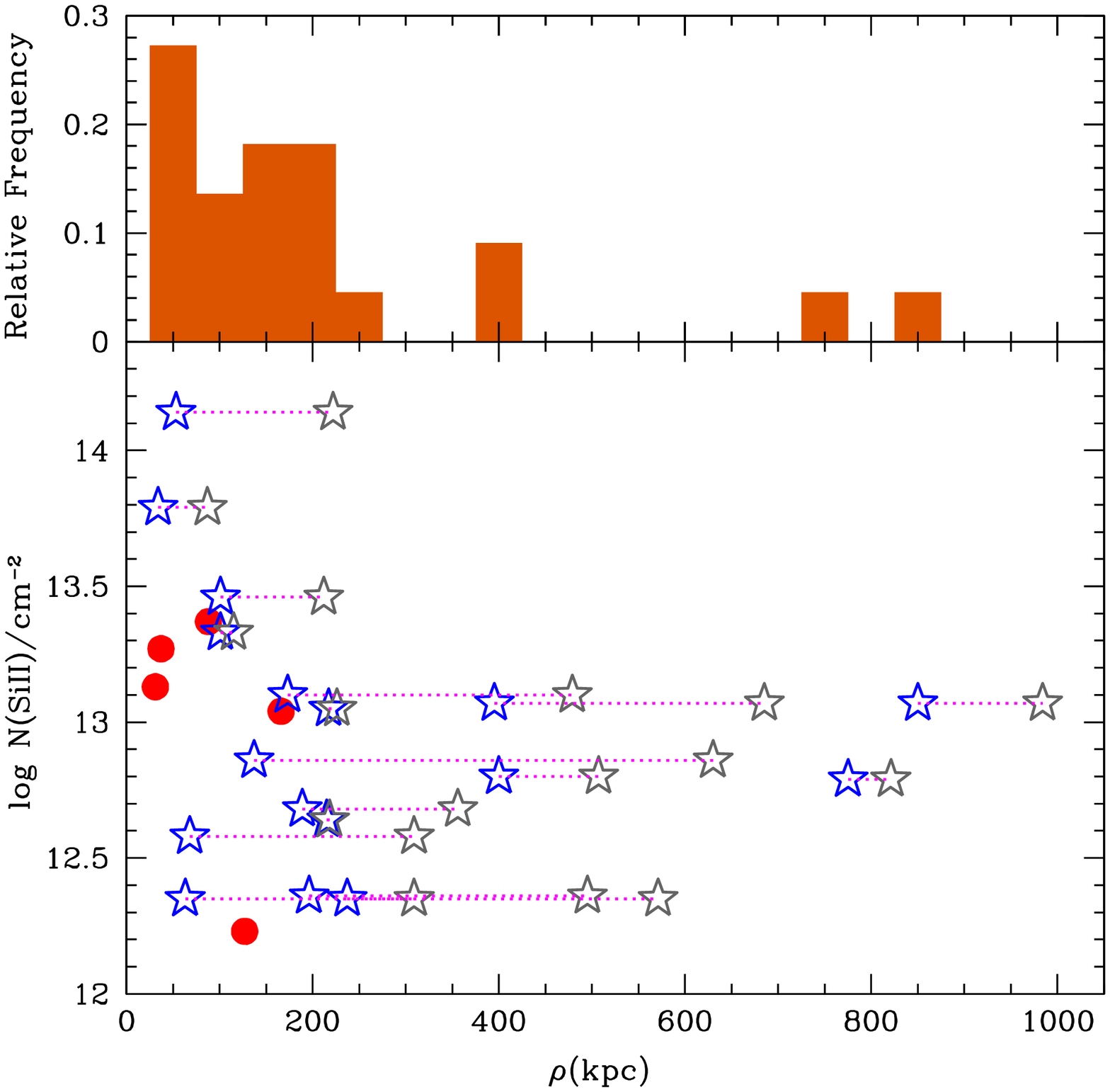}   
\includegraphics[width=0.50\textwidth]{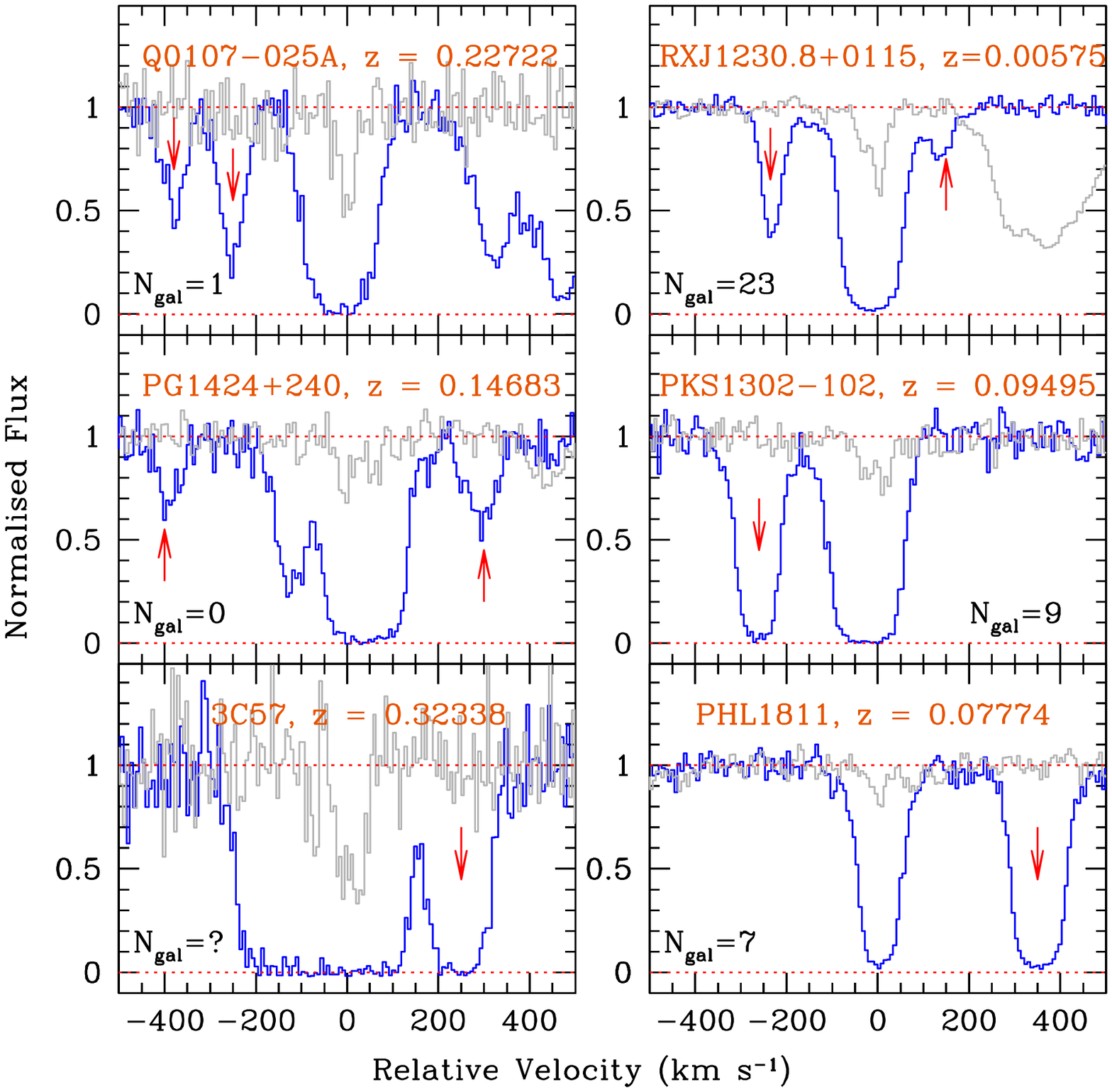}   
}} 
\vskip-0.2cm  
\caption{{\it Left:} The distribution of impact parameters of nearest known galaxies (top) for 22 of the weak absorbers. The impact parameter ranges from 31--850~kpc with a median value of 166~kpc. Only 3/22 absorbers show host-galaxy impact parameters $<50$~kpc. In 17/22 cases at least 2 galaxies are found within 1~Mpc from the QSO sightline and within a velocity window of $\pm$500~\kms\ around the absorption redshift. In the bottom panel, the \SiII\ column densities are plotted against the impact parameters of the nearest galaxies (blue star) and the second nearest galaxies (grey star) connected by the dotted (magenta) line. The (red) filled circles represent the galaxies that do not have a known companion. No significant correlation is seen between $N(\SiII)$ and $\rho$. {\it Right:} The \HI\ absorption profiles of 6 weak absorbers around which different number of galaxies ($N_{\rm gal}=$~0--23) have been identified. No galaxy information is available for the system towards 3C57. The \CII\ absorption profiles are shown in grey. Note that all of them show additional \HI\ absorption within $\pm$500~\kms\ of the absorption redshift (0~\kms) indicated by the arrows. The presence of additional absorption clouds further suggests that they arise from a group environment (see text).}            
\label{fig:gal}       
\end{figure*}

\subsection{Metal Abundance}  
\label{sec:dissmetal}

The most interesting physical properties of the weak absorbers is their unusual high metallicities. The median value of the $\rm [Si/H]$ distribution of $0.0$ indicates that $\approx$50\% of the weak absorbers in our sample show a solar/super solar $\rm Si$ abundance. Moreover, 27/32 absorbers ($\approx85$\%)  show $\rm [Si/H]>-1.0$. If we consider only $Q>2$ systems, then the fraction of absorbers showing $\rm [Si/H]\geq0.0$ \linebreak ($\geq-1.0$) is $\approx44$\% ($\approx87$\%). Here we note that the fraction of weak absorbers showing high metallicity is significantly higher compared to the \HI-selected sample of \cite{Wotta16} and the galaxy-selected sample of \cite{Prochaska17}. In Fig.~\ref{fig:logZdist} we compare the metallicity distributions of these three samples. The cumulative metallicity distribution, shown in the top panel, of our low-$z$ sample is significantly different compared to that of \cite{Wotta16}. A two-sided KS-test gives $D_{\rm KS} =0.51$ and $P_{\rm KS} = 2.1\times10^{-5}$ confirming that the difference has a 99.99\% significance. The significance is somewhat lower ($\approx98$\%) with a $D_{\rm KS}=0.37$ when compared with the COS-Halos sample \citep[]{Prochaska17}. The median metallicities of \cite{Wotta16} and \cite{Prochaska17} samples, i.e., $-1.1$ and $-0.5$ dex respectively, are much lower than our sample. Only 3/58 ($\approx$5\%) of the absorbers in \cite{Wotta16} and 7/31 ($\approx$22\%) of the absorbers in \cite{Prochaska17} have solar/super-solar metallicities. Clearly, the weak absorbers are significantly more metal-rich when compared to the galaxy-selected and \HI-selected samples of absorbers probing the CGM at, by and large, similar redshifts \footnote{Note that the systems in \cite{Wotta16} sample shows a wide range in redshift (0.08--1.09) with a median value of 0.57, which is somewhat higher than that of our sample. Nonetheless, there is no significant trend between the absorption redshift and metallicity in either sample. Moreover, the median metallicity of the 10 systems in \cite{Wotta16} sample with \zabs~$< 0.3$ is $-0.65$, which is much lower than the median metallicity of our sample. Therefore the observed difference in metallicity distribution can not be attributed to difference in the redshift distributions of the two samples.}.  

Here we point out that both \cite{Wotta16} and \cite{Prochaska17} have used an UVB calculated by \citet[][HM12 hereafter]{Haardt12}. The intensity of the HM12 UVB is somewhat lower than that of the KS15 UVB used in this study, for energies $>1$ Ryd \citep[see e.g., Fig.~1 of][]{Hussain17}. In order to assess the effect of the UVB on our metallicity estimates, we model an imaginary cloud at $z=0.1$ with column densities of \HI, \SiII, and \SiIII\ equal to the corresponding median values we obtained for our sample (i.e. $10^{16.0}$, $10^{12.9}$, and $10^{13.2}$~\sqcm, respectively). Grids of constant density {\sc cloudy} models were run with a solar metallicity and with a stopping $N(\HI)$ of $10^{16.0}$~\sqcm. The density derived for the HM12 UVB (i.e., $10^{-3.3} \rm cm^{-3}$) is $\approx0.4$ dex lower than that obtained for the KS15 UVB. However, the metallicity obtained for the HM12 UVB (i.e., $\rm [Si/H] = -0.14$) is only $\approx 0.1$ dex lower than that derived for the KS15 UVB. Thus, overall, our metallicity estimates are consistent with those expected from the HM12 UVB.   

Next, we evaluate the effect on the derived metallicity of the imaginary cloud, if some amount of \SiIII\ is contributed by the high ionization gas phase, as it might be the case for some of the absorbers studied here. We found that even if 90\% of the $N(\SiIII)$ is contributed by the high ionization phase (i.e. using $\log N(\SiIII)/\rm cm^{-2}$ of 12.2 instead of 13.2 in the model), the derived metallicity changes only by $\approx0.1$ dex for both KS15 and HM12 incident continua. The density in such a situation is increases by $\approx0.6$ dex, as expected.    

In order to investigate whether the inferred abundances of $\rm [Si/H] > 0.301$ ($>2$ times solar) for the 10 systems in our sample remain as high, we recalculate the metallicities and densities using a new {\sc cloudy} grid with a super-solar metallicity of $\rm [X/H] = 0.699$ (5 times solar). We find that the $\rm [Si/H]$ values are consistent within 0.2 dex. The densities, however, decrease by $\approx0.6$~dex leading to an increase in $N_{\rm H}$ by 0.3 dex and in line-of-sight thickness by 0.9 dex. The decrease in density for the super-solar {\sc cloudy} grid increases the column densities for the high-ions (i.e., \CIV, \NV, and \OVI). Nonetheless, the predicted high-ions column densities, particularly the $N(\OVI)$ and $N(\NV)$, are still very low (e.g., $\log N/\rm cm^{-2} < 13.5$). The predicted $N(\CIV)$ values, although increased by 0.8--1.0 dex, are still consistent with the $3\sigma$ upper limits or are lower than the observed values except for the two systems (\zabs~$=0.00575$ towards RXJ1230.8+0115  and \zabs~$ = 0.09784$ towards SDSSJ1357+1704) as already noted in Section~\ref{sec:highions}.   

Finally, besides the absolute abundances, the measurements of $N(\CII)$ allow us to calculate the relative abundance of $\rm Si$ relative to $\rm C$, i.e., the $\rm [C/Si]$. We find $\rm[C/Si]=0.13\pm0.18$(std), which is somewhat higher albeit consistent with solar value. Our estimates of $\rm [C/Si]$ are generally comparable to the $\rm [C/\alpha]$ measurements in solar-type stars in the Galactic disk \citep[]{Gustafsson99}.

\subsection{Galaxy Environments} 
\label{sec:dissgal} 

\input{tables/galaxies.tex}

In what types of galaxy environments do the low-$z$ weak absorbers reside? In order to investigate, we have searched for galaxies around these absorbers in SDSS and in the literature. We have found host-galaxy information for 22/34 absorbers. In our SDSS search we have queried for galaxies with spectroscopic redshifts consistent within $\pm500$~\kms\ of \zabs\ and within 1~Mpc projected separation from the QSO sightline. The SDSS spectroscopic database is 90\% complete down to an $r$-band apparent magnitude, $m_r$, of 17.77 \citep[]{Strauss02,Blanton03a}. This corresponds to a luminosity of $>1.2L_*$ at $z=0.1$ \citep[][no K-correction has been applied]{Blanton03}. The details of the galaxies around the weak absorbers are listed in Table~\ref{tab:gal}. In the left panel of Fig.~\ref{fig:gal} we show the impact parameter distribution of the nearest known galaxies which varies from 31--850~kpc. Interestingly, the median impact parameter of 166~kpc is in agreement with the halo radius ($R_h$) we estimated in Section~\ref{sec:cosmo}. Albeit having high metallicities, only $\approx14$\% (3/22) of the absorbers have nearest known bright galaxies within 50~kpc. However, for $\approx70$\% of the cases a galaxy is detected within 200~kpc. We did not find any trend between $N(\SiII)$, metallicity (or density) and the impact parameter of the nearest known galaxy.    

There are 26/34 fields that are covered in the SDSS footprint. We have found 75 galaxies  in total in these fields ($<1$~Mpc) around the weak absorbers ($\pm500$~\kms) in contrast to only 6 galaxies detected in random 26 fields\footnote{The random 26 fields were selected by shifting the RA and Dec of the 26 quasars by 10 degree at random but making sure that they do not fall outside the SDSS footprint.}. It indicates a significant galaxy-overdensity around the low-$z$ weak absorbers.  Furthermore, in about $\approx80$\% (17/22) of the cases we find 2 or more galaxies within 1~Mpc from the sightline and within a velocity window of $\pm500$~\kms\ (see Table~\ref{tab:gal}). All these facts suggest that {\it the majority of the weak absorbers live in galaxy groups.} Interestingly, as many as 23 galaxies are found around the absorber at \zabs~$=0.00575$ towards RXJ1230.8+0115, in which the sightline passes through the Virgo cluster \citep[]{Yoon17}. In 6/26 cases the SDSS search did not yield even one galaxy that satisfied our criteria. Nonetheless, galaxies with consistent photometric redshifts are identified in all but one case. A detail study of all the galaxies listed in Table~\ref{tab:gal} will be presented in future.       

If a line of sight passes through multiple galaxy halos in a group, then it is more likely to observe more than one absorption clump within the characteristic group velocity. We, thus, re-examine the $\pm500$~\kms\ velocity range around the \lya\ absorption in each of the weak absorbers to investigate any such possible signs of group environment. The total redshift path-length covered by this velocity range is $\approx0.127$. From the observed $d\mathcal{N}/dz$ \citep[]{Danforth16}, the expected number of random IGM \HI\ clouds with $N(\HI)>10^{13.5}$ at $z\approx0.1$ is 24. Intriguingly, we have found nearly 60 \HI\ absorbers (including the 34 weak absorbers) which is $\approx2.5$ times higher than the number expected from random chance coincidence. As an example, the $\pm500$~\kms\ spectral ranges for 6 of the weak absorbers are shown in the right panel of Fig.~\ref{fig:gal}. For all of the 6 cases additional \HI\ absorption has been detected, even when no galaxies have been identified or known in the field. It possibly suggests that all of these weak absorbers may reside in group environments even if the group environment is not apparent from the existing galaxy data. We recall here that the SDSS spectroscopic database is sensitive only to bright ($>1.2L_*$) galaxies. A systematic survey of galaxies around these fields is thus essential.  

Alternatively, or in addition, a group environment may be conducive to processes that give rise to these high metallicity clouds far from galaxy centers. A clue about the processes at work in galaxy groups is provided by the study of \OVI\ absorbers by \citet{Pointon17}. They find that \OVI\ is weaker and has a narrower velocity distribution for lines of sight through groups than for those near isolated galaxies. They interpret this to mean that a group is heated beyond the point where individual \OVI\ halos would be superimposed to produce stronger absorption. They suggest that the \OVI\ that is observed is at the interface between the hot, X-ray gas and the cooler CGM. Weak, low-ionization absorption could also be related to such an interface via thermal/hydrodynamical instabilities.          

\subsection{Possible Origin(s)}

The high metallicity ($\rm [Si/H] \approx 0.0$) and solar relative abundances of heavy elements ($\rm [C/Si]\approx0.13\pm0.18$) strongly suggest that the absorbing clouds are related to star-forming regions. The impact parameters of the nearest galaxies of $>50$~kpc imply that mechanisms such as galactic/AGN winds and/or tidal/ram pressure stripping of the ISM of satellite galaxies at early epochs could give rise to the cool clouds seen in absorption. Recall that the majority of the weak absorbers seems to live in a group environment in which interactions are common, and gas is often stripped from individual galaxies. 

The metal-enriched, cool clouds can form in outflows predominantly via two different channels. First, metal-rich ISM clouds can be swept-up by hot wind material to the CGM by means of ram pressure and radiation pressure \citep[e.g.,][]{Zubovas14,McCourt15,Schneider17,Heckman17}. Second, the clouds can form {\it in-situ}, condensing out of the hot wind due to thermal instabilities \citep[e.g.,][]{Field65,Costa15,Voit16,Ferrara16,Thompson16}. Recently, \cite{McCourt16} have suggested that an optically thin, $T\sim10^{6}$~K cooling perturbation with an initial size $\gg c_s t_{\rm cool}$ ($c_s$ is the internal sound speed, $t_{\rm cool}$ is the cooling time) will be fragmented quickly to a large number of cloudlets due a process called ``shattering'', giving rise to a ``fog''  like structure. These fragments are found to have a characteristic length-scale of $l_{\rm cloudlet} \sim 0.1 \rm pc ~ (n_{\rm H}/\rm cm^{-3})^{-1}$ as they reach a temperature of $\sim10^{4}$~K. The column density of an individual cloudlet is then $N_{\rm cloudlet}=n_{\rm H} l_{\rm cloudlet} \sim 10^{17.5}$~\sqcm. It is interesting to note that the median $N_{\rm H}$ of our sample, $\approx10^{18.1}$~\sqcm, is broadly in agreement with their prediction. Here we also recall that a large number of such clouds is indeed required to match the observed $d\mathcal{N}/dz$ (see Section~\ref{sec:cosmo}). These facts indicate that the ``fog'' like structure as suggested by \cite{McCourt16} is consistent with the properties of the weak absorbers. 

In both the situations discussed above the clouds will be metal-rich but will be subject to hydrodynamical (e.g., Rayleigh--Taylor, Kelvin--Helmholtz) instabilities \citep[]{Klein94}. Moreover, cloud-wind interactions will also disrupt/shred the clouds on a ``crushing'' time-scale which is roughly similar to the time-scales of hydrodynamical instabilities. Radiative cooling and the presence of magnetic field can extend the lifetime of the cloud considerably \citep[e.g.,][]{Cooper09,McCourt15}. Next generation cosmological hydrodynamical simulations that can resolve pc-scale structures are essential for better theoretical understanding of the origins of these exciting absorbers.

\section{summary} 
\label{sec:summary} 
Using archival $HST/$COS spectra we have conducted a survey (``COS-Weak'') of low-ionization weak metal line absorbers, with $W_r(\CII\lambda1334)<0.3$~\AA\ and $W_r(\SiII\lambda1260)<0.2$~\AA\ at $z<0.3$, that are analogous to weak \MgII\ absorbers which are mostly studied at high-$z$. We have constructed the largest sample of weak absorber analogs that comprises of 34 absorbers, increasing the number of known/reported weak absorbers by a factor of $>5$ at low-$z$. We have measured column densities of low- and high-ionization metal lines and of \HI, performed simple photoionization models using {\sc cloudy}, and have discussed the implications of the observed and inferred properties of these absorbers and their galaxy environments. Our main findings are as follows: 

\begin{itemize} 

\item We estimate $d\mathcal{N}/dz=0.8\pm0.2$ for our survey at $z<0.3$, consistent with the previous study of \cite{Narayanan05}. This implies that the processes that produce the weak absorber clouds are less active at low-$z$ than at $z\approx1$ (Section~\ref{sec:dndz} \& \ref{sec:dissred}).

\item Our simple photoionization models assumed \SiIII\ and \SiII\ arise in the same phase, and that most/all of the \HI\ absorption is also in this phase. Absorber densities range from $10^{-3.3}$ to $10^{-2.4}$~cm$^{-3}$, with a median of $10^{-2.8}$~cm$^{-3}$. Metallicities, represented as $\rm [Si/H]$, ranged from $-2.5$ to $1.6$, with a median of $0.0$. The line of sight thicknesses of the clouds ranged from 1~pc to 53~kpc. The neutral hydrogen column densities range from $\log N(\HI)/\rm cm^{-2}$ of 14--18, and the total hydrogen column densities $\log N_{\rm H}/\rm cm^{-2}$ from 16--20. There is an anti-correlation between $\rm [Si/H]$ and $N(\HI)$. Although this is partly a selection effect for small $N(\HI)$, it is clear that the highest metallicities do arise in environments that do not have a large amount of neutral hydrogen nearby to dilute the metals (Fig.~\ref{fig:model-param}).

\item At least two thirds of the absorbers have a separate, hotter and/or higher ionization phase with \CIV\ and/or \OVI\ absorption detected. In several cases \CIV\ can arise in the same phase with the \SiII/\CII, but generally when it is detected a separate phase is needed. \OVI\ must always arise in a different phase (Section~\ref{sec:highions}).

\item We find that the highest metallicity systems are the tiniest in size (Fig.~\ref{fig:hydrostatic}). The majority of these absorbers have (gas) masses \linebreak ($<10^{5}M_{\odot}$) too small to be in local hydrostatic equilibrium unless they are significantly dark matter dominated structures. In the absence of any confining medium these absorbers will be evaporated on the free expansion time-scale of $\sim10^{7}$~yr.

\item  The weak absorbers outnumber bright galaxies by tens of thousands in one, suggesting that the population must be huge. Nonetheless, they carry a negligible fraction of cosmic baryons. Adopting the geometrical model for CGM absorbers prescribed by \cite{Richter16}, we obtain a covering fraction of $\approx$30\% for the weak absorbers, assuming that the population extends out to the virial radii of galaxies with $L/L_*>0.001$ (Section~\ref{sec:cosmo}).

\item The weak absorber population is found to be significantly more metal-rich compared to the \HI-selected \citep[]{Wotta16} and galaxy-selected \citep[]{Prochaska17} samples of absorbers probing the CGM at similar redshift (Section~\ref{sec:dissmetal}).

\item We have searched for galaxies around these absorbers in SDSS and in the literature and found that they live in regions of significant galaxy over-density. In about 80\% of the cases, more than one galaxy is detected within 1~Mpc from the QSO sightline and within $\pm500$~\kms\ of the absorber's redshift. The impact parameters of the nearest known galaxies range from $\approx$31--850~kpc with a median of 166~kpc. In $\approx$70\% of the cases the nearest galaxy is detected within 200~kpc. It implies that the weak absorbers are abundant in the halos of galaxies that are in group environments (Section~\ref{sec:dissgal}).

The origin of these metal-rich, compact structures, apart from galaxies, remains unclear, but we suggest that they are transient structures in the halos of galaxies most of which live in group environments and they are related to outflows and/or to stripping of metal-rich gas from the galaxies. A systematic search for galaxies in these fields and next generation cosmological simulations with the capability of resolving pc-scale structures are essential for further insights into the nature and origin of these fascinating cosmic structures.

\end{itemize}

\noindent  
{\it Acknowledgements:} Support for this research was provided by NASA through grants HST AR-12644 from the Space Telescope Science Institute, which is operated by the Association of Universities for Research in Astronomy, Inc., under NASA contract NAS5-26555. SM thanks Tiago Costa for stimulating discussion on the theoretical aspects of galaxy outflows. SM thankfully acknowledges IUCAA (India) for providing hospitality while a part of the work was done. SM also acknowledges support from the European Research Council (ERC), Grant Agreement 278594-GasAroundGalaxies. This work is benefited from the SDSS. Funding for the Sloan Digital Sky Survey IV has been provided by the Alfred P. Sloan Foundation, the U.S. Department of Energy Office of Science, and the Participating Institutions. SDSS-IV acknowledges support and resources from the Center for High-Performance Computing at the University of Utah. The SDSS web site is www.sdss.org. 

SDSS-IV is managed by the Astrophysical Research Consortium for the Participating Institutions of the SDSS Collaboration including the Brazilian Participation Group, the Carnegie Institution for Science, Carnegie Mellon University, the Chilean Participation Group, the French Participation Group, Harvard-Smithsonian Center for Astrophysics, Instituto de Astrof\'isica de Canarias, The Johns Hopkins University, Kavli Institute for the Physics and Mathematics of the Universe (IPMU) / University of Tokyo, Lawrence Berkeley National Laboratory, Leibniz Institut f\"ur Astrophysik Potsdam (AIP), Max-Planck-Institut f\"ur Astronomie (MPIA Heidelberg), Max-Planck-Institut f\"ur Astrophysik (MPA Garching), Max-Planck-Institut f\"ur Extraterrestrische Physik (MPE), National Astronomical Observatories of China, New Mexico State University, New York University, University of Notre Dame, Observat\'ario Nacional / MCTI, The Ohio State University, Pennsylvania State University, Shanghai Astronomical Observatory, United Kingdom Participation Group, Universidad Nacional Aut\'onoma de M\'exico, University of Arizona, University of Colorado Boulder, University of Oxford, University of Portsmouth, University of Utah, University of Virginia, University of Washington, University of Wisconsin, Vanderbilt University, and Yale University.

\def\aj{AJ}%
\def\actaa{Acta Astron.}%
\def\araa{ARA\&A}%
\def\apj{ApJ}%
\def\apjl{ApJ}%
\def\apjs{ApJS}%
\def\ao{Appl.~Opt.}%
\def\apss{Ap\&SS}%
\def\aap{A\&A}%
\def\aapr{A\&A~Rev.}%
\def\aaps{A\&AS}%
\def\azh{AZh}%
\def\baas{BAAS}%
\def\bac{Bull. astr. Inst. Czechosl.}%
\def\caa{Chinese Astron. Astrophys.}%
\def\cjaa{Chinese J. Astron. Astrophys.}%
\def\icarus{Icarus}%
\def\jcap{J. Cosmology Astropart. Phys.}%
\def\jrasc{JRASC}%
\def\mnras{MNRAS}%
\def\memras{MmRAS}%
\def\na{New A}%
\def\nar{New A Rev.}%
\def\pasa{PASA}%
\def\pra{Phys.~Rev.~A}%
\def\prb{Phys.~Rev.~B}%
\def\prc{Phys.~Rev.~C}%
\def\prd{Phys.~Rev.~D}%
\def\pre{Phys.~Rev.~E}%
\def\prl{Phys.~Rev.~Lett.}%
\def\pasp{PASP}%
\def\pasj{PASJ}%
\def\qjras{QJRAS}%
\def\rmxaa{Rev. Mexicana Astron. Astrofis.}%
\def\skytel{S\&T}%
\def\solphys{Sol.~Phys.}%
\def\sovast{Soviet~Ast.}%
\def\ssr{Space~Sci.~Rev.}%
\def\zap{ZAp}%
\def\nat{Nature}%
\def\iaucirc{IAU~Circ.}%
\def\aplett{Astrophys.~Lett.}%
\def\apspr{Astrophys.~Space~Phys.~Res.}%
\def\bain{Bull.~Astron.~Inst.~Netherlands}%
\def\fcp{Fund.~Cosmic~Phys.}%
\def\gca{Geochim.~Cosmochim.~Acta}%
\def\grl{Geophys.~Res.~Lett.}%
\def\jcp{J.~Chem.~Phys.}%
\def\jgr{J.~Geophys.~Res.}%
\def\jqsrt{J.~Quant.~Spec.~Radiat.~Transf.}%
\def\memsai{Mem.~Soc.~Astron.~Italiana}%
\def\nphysa{Nucl.~Phys.~A}%
\def\physrep{Phys.~Rep.}%
\def\physscr{Phys.~Scr}%
\def\planss{Planet.~Space~Sci.}%
\def\procspie{Proc.~SPIE}%
\let\astap=\aap
\let\apjlett=\apjl
\let\apjsupp=\apjs
\let\applopt=\ao
\bibliographystyle{mn}
\bibliography{/home/sowgat/Work/Softwares/LaTeX/mybib.bib}
\bsp 

\appendix 
\section{Connection to weak \MgII\ absorbers}  
\label{MgIIconnection} 

In this Appendix we revisit the connection between \SiII\ and \MgII\ absorption and confirm that our selection criteria are indeed well suited for studying the weak \MgII\ absorbers at $z<0.3$ in the absence of spectral coverage for the \MgII\ doublet.

\begin{figure} 
\vbox{\hbox{
\includegraphics[width=0.45\textwidth]{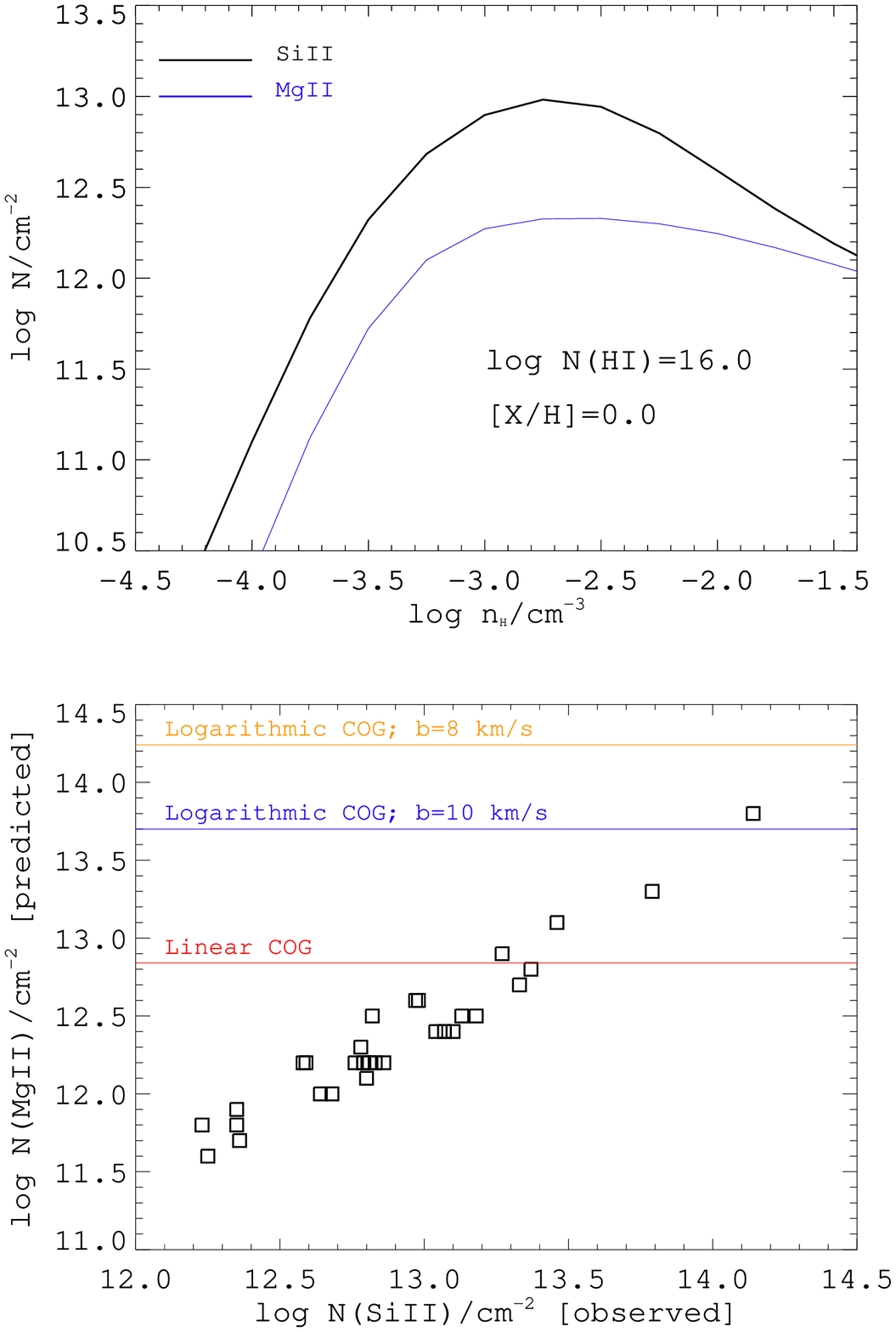}   
}} 
\vskip-0.2cm  
\caption{Top: {\sc cloudy} predicted column densities of \SiII\ (black) and \MgII\ (blue) 
as a function of density. The model is run with a stopping $N(\HI)$ of $10^{16}$~\sqcm\ and with 
a solar metallicity. \SiII\ and \MgII\ closely follow each other for a wide range in density. 
Bottom: The model predicted \MgII\ column densities, listed in the last column of Table~\ref{tab:summary}, 
against the observed \SiII\ column densities. As expected from the top panel, a tight correlation 
($\rho_s = 0.92, \rho_s/\sigma=5.2$) between $N(\MgII)$ and $N(\SiII)$ is seen. The horizontal lines 
correspond to the maximum $N(\MgII)$ for an weak \MgII\ absorber with $W_r(\MgII\lambda2796)<0.3$~\AA, 
provided the line falls on the linear part of the COG or logarithmic part of the COG with $b=$~10 and  
8 \kms. Clearly, all of these absorbers are consistent with being weak \MgII\ absorbers' analogs.}               
\label{fig:MgIIconnection}        
\end{figure}

In Section~\ref{sec:intro}, we have argued that \SiII\ is a good proxy for \MgII. The top panel of Fig.~\ref{fig:MgIIconnection} further strengthens our argument. For typical $N(\HI)$ and $\rm [X/H]$ it shows that \SiII\ traces \MgII\ for a wide range of densities relevant for these absorbers. In the bottom panel, the $N(\MgII)$ values predicted by our adopted PI models are plotted against the observed $N(\SiII)$. As expected, a statistically significant correlation is seen between these two quantities. The predicted $N(\MgII)$ values are in the range of $10^{11.6-13.8}$~\sqcm\ with a median of $10^{12.3}$~\sqcm. An equivalent width of 0.3~\AA\ corresponds to a column density of $10^{12.84}$~\sqcm\ (indicated by the red horizontal line) for the \MgII$\lambda2796$ line on the linear part of the COG. A vast majority (28/32) of the systems show $N(\MgII)<12.84$, indicating that these are truly analogous to the weak \MgII\ absorbers that are primarily studied at high-$z$.

\section{Supplementary Lists}       

\input{tables/sample-ordered-abridged.tex}  
\input{tables/sample-lowsnr-abriged.tex}

\input{tables/strong-systems.tex}

\input{tables/intrinsic-systems.tex}

\section{Velocity plots and system descriptions}    
\label{appendix:vplots} 
\input{tables/velocityplots}

\label{lastpage} 
\end{document}

%% file: tables/systems-ordered.tex
\begin{table*}
\begin{threeparttable}[t] 
\caption{List of 34 weak \MgII\ analogs.} 
\begin{tabular}{p{2.3cm}p{1.2cm}p{1.2cm}p{0.7cm}p{0.8cm}p{0.9cm}p{0.9cm}p{0.9cm}p{0.9cm}p{0.9cm}p{0.9cm}p{0.4cm}p{0.8cm}} 
\hline \hline 
QSO  & RA &  Dec  &  \zqso  &  \zabs  &  Gratings  &  $(\rm S/N)$  & $(\rm S/N)$ &  \multicolumn{3}{c}{$W_r$(m\AA)} &  Flag &  PID \\ \cline{9-11}
     & (J2000)  & (J2000)  &  &   &   &    (G130M)  &  (G160M)  &  \lya & \CII  &  \SiII &     \\ 
(1)  & (2) & (3)  & (4) & (5) & (6) & (7) & (8) & (9) & (10) & (11) & (12) &  (13) \\ 
\hline 
          PG0003+158 &      1.49683 &     16.16361 &    0.451 &   0.16512 &    3 &  15 &    9 &  696$\pm$3  &  55$\pm$ 5 &  22$\pm$ 4 &   1 &          12038  \\
          Q0107-025A &     17.55475 &     -2.33136 &    0.960 &   0.22722 &    3 &  10 &   12 &  696$\pm$19 &  87$\pm$20 & 111$\pm$13 &   1 &          11585  \\
         HE0153-4520 &     28.80500 &    -45.10333 &    0.451 &   0.22597 &    3 &  20 &   13 &  628$\pm$4  & 152$\pm$7  &  73$\pm$5  &   1 &          11541  \\
                3C57 &     30.48817 &    -11.54253 &    0.670 &   0.32338 &    3 &  17 &   10 &  636$\pm$5  & 183$\pm$16 &  78$\pm$12 &   1 &          12038  \\
      SDSSJ0212-0737 &     33.07633 &     -7.62217 &    0.174 &   0.01603 &    3 &   7 &    4 & 1016$\pm$26 & 244$\pm$17 & 161$\pm$22 &   0 &          12248  \\
      SDSSJ0212-0737 &     33.07633 &     -7.62217 &    0.174 &   0.13422 &    3 &   7 &    4 &  449$\pm$14 &  89$\pm$16 &  67$\pm$5  &   0 &          12248  \\
  UKS-0242-724$^{a}$ &     40.79000 &    -72.28011 &    0.102 &   0.06376 &    3 &  11 &    8 &  886$\pm$8  & 112$\pm$9  & 113$\pm$5  &   1 &          12263  \\
           Q0349-146 &     57.86900 &    -14.48556 &    0.616 &   0.07256 &    3 &  10 &    6 & 1155$\pm$8  & 108$\pm$15 & 136$\pm$11 &   0 &          13398  \\
         PKS0405-123 &     61.95180 &    -12.19344 &    0.573 &   0.16710 &    3 &  50 &   22 &  904$\pm$2  & 233$\pm$4  & 155$\pm$3  &   1 &    11508/11541  \\
    IRAS-F04250-5718 &     66.50321 &    -57.20031 &    0.104 &   0.00369 &    3 &  40 &   22 &  488$\pm$2  &  46$\pm$1  &  25$\pm$2  &   1 &    11686/11692  \\
      FBQS-0751+2919 &    117.80129 &     29.32733 &    0.916 &   0.20399 &    3 &  23 &   23 &  702$\pm$5  & 198$\pm$9  & 104$\pm$4  &   1 &          11741  \\
    VV2006J0808+0514 &    122.16187 &      5.24444 &    0.360 &   0.02930 &    1 &   5 &   -- &  741$\pm$17 & 206$\pm$18 & 177$\pm$19 &   0 &          12603  \\
          PG0832+251 &    128.89917 &     24.99472 &    0.329 &   0.02811 &    3 &   7 &    8 &  344$\pm$9  &  46$\pm$7  &  72$\pm$10 &   1 &          12025  \\
      SDSSJ0929+4644 &    142.29080 &     46.74000 &    0.240 &   0.06498 &    3 &  11 &    6 &  242$\pm$8  &  43$\pm$8  &  40$\pm$8  &   0 &          12248  \\
       PMNJ1103-2329 &    165.90667 &    -23.49167 &    0.186 &   0.08352 &    3 &   9 &    6 &  917$\pm$12 & 110$\pm$8  &  67$\pm$9  &   1 &          12025  \\
          PG1116+215 &    169.78583 &     21.32167 &    0.176 &   0.13850 &    3 &  25 &   18 &  512$\pm$4  &  82$\pm$4  &  61$\pm$1  &   1 &          12038  \\
      SDSSJ1122+5755 &    170.68704 &     57.92861 &    0.906 &   0.05319 &    3 &   5 &    4 &  346$\pm$28 &  61$\pm$20 &  61$\pm$15 &   0 &          12248  \\
          PG1121+422 &    171.16324 &     42.02917 &    0.225 &   0.19238 &    3 &  11 &    7 &  705$\pm$22 & 136$\pm$13 & 105$\pm$5  &   1 &          12024  \\
               3C263 &    174.98746 &     65.79700 &    0.646 &   0.06350 &    3 &  23 &   17 &  985$\pm$5  &  59$\pm$6  &  27$\pm$3  &   1 &          11541  \\
          PG1202+281 &    181.17599 &     27.90331 &    0.165 &   0.13988 &    3 &   4 &    6 &  757$\pm$15 &  76$\pm$15 &  56$\pm$10 &   0 &          12248  \\
         PG-1206+459 &    182.24171 &     45.67653 &    1.163 &   0.21439 &    3 &  16 &   17 &  597$\pm$2  & 160$\pm$12 & 119$\pm$5  &   1 &          11741  \\
      SDSSJ1210+3157 &    182.65650 &     31.95167 &    0.389 &   0.05974 &    3 &   5 &    5 &  567$\pm$24 & 128$\pm$10 & 145$\pm$26 &   0 &          12248  \\
      SDSSJ1210+3157 &    182.65650 &     31.95167 &    0.389 &   0.14964 &    3 &   5 &    5 &  674$\pm$20 & 213$\pm$24 & 108$\pm$15 &   0 &          12248  \\
SDSSJ1214+0825$^{a}$ &    183.62729 &      8.41892 &    0.585 &   0.07407 &    1 &   5 &   -- &  628$\pm$21 & 143$\pm$19 & 107$\pm$13 &   1 &          11698  \\
      RXJ1230.8+0115 &    187.70834 &      1.25597 &    0.117 &   0.00575 &    3 &  30 &   19 &  599$\pm$2  &  85$\pm$2  &  55$\pm$2  &   1 &          11686  \\
         PKS1302-102 &    196.38750 &    -10.55528 &    0.278 &   0.09495 &    3 &  18 &   11 &  703$\pm$5  &  70$\pm$6  &  54$\pm$6  &   1 &          12038  \\
      SDSSJ1322+4645 &    200.59450 &     46.75978 &    0.374 &   0.21451 &    3 &   5 &    4 & 1052$\pm$20 & 273$\pm$71 & 192$\pm$32 &   0 &          11598  \\
      SDSSJ1357+1704 &    209.30254 &     17.07892 &    0.150 &   0.09784 &    3 &  10 &    7 &  911$\pm$14 & 116$\pm$10 &  76$\pm$7  &   0 &          12248  \\
      SDSSJ1419+4207 &    214.79250 &     42.12969 &    0.873 &   0.17885 &    3 &   5 &    4 &  854$\pm$43 & 147$\pm$29 & 119$\pm$22 &   0 &          11598  \\
          PG1424+240 &    216.75163 &     23.80000 &   $>0.6$ &   0.12126 &    3 &  14 &   12 &  672$\pm$6  &  24$\pm$4  &  26$\pm$5  &   1 &          12612  \\
          PG1424+240 &    216.75163 &     23.80000 &   $>0.6$ &   0.14683 &    3 &  14 &   12 &  834$\pm$4  &  90$\pm$8  &  71$\pm$6  &   1 &          12612  \\
         PG-1630+377 &    248.00466 &     37.63055 &    1.479 &   0.17388 &    3 &  26 &   11 &  714$\pm$3  & 193$\pm$7  & 131$\pm$6  &   1 &          11741  \\
             PHL1811 &    328.75623 &     -9.37361 &    0.190 &   0.07774 &    3 &  25 &   14 &  426$\pm$5  &  31$\pm$2  &  32$\pm$3  &   1 &          12038  \\
             PHL1811 &    328.75623 &     -9.37361 &    0.190 &   0.08091 &    3 &  25 &   14 &  933$\pm$5  & 135$\pm$3  & 167$\pm$2  &   1 &          12038  \\
\hline 
\end{tabular} 
\label{tab:sample} 
\begin{tablenotes}[para, centering]    
Notes-- (1) QSO name; (2) Right-ascension (J2000); (3) Declination (J2000); (4) QSO redshift from NED except for QSO PG1424+240, which is from \cite{Furniss13}; (5) Absorption redshift; (6) COS gratings used for observations, 1: G130M, 2: G160M, 3: G130M+160M; (7) S$/$N of the G130M data estimated near 1400~\AA; (8) S$/$N of the G160M data estimated near 1600~\AA\ (blank when data are not available); (9) Rest-frame equivalent width of \lya\ in m\AA\ (10) Rest-frame equivalent width of \CII~$\lambda$1334\ in m\AA\ (11) Rest-frame equivalent width of \SiII~$\lambda$1260\ in m\AA\ (12) Spectrum flag, 1: part of the statistical sample, 0: not part of the statistical sample; (13) $HST$ proposal ID. Besides these systems we found two tentative systems that are not part of this study. These are: (a) \zabs~$=0.06206$ towards QSO-B2356-309 in which \SiII\ is noisy and barely detected at $3\sigma$ level. In addition, what would be the \CII$\lambda1334$ of this system has been identified as \lya\ of \zabs~$=0.1659$ by \cite{Fang14}. (b) \zabs~$=0.06206$ towards Q2251+155 in which the profiles of \CII$\lambda1334$ and $\lambda1036$ are not consistent with each other and with the \SiII$\lambda1260/\lambda1193$ lines. $^{a}$These systems are not used for photoionization model due to the lack of \SiIII\ coverage (see text).    
\end{tablenotes}       
\end{threeparttable}
\end{table*}

%% file: tables/summaryall.tex
\begin{landscape} 
\begin{table} 
\begin{threeparttable}[t] 
\caption{Summary of column density measurements and photoionization model parameters.} 
\begin{tabular}{lcccccccrccccccc} 
\hline \hline  
QSO & \zabs & $\log N(\HI)$ &  $\log N(\CII)$ & $\log N(\SiII)$ & $\log N(\SiIII)$ & $Q$ & $\log n_{\rm H} (U)$ & $\rm [Si/H]$ & $\log N_{\rm H}$ & $L$ &  \multicolumn{2}{c}{$\log N(\CIV)$} & \multicolumn{2}{c}{$\log N(\OVI)$} & $\log N(\MgII)$ \\  
\cline{12-13} \noalign{\smallskip} \cline{14-15}   
    &       &      &      &      &      &     & $\rm (1/cm^{-3})$ &          & $\rm (1/cm^{-2})$ & (kpc)  &   (observed)     &     (model)    &    (observed)   &    (model)     &  (model)  \\ 
\hline        
       PG0003+158 &  0.16512 & 18.16 $\pm$ 0.03 &  13.60 $\pm$ 0.05 &  12.23 $\pm$ 0.09 &  $<$12.60         & 3 & $-$2.9 ($-$3.1) & $-$2.5 & 19.9 &  2.02E+01 &          NA         & 12.2  &   13.82 $\pm$ 0.12 &   9.7 	&  11.8	\\  
       Q0107-025A &  0.22722 & 15.92 $\pm$ 0.08 &  13.85 $\pm$ 0.07 &  13.04 $\pm$ 0.06 &  12.90 $\pm$ 0.05 & 5 & $-$2.6 ($-$3.4) &    0.2 & 17.7 &  6.91E-02 &    14.24 $\pm$ 0.12$^{a}$ & 12.4  &   14.57 $\pm$ 0.23 &   9.2	&  12.4      \\ 
      HE0153-4520 &  0.22597 & 16.83 $\pm$ 0.07 &  14.09 $\pm$ 0.03 &  12.78 $\pm$ 0.04 &  $>$13.73         & 4 & $-$3.1 ($-$2.9) & $-$0.7 & 19.2 &  6.34E+00 &          NA         & 13.9  &   14.24 $\pm$ 0.02 &  11.9 	&  12.3      \\ 
             3C57 &  0.32338 & 16.14 $\pm$ 0.02 &  14.05 $\pm$ 0.02 &  12.83 $\pm$ 0.11 &  $>$13.61         & 5 & $-$3.3 ($-$2.7) &    0.0 & 18.6 &  2.47E+00 &          NA         & 14.2  &   $<$13.4          &  12.6	&  12.2      \\ 
   SDSSJ0212-0737 &  0.01603 & 18.27 $\pm$ 0.17 &  $>$15.50         &  14.14 $\pm$ 0.14 &  $>$14.65         & 3 & $-$3.1 ($-$2.9) & $-$0.7 & 20.2 &  5.33E+01 &    15.03 $\pm$ 0.25 & 14.5  &         NA         &  12.3	&  13.8      \\ 
   SDSSJ0212-0737 &  0.13422 & 14.95 $\pm$ 0.08 &  13.91 $\pm$ 0.10 &  12.81 $\pm$ 0.04 &  12.43 $\pm$ 0.15 & 3 & $-$2.5 ($-$3.5) &    0.9 & 16.6 &  3.59E-03 &    $<$13.4          & 11.8  &   $<$13.7          &   8.1	&  12.2      \\ 
     UKS-0242-724 &  0.06376 & 15.27$\pm$0.27   &  13.73 $\pm$ 0.09 &  12.90 $\pm$ 0.05 &  NA               & 1 &  ...            &   ...  & ...  &  ...      &    $<$13.3          &  ...  &         NA         &  ... 	&  ...       \\    
        Q0349-146 &  0.07256 & 16.24 $\pm$ 0.53 &  13.91 $\pm$ 0.09 &  13.18 $\pm$ 0.05 &  13.01 $\pm$ 0.09 & 1 & $-$2.6 ($-$3.4) &    0.0 & 18.0 &  1.34E-01 &    $<$13.5          & 12.5  &         NA         &   9.3	&  12.5      \\ 
      PKS0405-123 &  0.16710 & 16.45 $\pm$ 0.05 &  $>$14.40         &  13.33 $\pm$ 0.03 &  $>$13.40         & 5 & $-$2.8 ($-$3.2) & $-$0.1 & 18.4 &  4.56E-01 &    14.16 $\pm$ 0.05$^{b}$ & 13.1  &   14.73 $\pm$ 0.13 &  10.3	&  12.7      \\ 
 IRAS-F04250-5718 &  0.00369 & 15.85 $\pm$ 0.08 &  13.49 $\pm$ 0.04 &  12.25 $\pm$ 0.17 &  12.78 $\pm$ 0.06 & 1 & $-$3.1 ($-$2.9) & $-$0.4 & 18.1 &  4.92E-01 &    13.21 $\pm$ 0.15 & 13.0  &         NA         &  10.9	&  11.6      \\ 
   FBQS-0751+2919 &  0.20399 & 17.77 $\pm$ 0.05 &  14.23 $\pm$ 0.06 &  12.97 $\pm$ 0.07 &  $>$13.55         & 2 & $-$2.8 ($-$3.2) & $-$1.5 & 19.8 &  1.40E+01 &          NA         & 13.1  &   14.35 $\pm$ 0.07 &  10.5	&  12.6      \\ 
 VV2006J0808+0514 &  0.02930 & 17.56 $\pm$ 0.43 &  $<$14.17         &  13.46 $\pm$ 0.09 &  $>$14.11         & 1 & $-$2.8 ($-$3.2) & $-$0.8 & 19.7 &  9.66E+00 &          NA         & 13.7  &         NA         &  11.1	&  13.1      \\ 
       PG0832+251 &  0.02811 & 14.28 $\pm$ 0.14 &  13.56 $\pm$ 0.08 &  12.79 $\pm$ 0.10 &  12.51 $\pm$ 0.11 & 2 & $-$2.5 ($-$3.5) &    1.6 & 16.0 &  1.01E-03 &    $<$13.3          & 11.9  &         NA         &   8.5	&  12.2      \\ 
   SDSSJ0929+4644 &  0.06498 & 14.67 $\pm$ 0.71 &  13.50 $\pm$ 0.11 &  12.68 $\pm$ 0.06 &  12.50 $\pm$ 0.09 & 2 & $-$2.6 ($-$3.4) &    1.1 & 16.4 &  3.35E-03 &    $<$13.1          & 12.0  &         NA         &   8.7	&  12.0      \\ 
    PMNJ1103-2329 &  0.08352 & 17.74 $\pm$ 0.44 &  14.03 $\pm$ 0.10 &  12.82 $\pm$ 0.13 &  $>$13.37         & 1 & $-$2.8 ($-$3.2) & $-$1.6 & 19.7 &  1.15E+01 &    14.43 $\pm$ 0.06 & 12.9  &         NA         &  10.2	&  12.5      \\ 
       PG1116+215 &  0.13850 & 16.20 $\pm$ 0.05 &  13.84 $\pm$ 0.02 &  12.86 $\pm$ 0.06 &  12.87 $\pm$ 0.03 & 5 & $-$2.7 ($-$3.3) & $-$0.3 & 18.1 &  2.07E-01 &    13.19 $\pm$ 0.08 & 12.5  &   13.84 $\pm$ 0.02 &   9.6	&  12.2      \\ 
   SDSSJ1122+5755 &  0.05319 & 15.16 $\pm$ 1.43 &  13.56 $\pm$ 0.13 &  12.80 $\pm$ 0.08 &  12.85 $\pm$ 0.11 & 1 & $-$2.7 ($-$3.3) &    0.7 & 17.1 &  2.03E-02 &    13.71 $\pm$ 0.15 & 12.5  &         NA         &   9.6	&  12.1      \\ 
       PG1121+422 &  0.19238 & 15.64 $\pm$ 0.05 &  14.12 $\pm$ 0.02 &  13.07 $\pm$ 0.02 &  13.34 $\pm$ 0.06 & 2 & $-$2.9 ($-$3.1) &    0.5 & 17.7 &  1.26E-01 &          NA         & 13.2  &   $<$13.4          &  10.8	&  12.4      \\ 
            3C263 &  0.06350 & 15.40 $\pm$ 0.12 &  13.55 $\pm$ 0.06 &  12.35 $\pm$ 0.05 &  13.09 $\pm$ 0.02 & 4 & $-$3.2 ($-$2.8) &    0.3 & 17.8 &  3.66E-01 &    14.13 $\pm$ 0.15 & 13.6  &         NA         &  11.9	&  11.8      \\ 
       PG1202+281 &  0.13988 & 15.73 $\pm$ 0.10 &  14.04 $\pm$ 0.17 &  12.79 $\pm$ 0.09 &  13.41 $\pm$ 0.24 & 1 & $-$3.1 ($-$2.9) &    0.3 & 18.1 &  5.15E-01 &    $<$13.9          & 13.8  &   $<$13.6          &  11.8	&  12.2      \\ 
      PG-1206+459 &  0.21439 & 15.71 $\pm$ 0.06 &  14.24 $\pm$ 0.05 &  13.13 $\pm$ 0.05 &  $<$13.65         & 3 & $-$3.1 ($-$2.9) &    0.6 & 18.0 &  3.44E-01 &          NA         & 13.9  &         NA         &  11.8	&  12.5      \\ 
   SDSSJ1210+3157 &  0.05974 & 14.83 $\pm$ 0.27 &  14.06 $\pm$ 0.05 &  13.10 $\pm$ 0.08 &  13.23 $\pm$ 0.11 & 1 & $-$2.8 ($-$3.2) &    1.3 & 16.8 &  1.23E-02 &    14.50 $\pm$ 0.10 & 13.0  &         NA         &  10.2	&  12.4      \\ 
   SDSSJ1210+3157 &  0.14964 & 17.28 $\pm$ 0.12 &  14.23 $\pm$ 0.11 &  12.98 $\pm$ 0.12 &  $>$13.69         & 1 & $-$2.8 ($-$3.2) & $-$1.0 & 19.4 &  5.95E+00 &    14.35 $\pm$ 0.13 & 13.4  &   14.72 $\pm$ 0.11 &  10.8	&  12.6      \\ 
   SDSSJ1214+0825 &  0.07407 & 15.00 $\pm$ 0.40 &  14.13 $\pm$ 0.07 &  13.05 $\pm$ 0.06 &  NA               & 1 & ...             &  ...   & ...  &  ...      &          NA         & ...   &         NA         &  ... 	&  ...       \\ 
   RXJ1230.8+0115 &  0.00575 & 15.42 $\pm$ 0.43 &  13.79 $\pm$ 0.01 &  12.64 $\pm$ 0.02 &  13.47 $\pm$ 0.07 & 1 & $-$3.3 ($-$2.7) &    0.6 & 17.9 &  5.48E-01 &    13.32 $\pm$ 0.19 & 14.2  &         NA         &  12.6	&  12.0      \\ 
      PKS1302-102 &  0.09495 & 16.88 $\pm$ 0.03 &  13.65 $\pm$ 0.09 &  12.58 $\pm$ 0.11 &  13.02 $\pm$ 0.06 & 4 & $-$2.6 ($-$3.4) & $-$1.1 & 18.8 &  7.78E-01 &    $<$13.0          & 12.3  &   $<$13.9          &   9.2	&  12.2      \\ 
   SDSSJ1322+4645 &  0.21451 & 17.14 $\pm$ 0.05 &  14.40 $\pm$ 0.05 &  13.27 $\pm$ 0.09 &  $>$13.71         & 4 & $-$2.6 ($-$3.4) & $-$0.6 & 19.0 &  1.46E+00 &          NA         & 13.0  &   14.55 $\pm$ 0.08 &   9.9	&  12.9      \\ 
   SDSSJ1357+1704 &  0.09784 & 15.79 $\pm$ 0.52 &  13.97 $\pm$ 0.04 &  12.76 $\pm$ 0.05 &  $>$13.64         & 1 & $-$3.3 ($-$2.7) &    0.4 & 18.3 &  1.60E+00 &    13.92 $\pm$ 0.12 & 14.4  &   $<$14.1          &  13.0	&  12.2      \\ 
   SDSSJ1419+4207 &  0.17885 & 16.63 $\pm$ 0.30 &  14.32 $\pm$ 0.14 &  13.37 $\pm$ 0.09 &  $>$14.18         & 4 & $-$3.3 ($-$2.7) &    0.1 & 19.2 &  9.34E+00 &          NA         & 14.9  &   14.43 $\pm$ 0.05 &  13.2	&  12.8      \\ 
       PG1424+240 &  0.12126 & 14.96 $\pm$ 0.08 &  13.22 $\pm$ 0.08 &  12.36 $\pm$ 0.09 &  12.46 $\pm$ 0.10 & 3 & $-$2.8 ($-$3.2) &    0.4 & 16.9 &  1.50E-02 &    13.70 $\pm$ 0.06 & 12.2  &   14.44 $\pm$ 0.05 &   9.4	&  11.7      \\ 
       PG1424+240 &  0.14683 & 16.56 $\pm$ 0.96 &  13.71 $\pm$ 0.06 &  12.59 $\pm$ 0.07 &  13.04 $\pm$ 0.05 & 1 & $-$2.6 ($-$3.4) & $-$0.7 & 18.5 &  3.84E-01 &    14.18 $\pm$ 0.06 & 12.4  &   14.04 $\pm$ 0.13 &   9.3	&  12.2      \\ 
      PG-1630+377 &  0.17388 & 15.83 $\pm$ 0.08 &  14.20 $\pm$ 0.03 &  13.07 $\pm$ 0.02 &  13.14 $\pm$ 0.02 & 1 & $-$2.8 ($-$3.2) &    0.3 & 17.7 &  1.03E-01 &          NA         & 12.8  &   13.86 $\pm$ 0.03 &  10.0	&  12.4      \\ 
          PHL1811 &  0.07774 & 16.00 $\pm$ 0.05 &  13.26 $\pm$ 0.03 &  12.35 $\pm$ 0.05 &  12.48 $\pm$ 0.04 & 5 & $-$2.4 ($-$3.6) & $-$0.4 & 17.6 &  3.24E-02 &    $<$13.0          & 11.4  &         NA         &   6.6	&  11.9      \\ 
          PHL1811 &  0.08091 & 17.94 $\pm$ 0.07 &  $>$14.55         &  13.79 $\pm$ 0.04 &  $<$13.80         & 4 & $-$2.5 ($-$3.5) & $-$0.8 & 19.4 &  2.42E+00 &    14.04 $\pm$ 0.04 & 12.8  &         NA         &   8.9	&  13.3      \\  
\hline 
\end{tabular} 
\label{tab:summary} 
\begin{tablenotes}[para, centering]    
Notes-- \\ No ionization model is done for the \zabs~$=$~0.06376 towards UKS-0242-724 and \zabs~$=$~0.07407 towards SDSSJ1214+0825 systems due to the lack of the $N(\SiIII)$ measurements. \\ ``NA'' indicates that the line is either severely blended or not covered by the corresponding COS spectrum. \\ The last column gives the predicted \MgII\ column densities. The model predicted $N(\MgII)$ values reassure that these are indeed analogs of weak \MgII\ absorbers (see also Appendix~\ref{MgIIconnection}). \\ While assigning formal $1\sigma$ confidence intervals in the inferred densities and metallicities is not straightforward, we find that our metallicity estimates are adequate within $0.2$~dex, provided that the $N(\HI)$ estimates are accurate (see Section~\ref{sec:dissmetal}). The densities are very sensitive to the metallicity assumed in the {\sc cloudy} grid, particularly for the metal-rich systems. We find that the densities can be off by a factor of $\approx$3--4. \\ $^{a}$From \citet{Muzahid14}, $^{b}$From \citet{Savage10}.  
\end{tablenotes}       
\end{threeparttable}
\end{table}   
\end{landscape}

%% file: tables/galaxies.tex
\begin{table*}
\setcounter{table}{3}  
\begin{threeparttable}[t] 
\caption{List of galaxies around weak absorbers.}  
\begin{tabular}{lcrlrl} 
\hline \hline 
QSO  &  $z_{\rm abs}$  &  Galaxy & $z_{\rm gal}$  &  $\rho$(\rm kpc) &  Comments/References \\ 
\hline      
PG0003+158	& 0.16512 &   SDSSJ000600.63+160908.1	 &   0.16519	&   127	   &	 SDSS+ 2 Photo-z objects at 678 kpc and 876 kpc \\  
Q0107-025A	& 0.22722 &   17.56667,-2.32678          &   0.2272     &   166	   & 	 \citet{Crighton10} + 2 Photo-z objects at 170 kpc and 921 kpc  \\  	
HE0153-4520	& 0.22597 &          ...		 &   ...	&   ...	   &     Outside SDSS      \\  	
3C57       	& 0.32338 &          ...                 &   ...        &   ...    &     Outside SDSS      \\  
SDSSJ0212-0737  & 0.01603 &   SDSSJ021213.92-073500.9	 &   0.017471	&    53    &  	 SDSS			\\ 		
                &         &   SDSSJ021202.56-074747.7	 &   0.016501 	&   222    &	 SDSS			\\ 	
                &         &   SDSSJ021315.80-073942.6	 &   0.015942   &   278    &	 SDSS			\\ 	
                &         &   SDSSJ021323.81-074355.5	 &   0.016087	&   340    &	 SDSS			\\ 	
                &         &   SDSSJ021327.79-074150.0	 &   0.016701   &   358    &	 SDSS			\\ 
                &         &   SDSSJ021529.88-071741.5	 &   0.016032   &   994    &	 SDSS			\\ 
SDSSJ0212-0737  & 0.13422 &	     ...       		 &   ...	&   ...	   &	 2 Photo-z objects at 796~kpc and 827~kpc  \\ 	
UKS-0242-724    & 0.06376 &	     ...  		 &   ...  	&   ...	   &	 Outside SDSS		\\ 
Q0349-146       & 0.07256 & 	     ...		 &   ...	&   ...	   &	 Outside SDSS		\\  
PKS0405-123     & 0.16710 &   61.95084, -12.18361        &   0.1670     &   101    &     \cite{Prochaska06}     \\ 		
    		&	  &   61.83833, -12.19389        &   0.1670     &   115    &     \cite{Prochaska06}     \\ 	
IRAS-F04250-5718& 0.00369 &          ...                 &   ...        &   ...    &     Outside SDSS           \\  
FBQS-0751+2919  & 0.20399 &          ...                 &   ...        &   ...    &     ...			\\ 
VV2006J0808+0514& 0.02930 &   SDSSJ080839.55+051725.9	 &   0.030795   &   101    &	 SDSS			\\ 	
                &         &   SDSSJ080816.43+051258.0	 &   0.030754	&   212    &	 SDSS			\\ 
                &         &   SDSSJ080901.35+051832.7	 &   0.028036	&   227    &	 SDSS			\\ 
                &         &   SDSSJ080857.48+052138.0	 &   0.030555	&   303    &	 SDSS			\\ 
                &         &   SDSSJ080758.11+051223.6	 &   0.029165	&   360    &	 SDSS			\\ 
                &         &   SDSSJ081009.39+051749.6	 &   0.028946	&   783    &	 SDSS			\\ 
                &         &   SDSSJ080811.31+053514.4	 &   0.030753	&   790    &	 SDSS			\\  
PG0832+251      & 0.02811 &   SDSSJ083720.73+245627.2	 &   0.027126	&   775	   &	 SDSS			\\    	
                &         &   SDSSJ083720.63+250915.8    &   0.026954   &   821    &	 SDSS			\\    
                &         &   SDSSJ083726.34+250338.8    &   0.028623   &   862    &	 SDSS			\\    
                &         &   SDSSJ083736.98+245959.1    &   0.028897   &   943    &	 SDSS			\\  
SDSSJ0929+4644  & 0.06498 &   SDSSJ092855.37+464345.9    &   0.065031   &   189    &     SDSS			\\ 	
                &         &   SDSSJ092849.63+464057.5    &   0.064019   &   356    &	 SDSS			\\ 
                &         &   SDSSJ092849.95+463912.0    &   0.064096   &   454    &	 SDSS			\\ 
                &         &   SDSSJ092841.92+463841.5	 &   0.064556   &   548	   &	 SDSS			\\  
PMNJ1103-2329   & 0.08352 &  	     ... 		 &	...	&   ...	   &	 Outside SDSS		\\ 
PG1116+215      & 0.13850 &   169.77779, 21.30785        &   0.13814    &   137    &     \citet{Tripp98}, SDSS	\\  	 
                &         &   169.77117, 21.25083        &   0.13736    &   630    &     \citet{Tripp98}	\\  	 
                &         &   169.70834, 21.26972        &   0.13845    &   776    &     \citet{Tripp98}	\\ 	
SDSSJ1122+5755  & 0.05319 &   SDSSJ112225.88+580147.3	 &   0.052616	&   400    &	 SDSS			\\ 		 
                &         &   SDSSJ112248.77+580358.3    &   0.053144	&   507    &	 SDSS			\\  	
                &         &   SDSSJ112359.26+575248.2	 &   0.052292	&   622    &	 SDSS			\\  
                &         &   SDSSJ112227.55+580614.2    &   0.053132	&   661    &	 SDSS			\\ 
                &         &   SDSSJ112207.40+580604.3	 &   0.052983	&   703    &	 SDSS			\\  
PG1121+422  	& 0.19238 &   SDSS J112428.67+420543.8   &   0.193994   &   850    & 	 SDSS			\\ 	
                &         &   SDSS J112445.06+415641.9   &   0.192222   &   984    &	 SDSS			\\  
3C263           & 0.06350 &   175.02154, 65.80042	 &   0.06322    &    63    &     \citet{Savage12}       \\  
	        &         &   174.73408, 65.87654        &   0.06281    &   571    &     \citet{Savage12}       \\  	
	        &         &   174.74136, 65.88236        &   0.06285    &   576    &     \citet{Savage12}       \\  
	        &         &   175.52030, 65.81338	 &   0.06498    &   956    &     \citet{Savage12}       \\  
PG1202+281  	& 0.13988 &	    ...                  &   ...        &   ...    &     1 Photo-z object at 912~kpc 	\\  	
PG-1206+459     & 0.21439 &	    NA                   &   0.2144     &    31    &     Rosenwasser et. al, Submitted	\\   
SDSSJ1210+3157  & 0.05974 &   SDSS J121028.01+315838.1   &   0.059548   &   173    &	 SDSS			\\      
                &         &   SDSS J121012.30+320137.3   &   0.059628   &   479    &	 SDSS			\\ 
                &         &   SDSS J121004.05+320115.2   &   0.060117   &   567    &	 SDSS			\\ 
                &         &   SDSS J120957.35+320024.6   &   0.058957   &   619    &	 SDSS			\\ 
                &         &   SDSS J121013.63+320501.4   &   0.060975   &   657    &	 SDSS			\\ 
                &         &   SDSS J121015.27+320854.7   &   0.059088   &   862    &	 SDSS			\\ 
                &         &   SDSS J120942.22+320258.8   &   0.058983   &   888    &	 SDSS			\\  
SDSSJ1210+3157  & 0.14964 &         ...                  &   ...        &   ...    &     1 Photo-z object at 832 kpc   \\   
SDSSJ1214+0825  & 0.07407 &   SDSSJ121425.17+082251.8    &   0.073271   &   217    &	 SDSS			\\  
                &         &   SDSSJ121431.23+082225.6    &   0.074042   &   226    &	 SDSS			\\ 
                &         &   SDSSJ121428.14+082225.5    &   0.072431   &   227    &	 SDSS			\\ 
                &         &   SDSSJ121439.76+082157.4    &   0.073442   &   324    &	 SDSS			\\ 
                &         &   SDSSJ121511.53+082444.3    &   0.073495   &   842    &	 SDSS			\\ 
                &         &   SDSSJ121511.14+082556.7    &   0.073064   &   831    &	 SDSS			\\ 
\hline 
\end{tabular} 
\label{tab:gal}  
\begin{tablenotes}[para, centering]    
\end{tablenotes}       
\end{threeparttable}
\end{table*} 
\begin{table*}
\setcounter{table}{3}  
\begin{threeparttable}[t] 
\caption{Cont.}  
\begin{tabular}{lcrlrl} 
\hline \hline 
QSO  &  $z_{\rm abs}$  &  Galaxy & $z_{\rm gal}$  &  $\rho$(\rm kpc) &  Comments/References \\ 
\hline       
RXJ1230.8+0115  & 0.00575 &   SDSSJ123246.60+013408.1    &   0.005088   &   215    &     SDSS                   \\  
                &         &   SDSSJ123246.10+013407.8    &   0.005166   &   218    &     SDSS                   \\   
                &         &   SDSSJ123320.76+013117.7    &   0.005557   &   227    &     SDSS                   \\   
                &         &   SDSSJ122950.57+020153.7    &   0.005924   &   353    &     SDSS                   \\  
                &         &   SDSSJ123227.94+002326.2    &   0.005051   &   354    &     SDSS                   \\ 
                &         &   SDSSJ123013.38+023730.5    &   0.00544    &   549    &     SDSS                   \\ 
                &         &   SDSSJ123014.49+023717.6    &   0.005494   &   553    &     SDSS                   \\ 
                &         &   SDSSJ123821.70+011207.5    &   0.004152   &   574    &     SDSS                   \\ 
                &         &   SDSSJ122902.17+024323.8    &   0.005221   &   587    &     SDSS                   \\ 
                &         &   SDSSJ123422.01+021931.4    &   0.005871   &   596    &     SDSS                   \\ 
                &         &   SDSSJ123238.38+024015.6    &   0.005692   &   619    &     SDSS                   \\ 
                &         &   SDSSJ123236.14+023932.5    &   0.005867   &   632    &     SDSS                   \\ 
                &         &   SDSSJ123251.28+023741.6    &   0.005911   &   633    &     SDSS                   \\ 
                &         &   SDSSJ122803.19+025449.7    &   0.004871   &   642    &     SDSS                   \\ 
                &         &   SDSSJ122658.50+022939.4    &   0.005645   &   649    &     SDSS                   \\ 
                &         &   SDSSJ122803.67+025434.9    &   0.004994   &   657    &     SDSS                   \\ 
                &         &   SDSSJ123434.98+023407.7    &   0.006155   &   727    &     SDSS                   \\ 
                &         &   SDSSJ123902.48+005058.9    &   0.005316   &   815    &     SDSS                   \\ 
                &         &   SDSSJ123805.17+012839.9    &   0.006158   &   832    &     SDSS                   \\ 
                &         &   SDSSJ123642.07+030630.3    &   0.004861   &   842    &     SDSS                   \\ 
                &         &   SDSSJ122329.97+020029.0    &   0.006051   &   878    &     SDSS                   \\ 
                &         &   SDSSJ122815.88+024202.9    &   0.007397   &   885    &     SDSS                   \\ 
                &         &   SDSSJ122323.40+014854.1    &   0.006295   &   895    &     SDSS                   \\ 
PKS1302-102     & 0.09495 &   196.38375, -10.56555       &   0.09358    &    68    &     \citet{Cooksey08}      \\  
                &         &   196.34417, -10.58000       &   0.09328    &   309    &     \citet{Cooksey08}      \\  
                &         &   196.33708, -10.58083       &   0.09393    &   350    &     \citet{Cooksey08}      \\  
                &         &   196.34874, -10.50694       &   0.09331    &   386    &	 \citet{Cooksey08}      \\  
                &         &   196.30041, -10.57250       &   0.09531    &   548    &     \citet{Cooksey08}      \\  
                &         &   196.28833, -10.61417       &   0.09523    &   715    &     \citet{Cooksey08}      \\   
                &         &   196.31500, -10.46389       &   0.09442    &   728    &     \citet{Cooksey08}      \\    
                &         &   196.42292, -10.44000       &   0.09332    &   756    &     \citet{Cooksey08}      \\  
                &         &   196.36833, -10.42083       &   0.09402    &   852    &	 \citet{Cooksey08}      \\  
SDSSJ1322+4645  & 0.21451 &   		NA		 &   0.2142     &    37    &     \citet{Werk14} + 2 Photo-z objects at 395~kpc and 821~kpc  \\   	
SDSSJ1357+1704  & 0.09784 &		...		 &	... 	&   ...	   &	 2 Photo-z objects at 303~kpc and 491~kpc \\  
SDSSJ1419+4207  & 0.17885 &	        NA               &   0.1792     &    88    & 	 \citet{Werk14}         \\   
PG1424+240      & 0.12126 &   SDSSJ142701.72+234630.9    &   0.121177   &   196    &     SDSS			\\  	 
                &         &   SDSSJ142714.52+235007.4    &   0.119514   &   495    &     SDSS			\\  
PG1424+240 	& 0.14683 &		...		 &	...	&   ...	   &	 2 Photo-z objects at 940~kpc and 950~kpc \\  
PG-1630+377 	& 0.17388 &   SDSSJ163155.09+373556.5    &   0.175373   &   395    & 	 SDSS			\\    	
                &         &   SDSSJ163149.79+373438.7    &   0.17391    &   685    &	 SDSS			\\   
PHL1811         & 0.07774 &    J215450.8-092235          &   0.078822   &   237    &     \citet{Keeney17}       \\ 
                &         &    J215447.5-092254          &   0.077671   &   309    &     \citet{Keeney17}       \\ 
		&	  &    SDSSJ215506.02-091627.0   &   0.078655   &   526    &	 SDSS			\\ 	 
                &         &    SDSSJ215453.14-091559.2   &   0.077813   &   584    &     SDSS		        \\ 	 
                &         &    SDSSJ215434.62-091632.7   &   0.078369   &   766    &	 SDSS			\\ 
                &         &    SDSSJ215536.27-092017.7   &   0.077963   &   766    &     SDSS			\\ 
                &         &    SDSSJ215437.22-091534.4   &   0.078117   &   787    &     SDSS			\\ 
PHL1811         & 0.08091 &    J21545996-0922249	 &   0.0808     &    34    &     \citet{Jenkins05}      \\  
	        &         &    J21545870-0923061  	 &   0.0804	&    87    &     \citet{Jenkins05}      \\ 
\hline 
\end{tabular} 
\label{tab:gal}  
\begin{tablenotes}[para, centering]    
\end{tablenotes}      
Notes-- The impact parameters from the literature are corrected for the adopted cosmology.   
\end{threeparttable}
\end{table*}

%% file: tables/sample-ordered-abridged.tex
\begin{table*}
\setcounter{table}{0}  
\caption{List of QSO spectra with $\rm S/N>5$ per pixel searched for weak \MgII\ analogs. 
The full list is available as ``online only'' material.} 
\begin{tabular}{lrrcccccccc} 
\hline \hline 
QSO  & RA &  Dec  &  \zqso  &  Gratings  &  $z_{\rm min}$  &  $z_{\rm max}$  & $(\rm S/N)_{G130M}$  & $(\rm S/N)_{G160M}$ & Flag &  PID \\ 
(1)  & (2) & (3)  & (4) & (5) & (6) & (7) & (8) & (9) & (10) & (11) \\ 
\hline 
             PHL2525  &      0.10167  &    -12.76333  &    0.199  &    3  &    0.000  &    0.179  &   12  &    8  &  1  &          12604   \\
          PG0003+158  &      1.49683  &     16.16361  &    0.451  &    3  &    0.000  &    0.343  &   15  &    9  &  1  &          12038   \\
              MRK335  &      1.58125  &     20.20278  &    0.025  &    3  &    0.000  &    0.008  &   16  &   11  &  1  &          11524   \\
      SDSSJ0012-1022  &      3.10004  &    -10.37403  &    0.228  &    3  &    0.000  &    0.208  &    5  &    3  &  0  &          12248   \\
       QSO-B0026+129  &      7.30712  &     13.26778  &    0.142  &    1  &    0.000  &    0.086  &   11  &   -1  &  1  &          12569   \\
        HS-0033+4300  &      9.09575  &     43.27786  &    0.120  &    3  &    0.025  &    0.101  &    6  &    6  &  1  &          11632   \\
      SDSSJ0042-1037  &     10.59287  &    -10.62883  &    0.424  &    3  &    0.000  &    0.343  &    6  &    5  &  0  &          11598   \\
          PG0044+030  &     11.77458  &      3.33194  &    0.623  &    1  &    0.000  &    0.099  &    6  &   -1  &  1  &          12275   \\
              IO-AND  &     12.07908  &     39.68656  &    0.134  &    3  &    0.025  &    0.115  &    8  &    7  &  1  &          11632   \\
       QSO-B0050+124  &     13.39567  &     12.69331  &    0.059  &    1  &    0.000  &    0.040  &   10  &   -1  &  1  &          12569   \\
         HE0056-3622  &     14.65567  &    -36.10133  &    0.164  &    3  &    0.000  &    0.145  &   17  &   13  &  1  &          12604   \\
              RBS144  &     15.11296  &    -51.23172  &    0.062  &    3  &    0.000  &    0.044  &   13  &   10  &  1  &          12604   \\
          J0101+4229  &     15.37971  &     42.49322  &    0.190  &    3  &    0.025  &    0.170  &    6  &    5  &  1  &          11632   \\
          Q0107-025A  &     17.55475  &     -2.33136  &    0.960  &    3  &    0.000  &    0.347  &   10  &   12  &  1  &          11585   \\
          Q0107-0232  &     17.56012  &     -2.28267  &    0.728  &    2  &    0.135  &    0.349  &   -1  &   10  &  1  &          11585   \\
          Q0107-025B  &     17.56771  &     -2.31417  &    0.956  &    3  &    0.000  &    0.346  &    9  &   11  &  1  &          11585   \\
          B0117-2837  &     19.89871  &    -28.35872  &    0.349  &    3  &    0.000  &    0.345  &   15  &   17  &  1  &          12204   \\
             TONS210  &     20.46463  &    -28.34939  &    0.116  &    3  &    0.000  &    0.097  &   28  &   18  &  1  &          12204   \\
            B0120-28  &     20.65317  &    -28.72258  &    0.436  &    3  &    0.086  &    0.342  &   12  &    8  &  1  &          12204   \\
            FAIRALL9  &     20.94071  &    -58.80578  &    0.047  &    3  &    0.000  &    0.030  &   27  &   26  &  1  &          12604   \\
\hline 
\end{tabular} 
\label{tab:COSlos}
\flushleft      
Notes-- (1) QSO name sorted by increasing RA; (2) Right-ascension (J2000); (3) Declination (J2000); (4) QSO redshift; (5) COS gratings used for observations, 1: G130M, 2: G160M, 3: G130M+160M; (6) Minimum search redshift; (7) Maximum search redshift; (8) S$/$N of the G130M data estimated near 1400~\AA\ ($-1$ when data are not available); (9) S$/$N of the G160M data estimated near 1600~\AA\ ($-1$ when data are not available); (10) Spectrum flag, 1: part of the statistical sample, 2: not part of the statistical sample; (11) $HST$ proposal ID.             
\end{table*} 

%% file: tables/sample-lowsnr-abriged.tex
\begin{table*}
\setcounter{table}{1}  
\caption{List of QSO spectra with $\rm S/N<5$ per pixel that are not used in this study. 
The full list is available as ``online only'' material.}   
\begin{tabular}{lrrcccccccc} 
\hline \hline 
QSO  & RA &  Dec  &  \zqso  &  Gratings  &  $z_{\rm min}$  &  $z_{\rm max}$  & $(\rm S/N)_{G130M}$  & $(\rm S/N)_{G160M}$ & Fl
ag &  PID \\ 
(1)  & (2) & (3)  & (4) & (5) & (6) & (7) & (8) & (9) & (10) & (11) \\ 
\hline 
       QSO-B0007+107  &      2.62925  &     10.97486  &    0.089  &    1  &    0.000  &    0.071  &    2  &   -1  &  1  &          12569   \\
      SDSSJ0242-0759  &     40.71188  &     -7.98728  &    0.172  &    3  &    0.000  &    0.152  &    4  &    4  &  0  &          12248   \\
      SDSSJ0843+4117  &    130.95620  &     41.29489  &    0.990  &    3  &    0.000  &    0.330  &    3  &    3  &  0  &          12248   \\
    VV2006J0852+0313  &    133.24675  &      3.22239  &    0.297  &    1  &    0.000  &    0.099  &    4  &   -1  &  0  &          12603   \\
          J0904P4007  &    136.09713  &     40.11797  &    0.411  &    2  &    0.108  &    0.323  &   -1  &    4  &  1  &          13423   \\
      SDSSJ0910+1014  &    137.62396  &     10.23711  &    0.463  &    3  &    0.000  &    0.345  &    4  &    4  &  0  &          11598   \\
      SDSSJ0912+2957  &    138.14758  &     29.95706  &    0.305  &    3  &    0.000  &    0.283  &    4  &    3  &  0  &          12248   \\
    VV2006J0914+0837  &    138.63238  &      8.62853  &    0.648  &    1  &    0.000  &    0.100  &    4  &   -1  &  0  &          12603   \\
      SDSSJ0930+2848  &    142.50791  &     28.81622  &    0.487  &    1  &    0.000  &    0.102  &    3  &   -1  &  0  &          12603   \\
      SDSSJ0935+0204  &    143.82578  &      2.07097  &    0.650  &    3  &    0.000  &    0.345  &    4  &    3  &  0  &          11598   \\
    VV2000J0936+3207  &    144.01617  &     32.11925  &    1.150  &    1  &    0.000  &    0.097  &    2  &   -1  &  0  &          12603   \\
      SDSSJ0937+1700  &    144.27854  &     17.00597  &    0.506  &    1  &    0.000  &    0.098  &    4  &   -1  &  0  &          12603   \\
      SDSSJ0943+0531  &    145.88170  &      5.52539  &    0.564  &    3  &    0.000  &    0.344  &    4  &    4  &  0  &          11598   \\
      SDSSJ0951+3307  &    147.78800  &     33.12939  &    0.644  &    1  &    0.000  &    0.102  &    3  &   -1  &  1  &          12486   \\
       BZBJ1001+2911  &    150.29254  &     29.19375  &    0.558  &    3  &    0.000  &    0.324  &    3  &    3  &  1  &          13008   \\
      VV96J1002+3240  &    150.72713  &     32.67753  &    0.829  &    1  &    0.000  &    0.100  &    3  &   -1  &  0  &          12603   \\
     2MASXJ1005-2417  &    151.38628  &    -24.28781  &    0.154  &    3  &    0.000  &    0.135  &    3  &    2  &  0  &          13347   \\
          J1010P2559  &    152.62725  &     25.99706  &    0.512  &    2  &    0.108  &    0.319  &   -1  &    4  &  1  &          13423   \\
     2MASSJ1013+0500  &    153.32487  &      5.00950  &    0.266  &    1  &    0.000  &    0.101  &    3  &   -1  &  0  &          12603   \\
     2MASSJ1015-2748  &    153.99683  &    -27.80800  &    0.102  &    3  &    0.000  &    0.084  &    4  &    2  &  0  &          13347   \\
\hline 
\end{tabular} 
\label{tab:nCOSlos}  
\flushleft      
Notes-- (1) QSO name; (2) Right-ascension (J2000); (3) Declination (J2000); (4) QSO redshift; (5) COS gratings used for observations, 1: G130M, 2: G160M, 3: G130M+160M; (6) Minimum search redshift; (7) Maximum search redshift; (8) S$/$N of the G130M data estimated near 1400~\AA\ ($-1$ when data are not available); (9) S$/$N of the G160M data estimated near 1600~\AA\ ($-1$ when data are not available); (10) Spectrum flag, 1: part of the statistical sample, 2: not part of the statistical sample; (11) $HST$ proposal ID.             
\end{table*}  

%% file: tables/strong-systems.tex
\begin{table*}
\begin{threeparttable}[t]  
\caption{List of strong \CII$/$\SiII\ systems. These systems are {\em not} part of the sample studied here.} 
\begin{tabular}{lrrccccccccc} 
\hline \hline 
QSO  & RA &  Dec  &  \zqso  &  \zabs  &  Gratings  &  $z_{\rm min}$  &  $z_{\rm max}$  & $(\rm S/N)_{G130M}$  & $(\rm 
S/N)_{G160M}$ & Flag &  PID \\ 
(1)  & (2) & (3)  & (4) & (5) & (6) & (7) & (8) & (9) & (10) & (11) & (12) \\ 
\hline 
      SDSSJ0012-1022 &      3.10004 &    -10.37403 &    0.228 &   0.22936 &    3 &    0.000 &    0.208 &    5 &    3 &    0 &          12248  \\ 
      SDSSJ0042-1037 &     10.59287 &    -10.62883 &    0.424 &   0.09487 &    3 &    0.000 &    0.343 &    6 &    5 &    0 &          11598  \\ 
            B0120-28 &     20.65317 &    -28.72258 &    0.436 &   0.18538 &    3 &    0.086 &    0.342 &   12 &    8 &    1 &          12204  \\
             PHL1226 &     28.61662 &      4.80519 &    0.404 &   0.01795 &    1 &    0.000 &    0.101 &    8 &   -1 &    1 &          12536  \\
      SDSSJ0155-0857 &     28.87509 &     -8.95111 &    0.165 &   0.16412 &    3 &    0.000 &    0.146 &    6 &    5 &    0 &          12248  \\
    VV2006J0159+1345 &     29.97067 &     13.76511 &    0.503 &   0.04419 &    1 &    0.000 &    0.100 &    6 &   -1 &    0 &          12603  \\
           Q0244-303 &     41.70779 &    -30.12814 &    0.530 &   0.00461 &    1 &    0.000 &    0.099 &    5 &   -1 &    1 &          12988  \\
          PG0832+251 &    128.89917 &     24.99472 &    0.329 &   0.01750 &    3 &    0.000 &    0.307 &    7 &    8 &    1 &          12025  \\
           Q0850+440 &    133.39266 &     43.81730 &    0.515 &   0.16368 &    3 &    0.000 &    0.329 &   13 &    6 &    0 &          13398  \\
      SDSSJ0925+4004 &    141.47793 &     40.07058 &    0.471 &   0.24766 &    3 &    0.000 &    0.344 &    4 &    5 &    0 &          11598  \\
      SDSSJ0928+6025 &    142.15824 &     60.42250 &    0.295 &   0.15384 &    3 &    0.000 &    0.273 &    4 &    5 &    0 &          11598  \\
      SDSSJ0930+2848 &    142.50791 &     28.81622 &    0.487 &   0.02284 &    1 &    0.000 &    0.102 &    3 &   -1 &    0 &          12603  \\
      SDSSJ0950+4831 &    147.50304 &     48.52481 &    0.589 &   0.21168 &    3 &    0.000 &    0.345 &    7 &    5 &    0 &          11598  \\
      SDSSJ0951+3307 &    147.78800 &     33.12939 &    0.644 &   0.00534 &    1 &    0.000 &    0.102 &    3 &   -1 &    1 &          12486  \\
     2MASSJ0951+3542 &    147.84971 &     35.71361 &    0.398 &   0.04435 &    1 &    0.009 &    0.103 &    7 &   -1 &    0 &          12603  \\
      SDSSJ0959+0503 &    149.81521 &      5.06531 &    0.162 &   0.05888 &    3 &    0.000 &    0.143 &    6 &    5 &    0 &          12248  \\
      SDSSJ1001+5944 &    150.26062 &     59.73730 &    0.746 &   0.30358 &    3 &    0.000 &    0.345 &   10 &    5 &    0 &          12248  \\
      SDSSJ1009+0713 &    152.25858 &      7.22883 &    0.456 &   0.11398 &    3 &    0.000 &    0.345 &    5 &    5 &    0 &          11598  \\
          J1010P2559 &    152.62725 &     25.99706 &    0.512 &   0.24464 &    2 &    0.108 &    0.319 &   -1 &    4 &    1 &          13423  \\
      SDSSJ1016+4706 &    154.09417 &     47.11203 &    0.821 &   0.16582 &    3 &    0.000 &    0.345 &    5 &    5 &    0 &          11598  \\
      SDSSJ1016+4706 &    154.09417 &     47.11203 &    0.821 &   0.25257 &    3 &    0.000 &    0.345 &    5 &    5 &    0 &          11598  \\ 
           Q1022+393 &    155.65591 &     39.53069 &    0.603 &   0.25225 &    2 &    0.121 &    0.328 &    4 &   -1 &    0 &          12593  \\
      SDSSJ1047+1304 &    161.79096 &     13.08183 &    0.400 &   0.00279 &    1 &    0.000 &    0.079 &    4 &   -1 &    1 &          12198  \\
      SDSSJ1112+3539 &    168.16296 &     35.65783 &    0.653 &   0.02480 &    3 &    0.000 &    0.345 &    4 &    4 &    0 &          11598  \\
      SDSSJ1117+2634 &    169.47630 &     26.57128 &    0.421 &   0.04747 &    3 &    0.000 &    0.326 &    6 &    5 &    0 &          12248  \\
    VV2006J1120+0413 &    170.02078 &      4.22314 &    0.544 &   0.04993 &    1 &    0.000 &    0.101 &    5 &   -1 &    0 &          12603  \\
      SDSSJ1133+0327 &    173.36576 &      3.45531 &    0.524 &   0.23733 &    3 &    0.000 &    0.345 &    5 &    3 &    0 &          11598  \\ 
          PG1216+069 &    184.83720 &      6.64403 &    0.331 &   0.00634 &    3 &    0.000 &    0.309 &   14 &    7 &    1 &          12025  \\   
      SDSSJ1233+4758 &    188.39613 &     47.96678 &    0.381 &   0.22193 &    3 &    0.000 &    0.344 &    8 &    4 &    0 &          11598  \\
      SDSSJ1241+2852 &    190.37350 &     28.87000 &    0.589 &   0.06667 &    1 &    0.000 &    0.098 &    5 &   -1 &    0 &          12603  \\
      SDSSJ1241+5721 &    190.47508 &     57.35203 &    0.583 &   0.20558 &    3 &    0.000 &    0.345 &    7 &    6 &    0 &    11598/13033  \\
    VV2006J1305+0357 &    196.35108 &      3.95858 &    0.545 &   0.00144 &    1 &    0.000 &    0.101 &    7 &   -1 &    0 &          12603  \\
            CSO-0873 &    199.98433 &     27.46894 &    1.015 &   0.28899 &    2 &    0.252 &    0.329 &   10 &   -1 &    0 &          11667  \\
      SDSSJ1330+2813 &    202.68813 &     28.22261 &    0.417 &   0.19227 &    3 &    0.000 &    0.345 &    6 &    5 &    0 &          11598  \\  
      SDSSJ1342+0505 &    205.52733 &      5.08994 &    0.266 &   0.13943 &    3 &    0.000 &    0.245 &    7 &    5 &    0 &          12248  \\
      SDSSJ1342-0053 &    205.71500 &      0.89592 &    0.326 &   0.07133 &    3 &    0.000 &    0.304 &    7 &    6 &    0 &          11598  \\
      SDSSJ1342-0053 &    205.71500 &      0.89592 &    0.326 &   0.22745 &    3 &    0.000 &    0.304 &    7 &    6 &    0 &          11598  \\
      SDSSJ1407+5507 &    211.88437 &     55.12378 &    1.026 &   0.00477 &    1 &    0.000 &    0.101 &    8 &   -1 &    1 &          12486  \\
      SDSSJ1410+2304 &    212.66001 &     23.07978 &    0.795 &   0.20962 &    3 &    0.000 &    0.340 &    7 &    6 &    1 &          12958  \\
      QSO-B1411+4414 &    213.45137 &     44.00389 &    0.089 &   0.08286 &    1 &    0.000 &    0.071 &   13 &   -1 &    1 &          12569  \\
      SDSSJ1415+1634 &    213.92876 &     16.57050 &    0.742 &   0.00770 &    1 &    0.000 &    0.085 &   10 &   -1 &    1 &          12486  \\
      SDSSJ1435+3604 &    218.79803 &     36.07700 &    0.428 &   0.20267 &    3 &    0.000 &    0.345 &    5 &    3 &    0 &          11598  \\
      SDSSJ1512+0128 &    228.15438 &      1.47944 &    0.265 &   0.02938 &    1 &    0.000 &    0.100 &    3 &   -1 &    0 &          12603  \\
      SDSSJ1553+3548 &    238.27050 &     35.80795 &    0.721 &   0.08292 &    3 &    0.000 &    0.344 &    7 &    4 &    0 &          11598  \\
      SDSSJ1555+3628 &    238.76830 &     36.48000 &    0.714 &   0.18915 &    3 &    0.000 &    0.345 &    5 &    3 &    0 &          11598  \\  
      SDSSJ1616+4154 &    244.20590 &     41.90453 &    0.440 &   0.32098 &    3 &    0.077 &    0.345 &    6 &    5 &    0 &          11598  \\
      SDSSJ1619+3342 &    244.81894 &     33.71067 &    0.470 &   0.09638 &    3 &    0.000 &    0.345 &   10 &   10 &    0 &          11598  \\
    IRAS-F22456-5125 &    342.17165 &    -51.16475 &    0.100 &   0.09937 &    3 &    0.000 &    0.082 &   32 &   16 &    1 &          11686  \\ 
\hline 
\end{tabular} 
\label{tab:strong}  
\begin{tablenotes}[para, centering]    
Notes-- (1) QSO name; (2) Right-ascension (J2000); (3) Declination (J2000); (4) QSO redshift; (5) Absorption redshift; (6) COS gratings used for observations, 1: G130M, 2: G160M, 3: G130M+160M; (7) Minimum search redshift; (8) Maximum search redshift; (9) S$/$N of the G130M data estimated near 1400~\AA; (10) S$/$N of the G160M data estimated near 1600~\AA. S$/$N is $-1$ when data are not available; (11) Spectrum flag, 1: part of the statistical sample, 0: not part of the statistical sample; (12) $HST$ proposal ID. 
\end{tablenotes}       
\end{threeparttable}  
\end{table*}

%% file: tables/intrinsic-systems.tex
\begin{table*}
\begin{threeparttable}[t]   
\caption{List of weak \MgII\ analogs that are intrinsic to the background QSO. These systems are {\em not} part of the sample studied here.}    
\begin{tabular}{lrrccccccccc} 
\hline \hline 
QSO  & RA &  Dec  &  \zqso  &  \zabs  &  Gratings  &  $z_{\rm min}$  &  $z_{\rm max}$  & $(\rm S/N)_{G130M}$  & $(\rm 
S/N)_{G160M}$ & Flag &  PID \\ 
(1)  & (2) & (3)  & (4) & (5) & (6) & (7) & (8) & (9) & (10) & (11) & (12) \\ 
\hline 
      SDSSJ0012-1022 &      3.10004 &    -10.37403 &    0.228 &   0.22936 &    3 &    0.000 &    0.208 &    5 &    3 &    0 &          12248  \\
      SDSSJ0155-0857 &     28.87509 &     -8.95111 &    0.165 &   0.16412 &    3 &    0.000 &    0.146 &    6 &    5 &    0 &          12248  \\
      QSO-B1411+4414 &    213.45137 &     44.00389 &    0.089 &   0.08286 &    1 &    0.000 &    0.071 &   13 &   -- &    1 &          12569  \\
    IRAS-F22456-5125 &    342.17165 &    -51.16475 &    0.100 &   0.09937, 0.09592 &    3 &    0.000 &    0.082 &   32 &   16 &    1 &          11686  \\  
\hline 
\end{tabular} 
\label{tab:intrinsic} 
\begin{tablenotes}[para, centering]    
Notes-- (1) QSO name; (2) Right-ascension (J2000); (3) Declination (J2000); (4) QSO redshift; (5) Absorption redshift; (6) COS gratings used for observations, 1: G130M, 2: G160M, 3: G130M+160M; (7) Minimum search redshift; (8) Maximum search redshift; (9) S$/$N of the G130M data estimated near 1400~\AA; (10) S$/$N of the G160M data estimated near 1600~\AA\ (blank when data are not available); (11) Spectrum flag, 1: part of the statistical sample, 0: not part of the statistical sample; (12) $HST$ proposal ID.    
\end{tablenotes}       
\end{threeparttable}  
\end{table*}

%% file: tables/velocityplots.tex
\clearpage 
\begin{figure*} 
\includegraphics[width=0.98\textwidth]{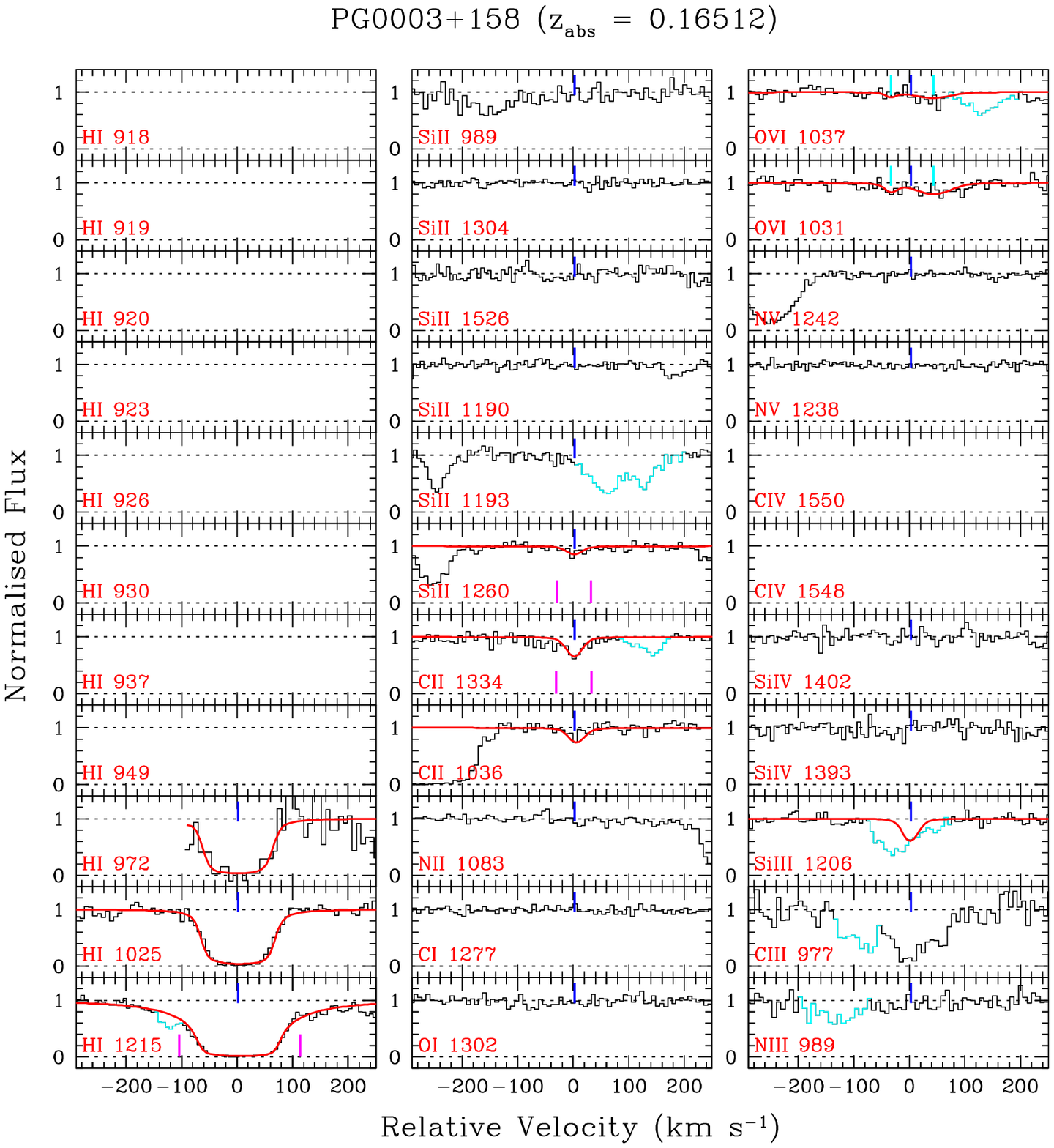} 
\vskip-0.8cm  
\caption{Velocity plot of the \zabs\ $=0.16512$ system towards QSO PG~0003$+$158. Relevant parts of the COS spectrum covering various important transitions from the absorber are shown on a velocity scale. The zero velocity corresponds to the absorption redshift, mentioned on the top. The smooth red curves overplotted on top of the data (black histogram) are the best fitting Voigt profile models. No data are displayed when a line is not covered by the COS spectrum. The blue vertical ticks mark the line centroids of the model Voigt profiles and the pairs of vertical magenta lines in \lya, \SiII$\lambda$1260, and \CII~$\lambda$1334 panels mark the region over which $W_r$ of the corresponding line is measured. Unrelated absorption lines are shown in cyan. Very weak \SiII$\lambda1260$ and \CII$\lambda1334$ absorption lines are seen. The \SiIII$\lambda1206$ line is heavily blended with the \lyb\ from the \zabs~$=0.37037$ system. The \lyg\ and \CIII$\lambda977$ lines fall at the blue edge of the spectrum. Both the lines require a velocity shift of $\sim +30$ \kms\ in order to be properly aligned with the other lines. The \lyb, \CII$\lambda1036$, and \OVI\ doublets, on the other hand, require a velocity shift of $\sim-10$~\kms. The \CII\ and \SiII\ lines are fitted simultaneously keeping the $z$ and Doppler-parameter ($b$) tied to each other. Since the \SiIII\ line is blended, we estimate the maximum allowed column density, using a $z$ and $b$-parameter same as the \CII/\SiII\ line, that do not exceed the observed profile. The simultaneous and single component fit to the Lyman series lines explains the observed profiles adequately. We believe that the $N(\HI)$ of $10^{18.16\pm0.03}$~\sqcm\ is adequately constrained from the wing of the \lya\ line along with the \lyb\ and \lyg\ profiles. We thus assign a quality factor of $Q=3$ for this system.}         
\label{PG0003_158_0.16512} 
\end{figure*}

\begin{figure*} 
\includegraphics[width=0.98\textwidth]{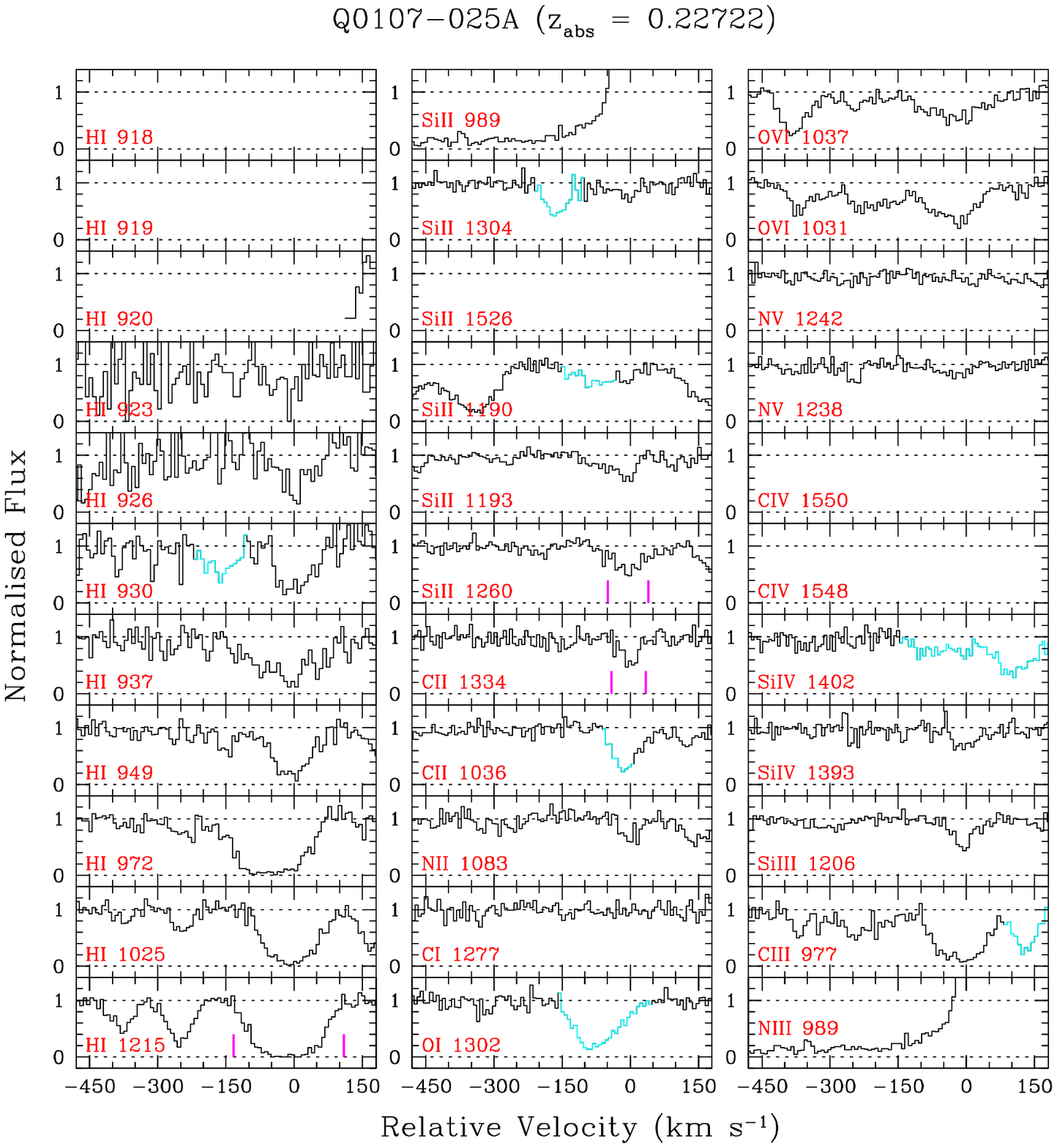} 
\vskip-0.8cm  
\caption{Similar to Fig.~\ref{PG0003_158_0.16512} but for the \zabs\ $=0.22722$ system towards Q0107--025A. The Lyman series lines below \HI$\lambda937$ fall at the blue edge of the spectrum. These lines require 10--20 \kms\ velocity shifts in order to be properly aligned with other members of the Lyman series lines. The blue wing of the \lyg\ line is blended with Galactic \SiII$\lambda1193$ absorption. The wavelength range corresponding to the \SiII$\lambda989$ and \NIII$\lambda989$ is affected by geo-coronal \lya\ emission. A full-blown multiphase analysis of this absorber is presented in \citet{Muzahid14}. Here we adopt the column densities of relevant ions from Table~1 of \citet{Muzahid14}. Since several unsaturated higher order Lyman series lines are available for this system, the measured $\log N(\HI)=15.92\pm0.08$ has the highest quality factor of $Q=5$.}          
\label{Q0107_025A_0.22722} 
\end{figure*}

\begin{figure*} 
\includegraphics[width=0.98\textwidth]{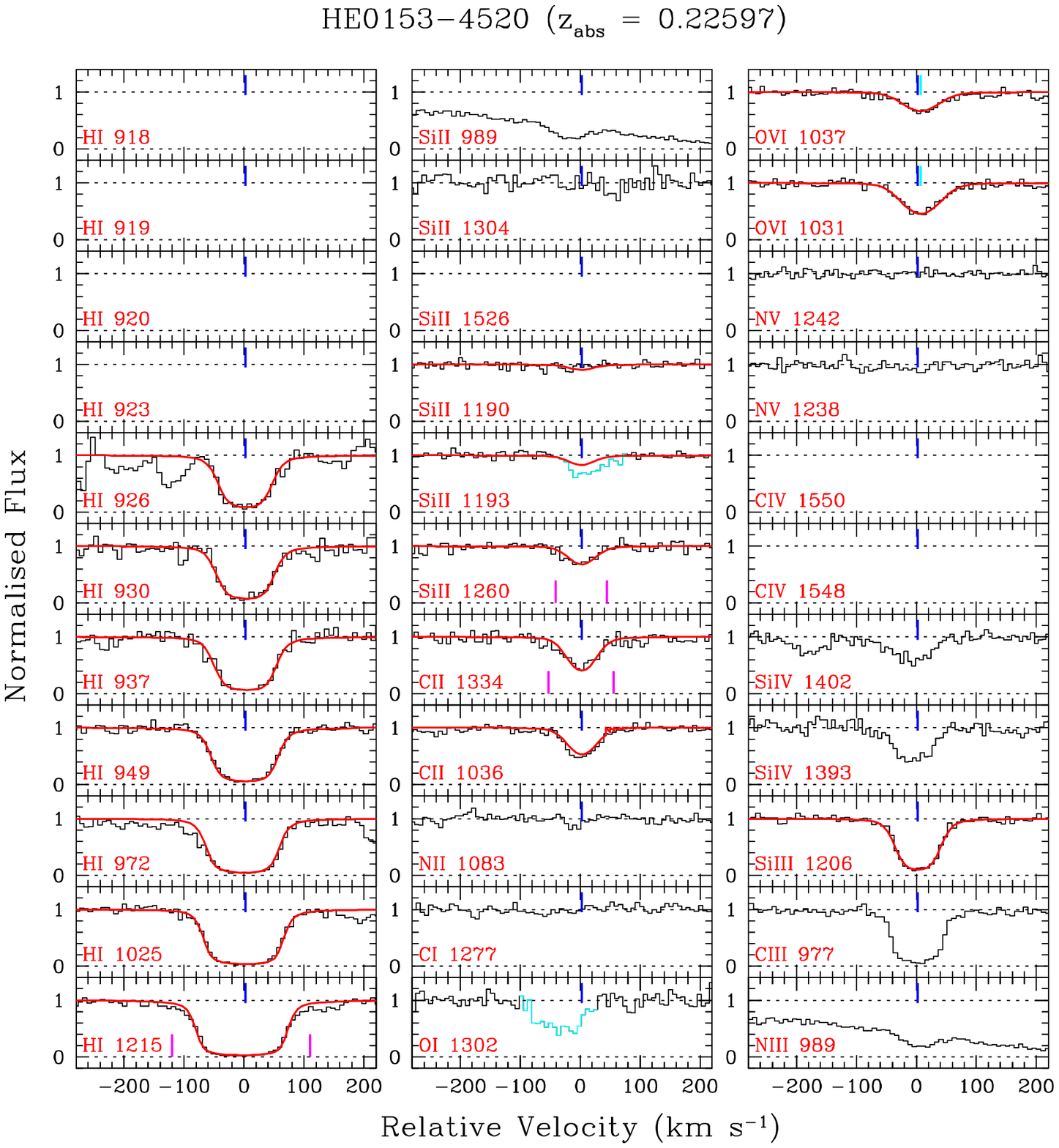} 
\vskip-0.8cm  
\caption{Similar to Fig.~\ref{PG0003_158_0.16512} but for the \zabs\ $=0.22597$ system towards HE0153--4520. The wavelength range corresponding to the \SiII$\lambda989$ and \NIII$\lambda989$ is affected by Galactic \lya\ absorption. The \SiII$\lambda1193$ line is blended with an unknown contaminant. The \HI$\lambda937$, $\lambda930$, and $\lambda926$ lines fall at the extreme edge of the spectrum. These lines require $\sim$$7$, $\sim$$15$, and $\sim$$30$ \kms\ velocity shifts, respectively, in order to be aligned with the other Lyman series lines. Voigt profile fitting of the metal lines (i.e., \CII$\lambda1036$, $\lambda1334$, \SiII$\lambda1260$, and \SiIII$\lambda1206$) is performed simultaneously by keeping the $z$ and $b$-parameter of each line tied to the others. Note that the \SiIII$\lambda1206$ line is strongly saturated. We consider the measured $N(\SiIII)$ to be a lower limit. A full-blown multiphase analysis of this absorber is presented in \citet{Savage11b}. The column densities we obtained for the metal lines are consistent with the values reported in \citet{Savage11b}. From the partial Lyman limit break seen in the FUSE spectrum they have constrained $\log N(\HI)=16.61^{+0.12}_{-0.17}$. Our single-component, simultaneous fit to the available Lyman series line provides a somewhat higher $N(\HI)$ value (i.e., $\log N(\HI)=16.83\pm0.07$) but the two values are consistent within $1.6\sigma$. We thus assign a quality factor of $Q=4$ for the $N(\HI)$ that we obtained.  }     
\label{HE0153_4520_0.22597} 
\end{figure*}

\begin{figure*} 
\includegraphics[width=0.98\textwidth]{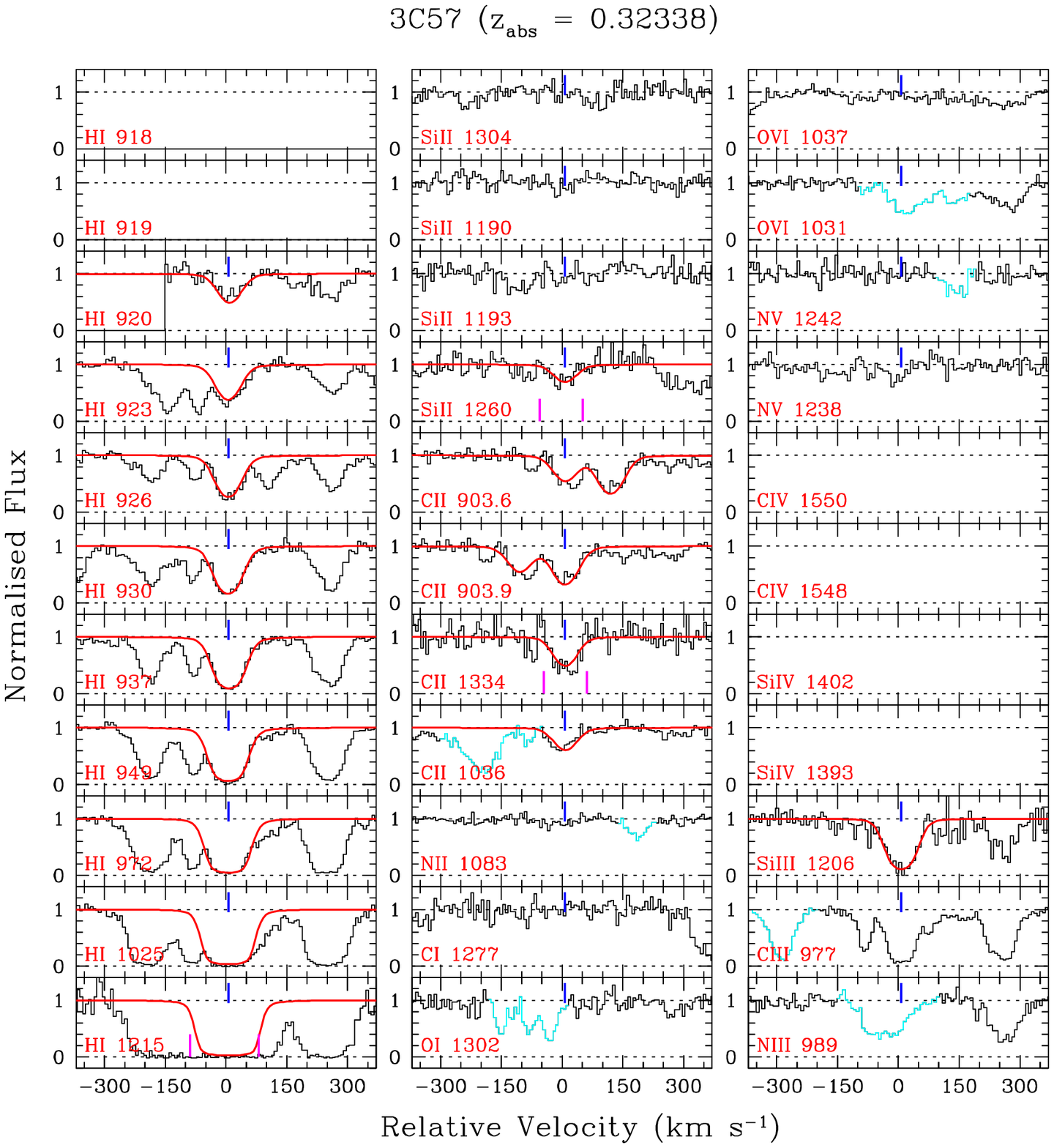} 
\vskip-0.8cm  
\caption{Similar to Fig.~\ref{PG0003_158_0.16512} but for the \zabs\ $=0.32338$ system towards 3C57. The presence of several unsaturated Lyman series lines permits an accurate $N(\HI)$ measurement for the component that shows weak \SiII\ and \CII\ absorption.  We assign a quality factor of $Q=5$ for this system. The lack of any detectable absorption in the \OVI$\lambda1037$ line near $0$~\kms\ indicates that the absorption seen in the \OVI$\lambda1031$ must be a blend. Voigt profile fitting for the metal lines is done simultaneously keeping the $z$ and $b$-parameter tied. Note that the \SiIII\ line is strongly saturated and hence the measured $N(\SiIII)$ is considered to be a conservative lower limit.}     
\label{3C57_0.32338} 
\end{figure*}

\begin{figure*} 
\includegraphics[width=0.98\textwidth]{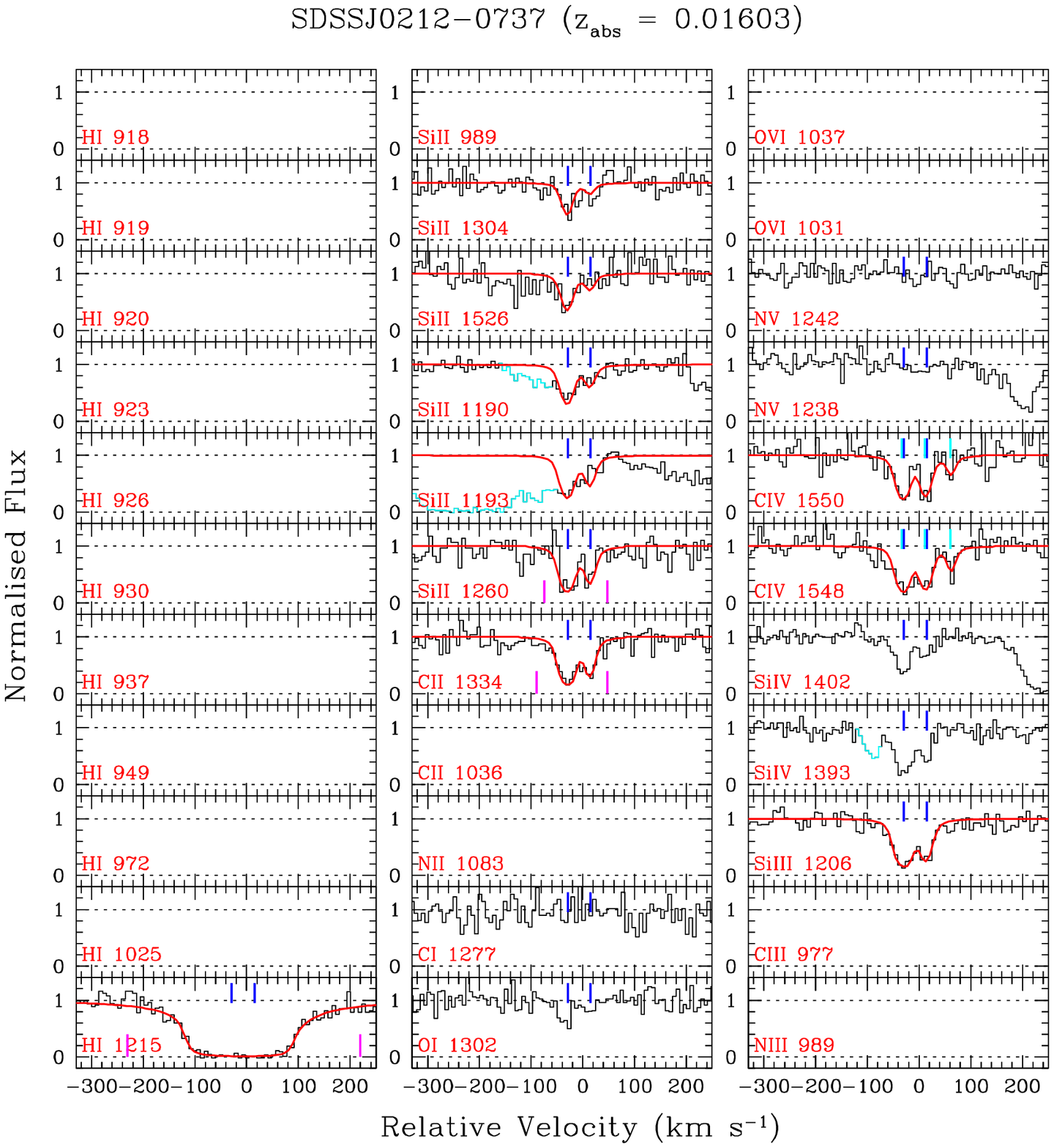} 
\vskip-0.8cm  
\caption{Similar to Fig.~\ref{PG0003_158_0.16512} but for the \zabs\ $=0.01603$ system towards SDSSJ0212--0737. Besides low-ions, strong \CIV\ and \SiIV\ lines are detected. The \SiII$\lambda1193$ line is blended with the strong \OVI~$\lambda1031$ absorption at \zabs~$=0.17403$, intrinsic to the background QSO. We note that the \SiII$\lambda1304$ and \OI$\lambda1302$ lines require a velocity shift of $\sim-7$~\kms\ with respect to the \SiIII. Additionally, the \CII$\lambda1334$ line is shifted by $\sim-3$~\kms. All the metal lines are fitted simultaneously using two Voigt profile components. The component at $\sim-30$~\kms\ is strongly saturated for both the \CII$\lambda1334$ and \SiIII$\lambda1206$ lines. In fact, the \CII$\lambda1334$ line is heavily saturated in both components. However, $N(\SiII)$ is adequately determined due to the presence of weaker \SiII\ lines (e.g. $\lambda1304$). The \lya\ profile and the presence of the \OI$\lambda1302$ line indicate a high $N(\HI)$. We believe that the estimated $N(\HI)$, using a two-component fit, is reasonably constrained from the wings of the \lya\ absorption. We assign $Q=3$ for this system.}       
\label{SDSSJ0212_0737_0.01603} 
\end{figure*}

\begin{figure*} 
\includegraphics[width=0.98\textwidth]{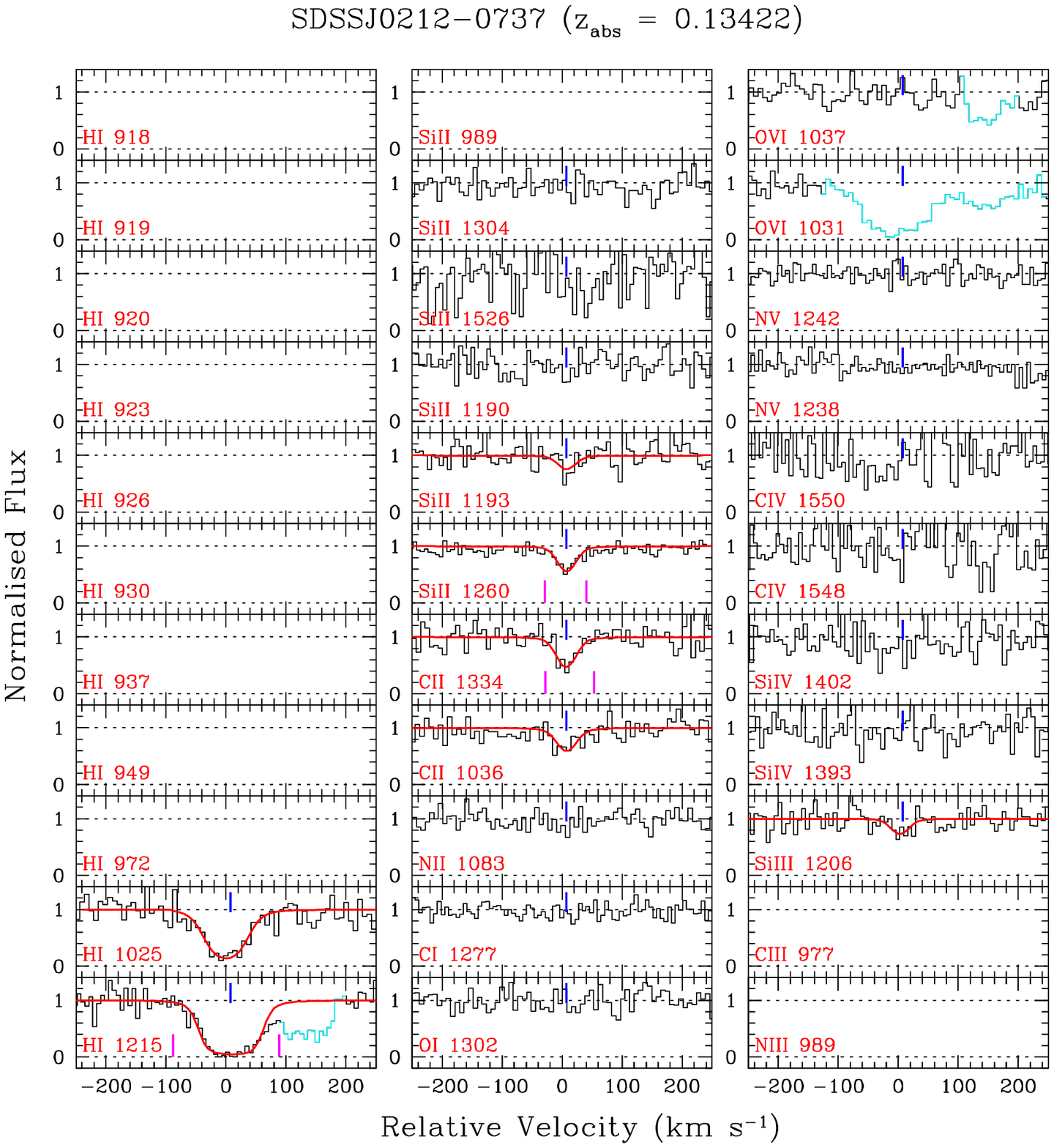} 
\vskip-0.8cm  
\caption{Similar to Fig.~\ref{PG0003_158_0.16512} but for the \zabs\ $=0.13422$ system towards SDSSJ0212--0737. The absorption seen in the \OVI$\lambda1031$ panel is actually \lyb\ from the \zabs\ $= 0.14107$ system. We note that there is a velocity offset of $\sim$$10$ \kms\ between the \lya\ and \lyb\ absorption lines. The metal lines are fitted simultaneously keeping $z$ and $b$-parameter tied. Similarly, we perform a simultaneous fit to the \lya\ and \lyb\ lines. Because the \lyb\ absorption is only mildly saturated, with total $\log N(\HI)=14.95\pm0.08$, we assign a quality factor of $Q=3$.}    
\label{SDSSJ0212_0737_0.13422} 
\end{figure*}

\begin{figure*} 
\includegraphics[width=0.98\textwidth]{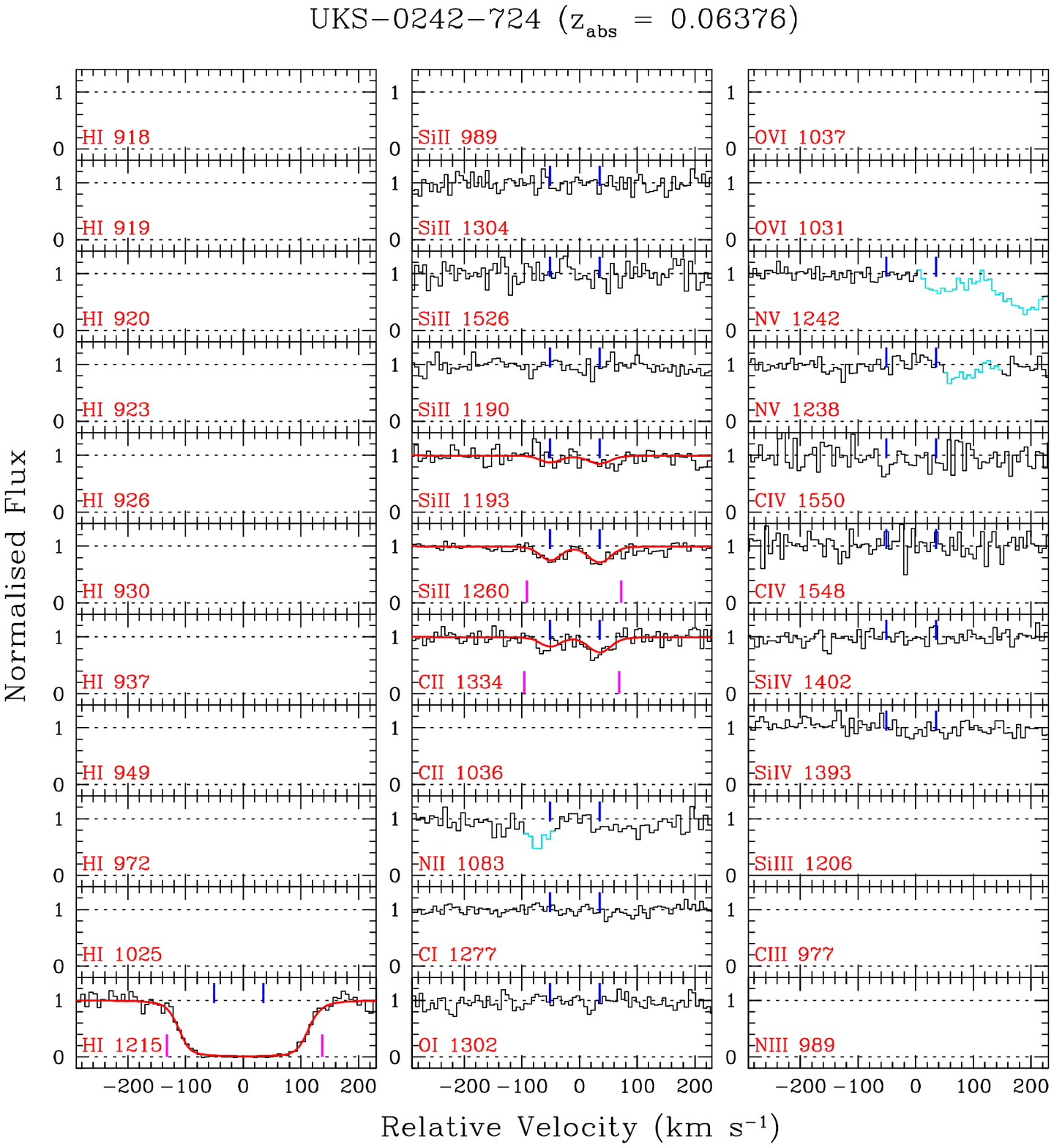} 
\vskip-0.8cm  
\caption{Similar to Fig.~\ref{PG0003_158_0.16512} but for the \zabs\ $=0.06376$ system towards UKS--0242--724. The \SiIII$\lambda1206$ line falls on a spectral gap. Both the \SiII\ and \CII\ lines show a two-component structure. All three metal transitions are fitted simultaneously keeping the $z$ and $b$-parameter tied to each other. Apart for \lya, no other Lyman series lines are covered by the spectrum. We have used two components to fit the \lya\ line using component redshifts obtained from the metal line fitting.  Since the \lya\ profile is heavily saturated and no higher order Lyman series lines are available, the estimated $N(\HI)$ is assigned the lowest quality factor of $Q=1$.}    
\label{UKS0242_724_0.06376} 
\end{figure*}

\begin{figure*} 
\includegraphics[width=0.98\textwidth]{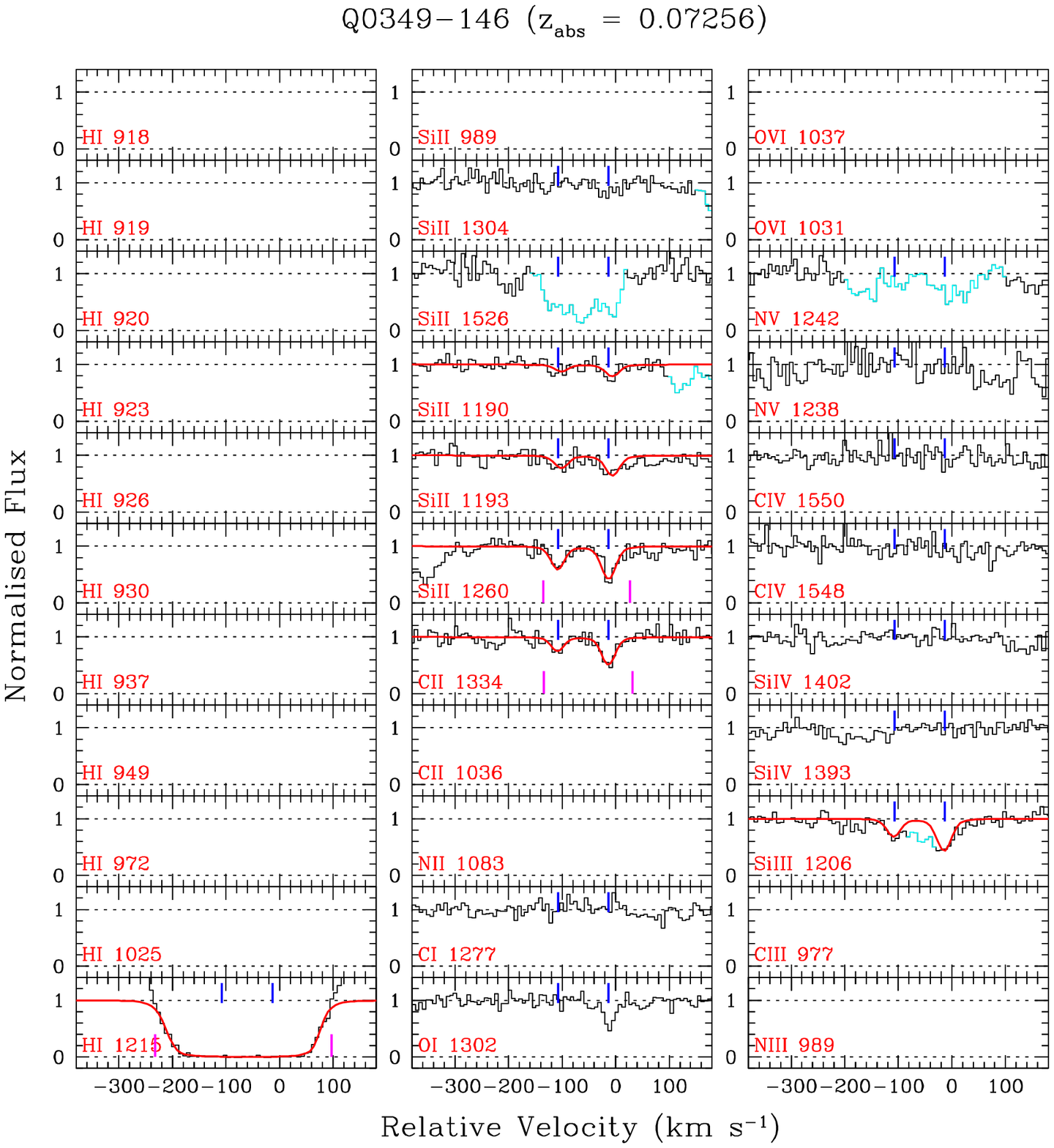} 
\vskip-0.8cm  
\caption{Similar to Fig.~\ref{PG0003_158_0.16512} but for the \zabs\ $=0.07256$ system towards Q0349--146. The absorption seen in the \SiII$\lambda1526$ panel is the \SiIII$\lambda1206$ line of the \zabs~$=0.35687$ system. The \SiII$\lambda1260$ and \CII$\lambda1334$ lines are $\sim+8$~\kms\ and $\sim-13$~\kms\ offset with respect to the \SiIII\ line, respectively. The \SiII\ and \CII\ lines show two-component absorption kinematics. There is some absorption in-between the two components in the \SiIII\ line which is likely to be a blend. All the metal lines are fitted simultaneously. The presence of narrow \OI\ line in the stronger component indicates a high $N(\HI)$ value. The \lya\ line falls in between the geo-coronal \OI$\lambda1302$ and \SiII$\lambda1304$ emission lines. We used two Voigt profile components, as seen in the \CII/\SiII\ lines, with redshifts locked at the values obtained from the metal-line fitting, in order to fit the \lya\ profile. This system is assigned a quality factor of $Q=1$, since we could not determine the \HI\ column density adequately.}      
\label{Q0349_146_0.07256} 
\end{figure*}

\begin{figure*} 
\includegraphics[width=0.98\textwidth]{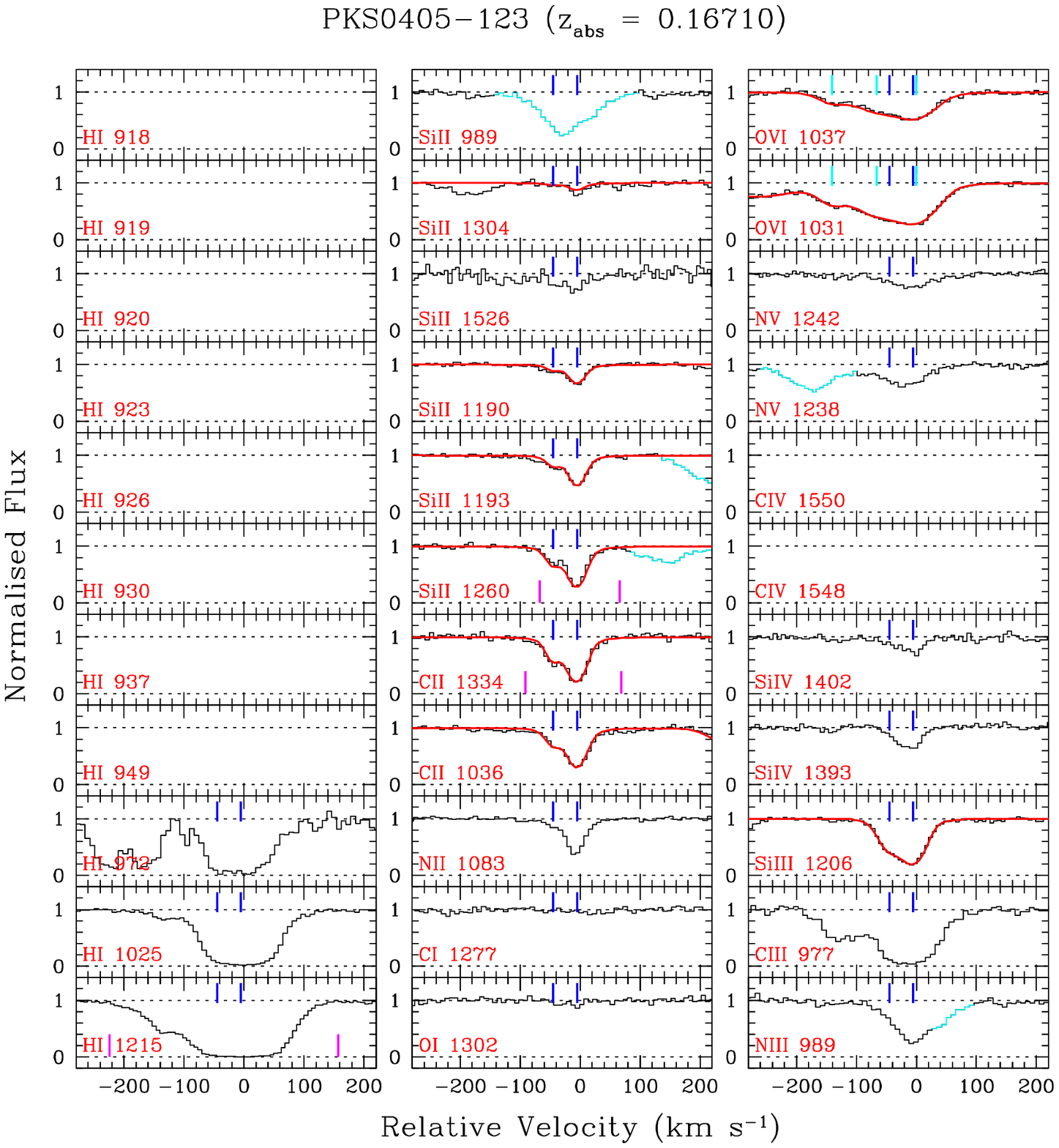} 
\vskip-0.8cm  
\caption{Similar to Fig.~\ref{PG0003_158_0.16512} but for the \zabs~$= 0.16710$ system towards PKS0405--123. This system has been studied by several authors \citep[e.g.,][]{Prochaska04b,Savage10}. A full-blown multiphase analysis of this absorber is presented in \citet{Savage10}. This particular spectrum has a serious wavelength-alignment problem. However, the high spectral $\rm S/N$, and relatively simple and uncontaminated absorption kinematics make it easy to understand the nature of the wavelength calibration uncertainty. From the partial Lyman limit break observed in a {\sc fuse} spectrum, \citet{Prochaska04b} have constrained the total $\log N(\HI)=16.45\pm0.05$. Here we adopt this $N(\HI)$ value and do not attempt to estimate $N(\HI)$ from the COS spectrum since all three Lyman series lines are heavily saturated. We assign $Q=5$ for this absorber. The different transitions of \CII\ and \SiII\ are fitted simultaneously keeping $z$ and $b$-parameters tied. \SiIII, however, requires somewhat higher $b$-values. The total column densities we estimate for the metal lines are consistent with the measurements of \citet{Savage10}. Both the \CII\ and \SiIII\ lines are strongly saturated. We consider both $N(\CII)$ and $N(\SiIII)$  as conservative lower limits \citep[see also][]{Savage10}.}    
\label{PKS0405_123_0.16710} 
\end{figure*}

\begin{figure*} 
\includegraphics[width=0.98\textwidth]{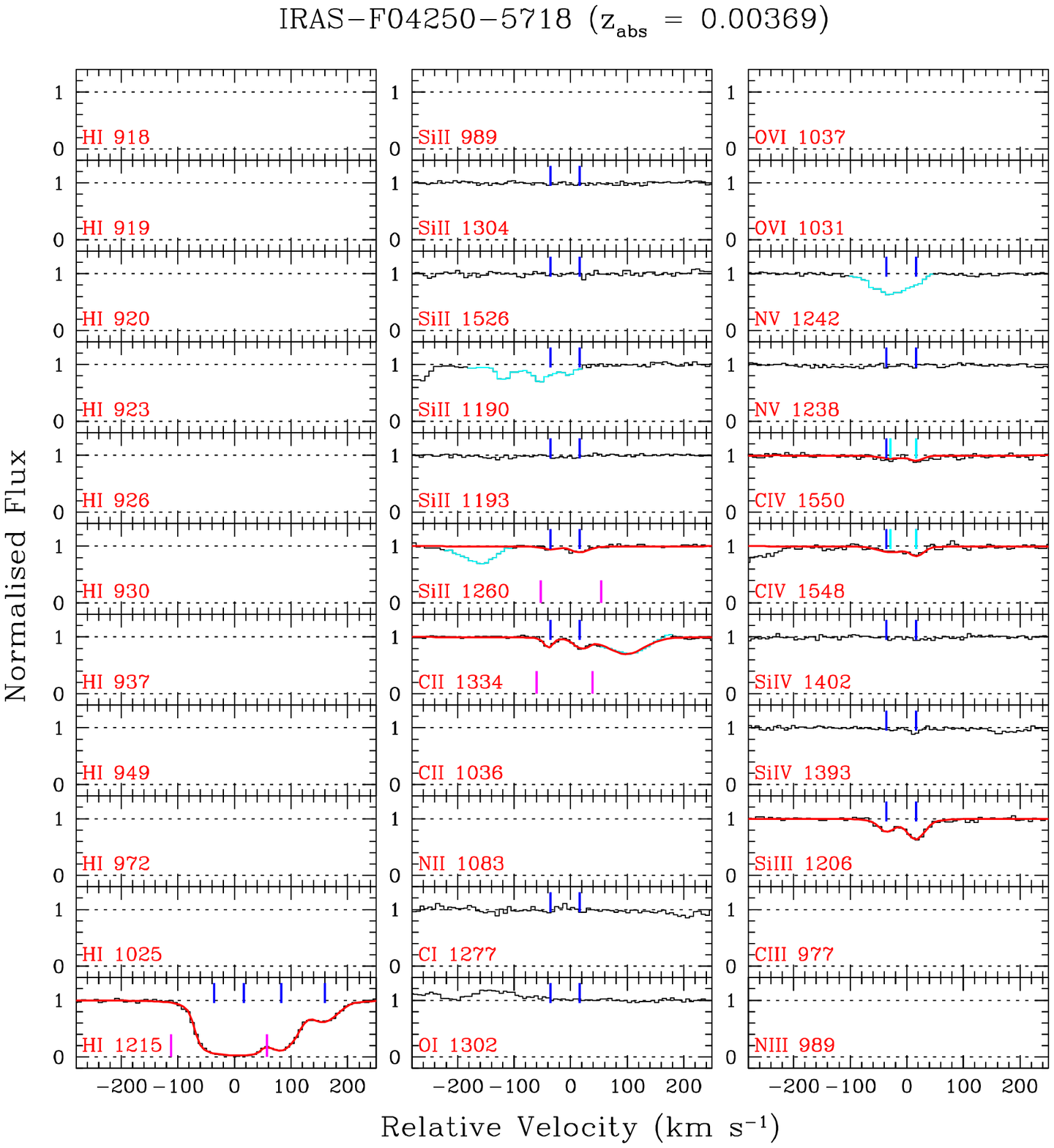} 
\vskip-0.8cm  
\caption{Similar to Fig.~\ref{PG0003_158_0.16512} but for the \zabs~$=0.00369$ system towards IRAS--F04250--5718. Very weak \CII$\lambda1334$, \SiII$\lambda1260$, \SiIII, and \CIV\ doublet lines are detected. The red wing of the \CII$\lambda1334$ line is blended, possibly with the \lya\ of an associated absorber at \zabs~$=0.10220$. Our fit to the \CII$\lambda1334$ line is corrected for the contaminating absorption. Except for the \lya, no other Lyman series lines are covered. We use the line centroids and $b$-parameters of the \SiIII\ lines to fit the main \lya\ absorption clump. In addition, two weak \lya\ components are required to fit the red wing. Since the main \lya\ absorption clump which is related to the metal lines is heavily saturated, we assign a quality factor of $Q=1$ for this system.}    
\label{IRASF04250_5718_0.00369} 
\end{figure*}

\begin{figure*} 
\includegraphics[width=0.98\textwidth]{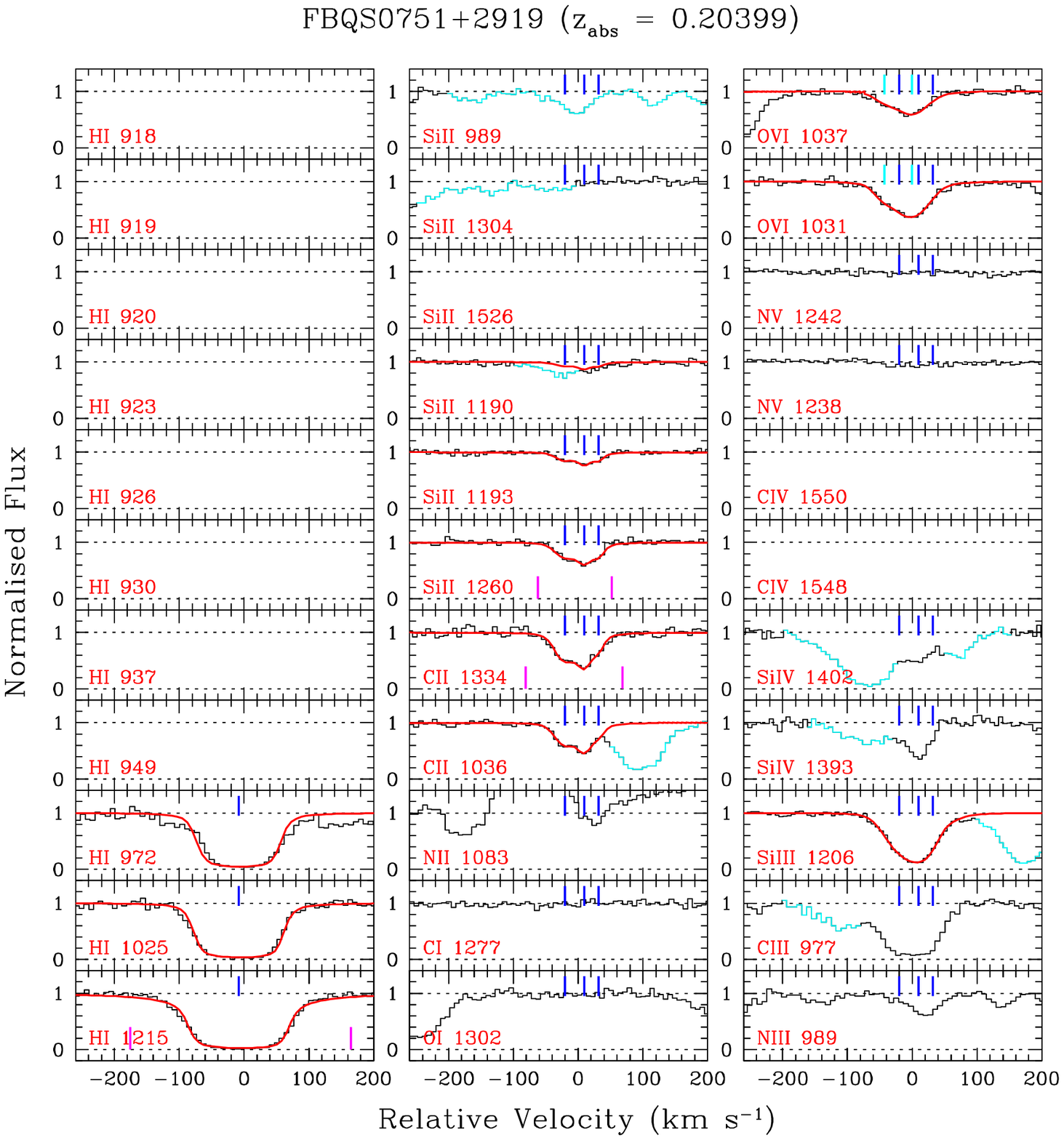} 
\vskip-0.8cm  
\caption{Similar to Fig.~\ref{PG0003_158_0.16512} but for the \zabs~$=0.20399$ system towards FBQS~0751+2919. Besides several weak low-ionization lines, strong \OVI\ is detected, however, no \NV\ is present. The \NII$\lambda1083$ absorption is severely affected by the geo-coronal \OI\ emission. The blue wing of the \CIII\ line is possibly blended with the \lyb\ of the \zabs~$=0.14644$ system. The strong absorption redward of the \CII$\lambda1036$ line is the \lya\ from the \zabs~$=0.02673$ system. The blue wings of the \SiIV$\lambda1393$ and $\lambda1402$ are blended by the higher order Lyman series lines (i.e., \HI$\lambda917$ and $\lambda923$, respectively) from the \zabs~$=0.82912$ system. We note that the lines with rest-frame wavelength $\lesssim1050$~\AA\ require a velocity shift of $\sim-12$~\kms. A minimum of three components is required to fit the \CII\ and \SiII\ lines simultaneously. The \SiIII\ line is strongly saturated and required larger component $b$-values compared to the singly ionized metal ions, suggesting an additional contribution from other gas phase(s). The measured $N(\SiIII)$ should be treated as a lower limit. The available Lyman series lines are heavily saturated. We have performed both single-component and three-component fits to the Lyman series lines. A single-component fit gives $\log N(\HI) = 17.77\pm0.05$ and $b(\HI)=24$~\kms. A three-component fit, on the other hand, gives an order of magnitude lower $N(\HI)$ of $10^{16.67\pm0.08}$~\sqcm. We adopted the single component fit results keeping in mind that the true $N(\HI)$ could only be less. We assign a quality factor of $Q=2$ for the system.}     
\label{FBQS0751_2919_0.20399} 
\end{figure*}

\begin{figure*} 
\includegraphics[width=0.98\textwidth]{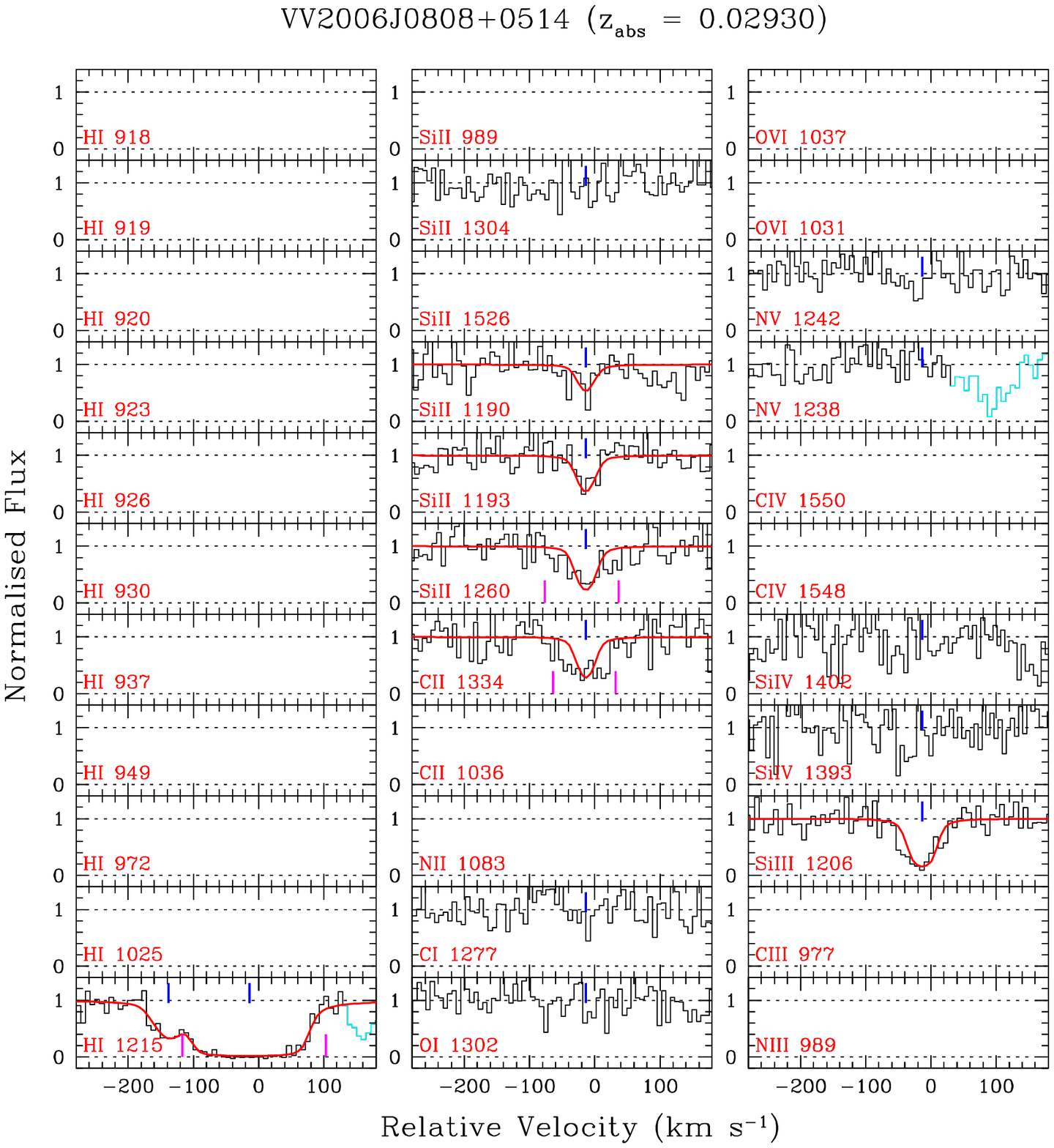} 
\vskip-0.8cm  
\caption{Similar to Fig.~\ref{PG0003_158_0.16512} but for the \zabs~$=0.02930$ system towards VV2006~J0808+0514. The \CIV\ doublet is not covered since the spectrum is taken only with the G130M grating covering up to $1460$~\AA\ (observed). The spectrum looks noisy, however, consistent metal absorption lines are detected. First we fit the \SiII\ lines ($\rm \lambda1190, \lambda1193, and \lambda1260$) simultaneously. We then use the best-fitting $z$ and $b$-parameter to fit the \CII$\lambda1334$ and \SiIII$\lambda1206$ lines. The \CII\ line might have contributions from other unknown and unrelated absorption. We consider the measured $N(\CII)$ as a conservative upper limit. The \SiIII\ line is heavily saturated and hence $N(\SiIII)$ should be treated as a lower limit. The $N(\HI)$ we estimate from the \lya\ profile alone ($\log N(\HI)=17.57\pm0.43$) is highly uncertain. We assign a quality factor of $Q=1$ for this system.}    
\label{VV2006J0808_0514_0.02930} 
\end{figure*}

\begin{figure*} 
\includegraphics[width=0.98\textwidth]{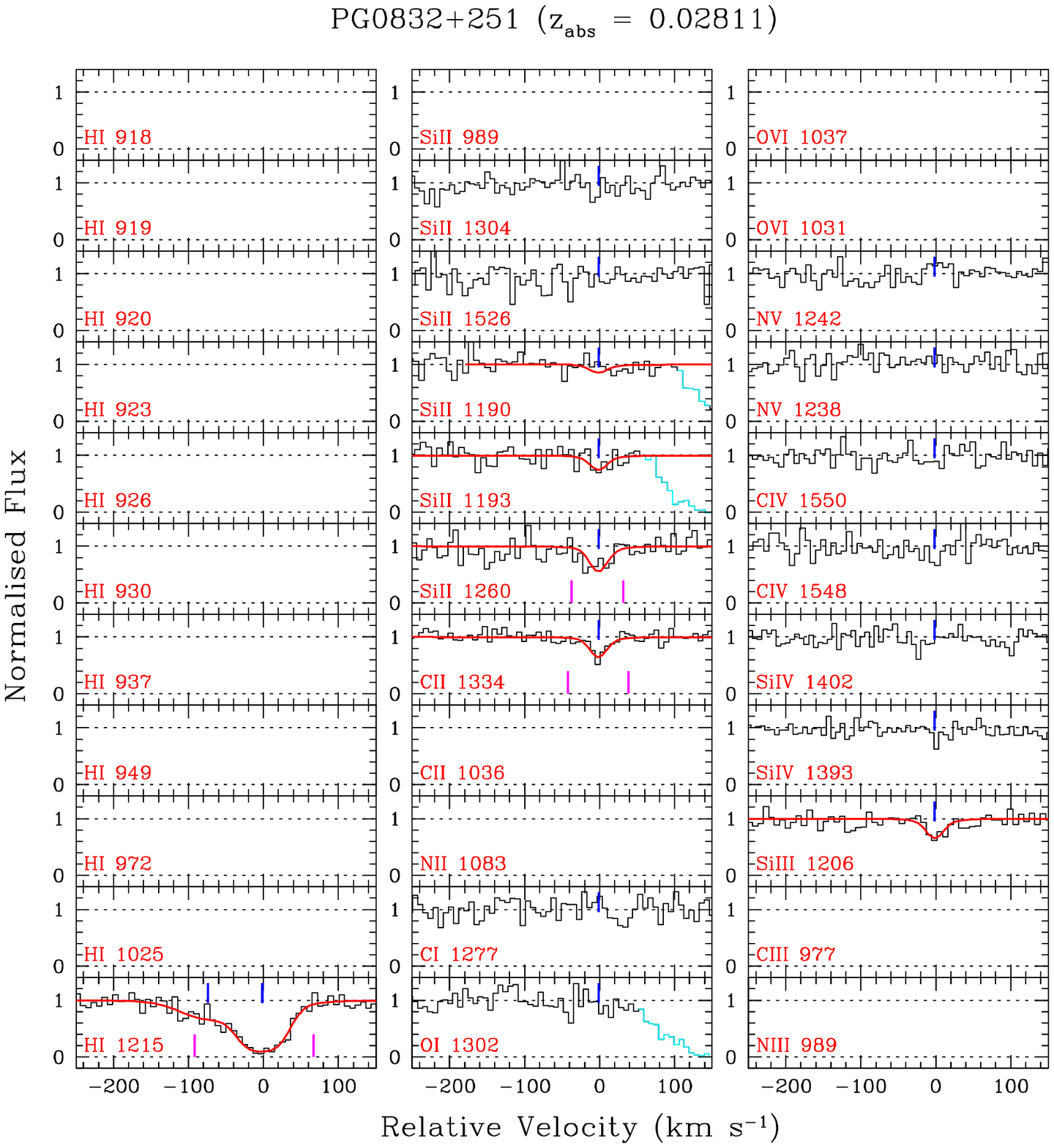} 
\vskip-0.8cm  
\caption{Similar to Fig.~\ref{PG0003_158_0.16512} but for the \zabs~$=0.02811$ system towards PG~0832+251. This is a very weak metal line absorber. Except for the \CII$\lambda1334$, which requires a velocity shift of $\sim+5$~\kms, all other lines are well aligned. The \lya\ line, which is only mildly saturated, exhibits an extra broad, weak component in the blue wing. Since the \lya\ is not heavily saturated we assign a quality factor of $Q=2$ for this system.}    
\label{PG0832_251_0.02811} 
\end{figure*}

\begin{figure*} 
\includegraphics[width=0.98\textwidth]{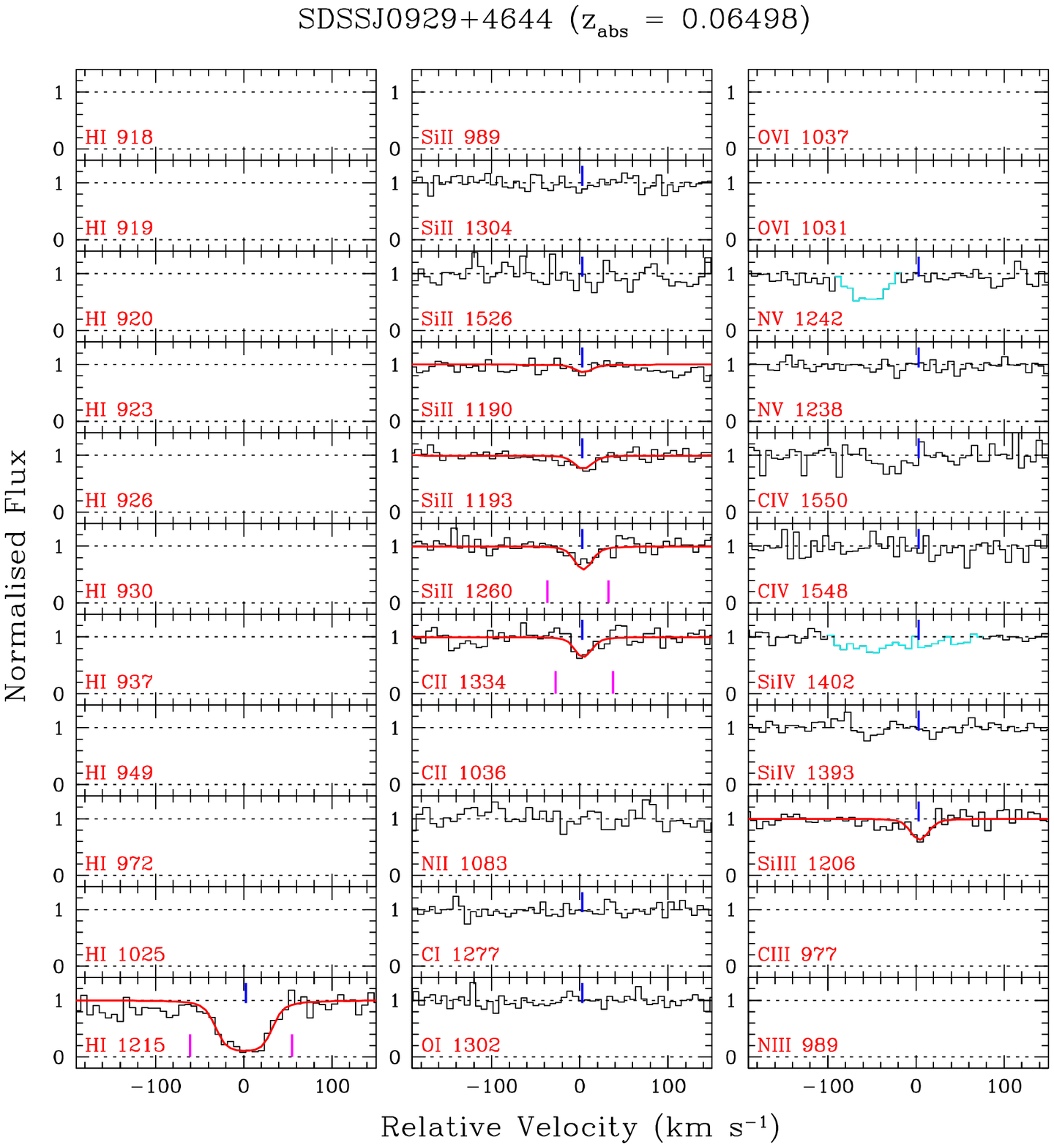} 
\vskip-0.8cm  
\caption{Similar to Fig.~\ref{PG0003_158_0.16512} but for the \zabs~$=0.06498$ system towards SDSS~J0929+4644. Single-component, weak metal lines are fitted simultaneously keeping $z$ and $b$-parameter tied. The \lya\ is narrow, only mildly saturated, and off center by 8 \kms\ with respect to \CII/\SiII. We assign a quality factor of $Q=2$ for this system. }     
\label{SDSSJ0929_4644_0.06498} 
\end{figure*}

\begin{figure*} 
\includegraphics[width=0.98\textwidth]{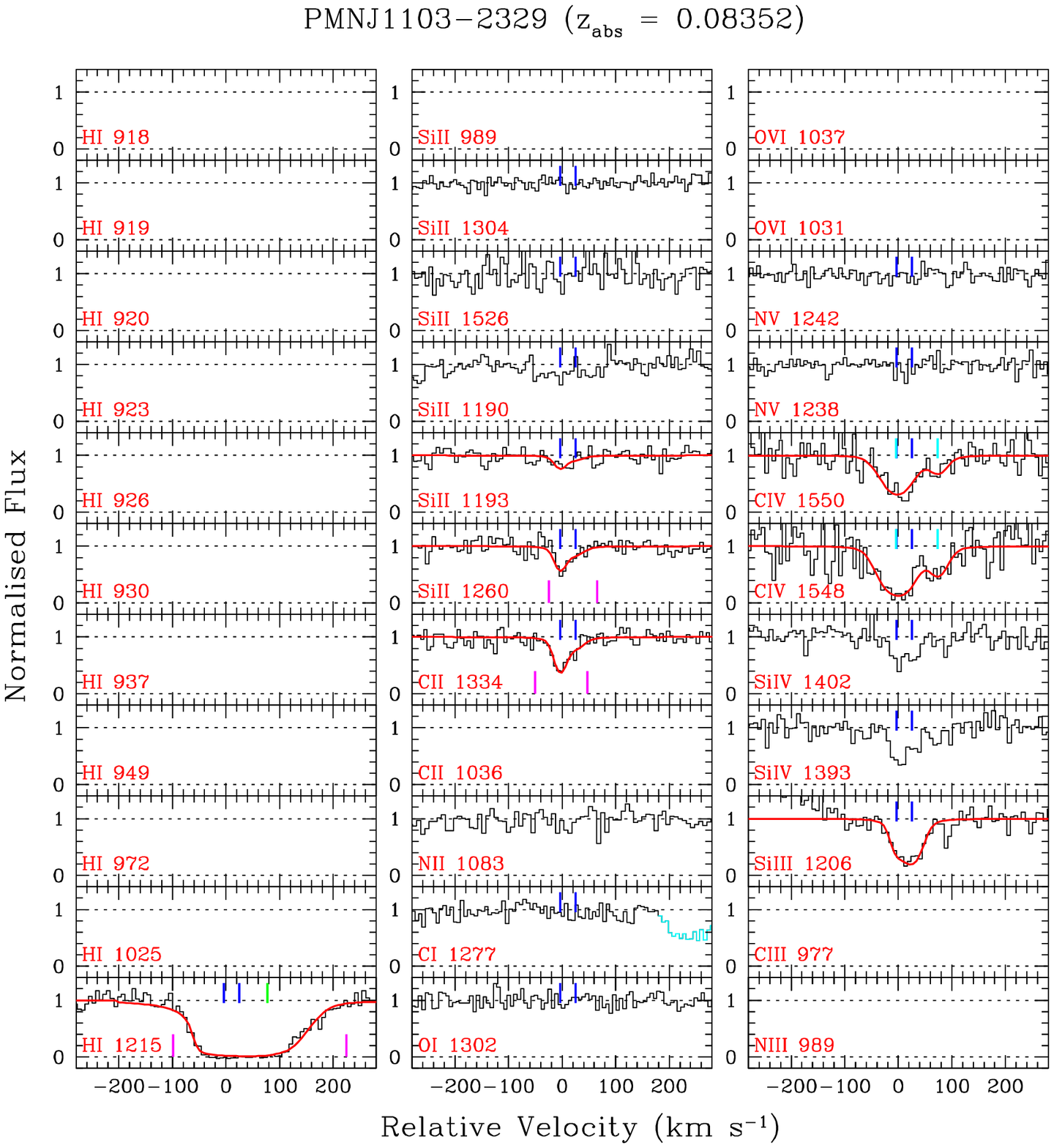} 
\vskip-0.8cm  
\caption{Similar to Fig.~\ref{PG0003_158_0.16512} but for the \zabs~$=0.08352$ system towards PMNJ1103--2329. Besides the low-ionization lines, multi-component, strong \CIV\ and \SiIV\ lines are present. The spectrum at $\sim-200$~\kms\ in the \SiIII$\lambda1206$ panel is affected by the geo-coronal \OI+\SiII\ emission lines. A minimum of two components is required to fit the metal lines simultaneously. The strength of the \SiIII\ absorption indicates that the line is heavily saturated. Therefore, the $N(\SiIII)$ should treated as a lower limit. In addition to the two components seen in the low-ionization lines, the heavily saturated \lya\ line requires a broad component at $\sim+80$~\kms, corresponding to the \CIV$/$\SiIV\ component seen at the same velocity. Some amount of \SiIII\ may be present in this component as indicated by the spiky absorption we ignored in our analysis. However, the contribution of such a weak component to the total $N(\SiIII)$ would be negligible. A free $b$-parameter fit to the \lya\ profile was unstable. Thus, for estimating $N(\HI)$ associated with the two components that are aligned with the low-ions, we assumed $b(\HI)=18.2$, corresponding to $T=2\times10^{4}$~K, for each component and determined the maximum $N(\HI)$ that can be accommodated within the observed profile, one-by-one. This allow us to put an upper limit on the $N(\HI)$. Due to the absence of the higher order Lyman series lines, we assign the lowest quality factor (i.e. $Q=1$) for this system.}   
\label{PMNJ1103_2329_0.08352} 
\end{figure*}

\begin{figure*} 
\includegraphics[width=0.98\textwidth]{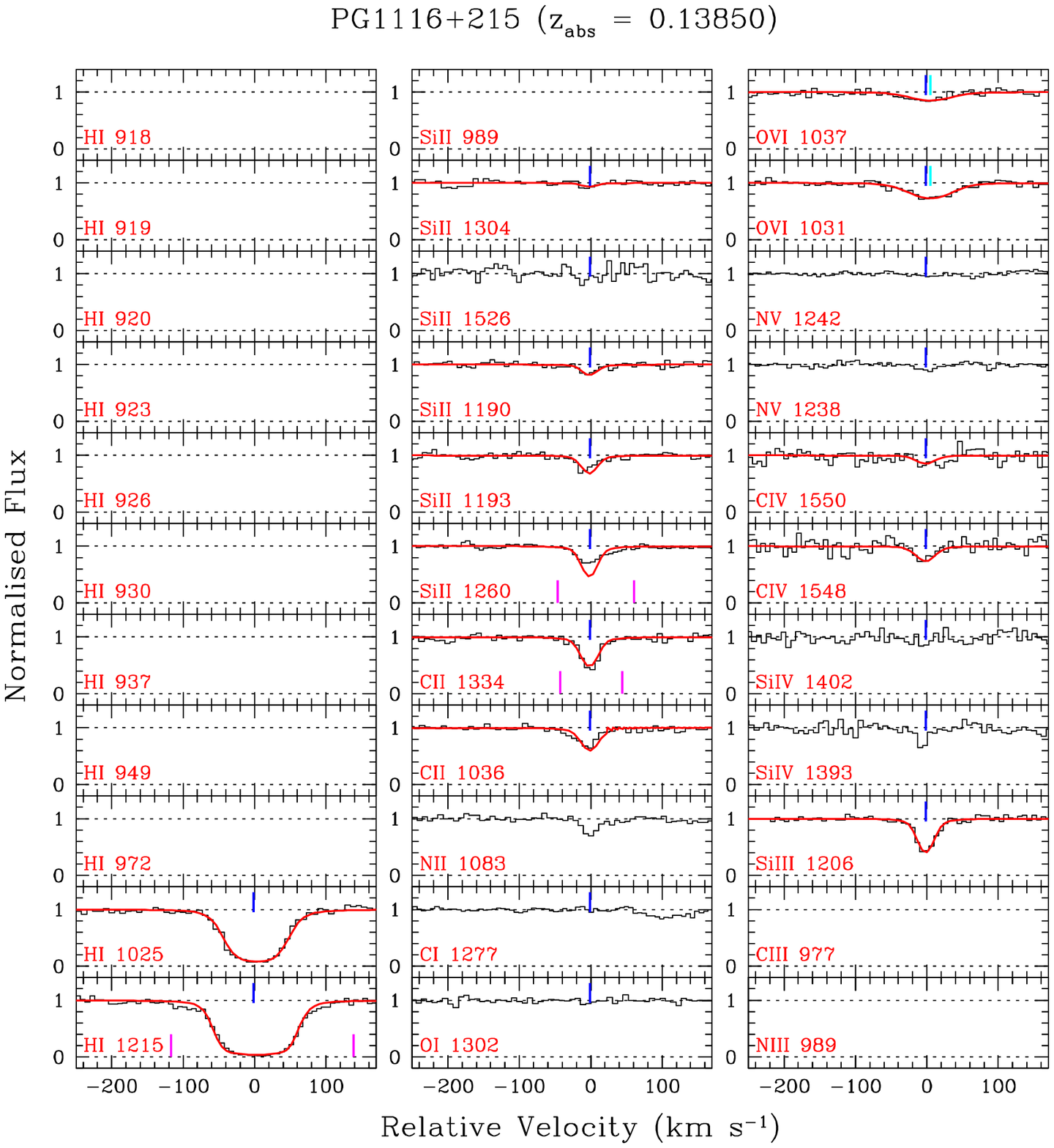} 
\vskip-0.8cm  
\label{PG1116_215_0.13850} 
\caption{Similar to Fig.~\ref{PG0003_158_0.16512} but for the \zabs~$=0.13850$ system towards PG1116+215. This is a narrow, weak, single component system. We note that the lines with rest-frame wavelength $<1040$~\AA\ require a $\sim-9$~\kms\ velocity shift. The metal lines are fitted simultaneously keeping the $z$ and $b$-parameter tied. We notice that the \SiII$\lambda1260$ line falls on top of the QSO's \lya+\NV\ emission line. The apparent mismatch between the data and the model profile for the \SiII$\lambda1260$ line could be due to the fact that the absorber is not fully covering the broad emission line region of the background QSO (i.e., the partial coverage effect, see e.g., \citet{Muzahid12b}). This would indicate that the size of the absorber is very small. However, the possibility of unresolved saturation of the \SiII$\lambda1260$ line and/or uncertainty in the continuum placement also cannot be ruled out. The constraint on $N(\SiII)$ comes from the other weaker transitions, in particular, from the \SiII$\lambda1190$ line. FUSE and STIS observations of the full Lyman series lines (i.e., \lya--\HI$\lambda916$) are presented by \citet{Sembach04}. Here we adopted their best-fitting COG value of $\log N(\HI)=16.20\pm0.05$ and assign the highest quality factor of $Q=5$.}     
\end{figure*}

\begin{figure*} 
\includegraphics[width=0.98\textwidth]{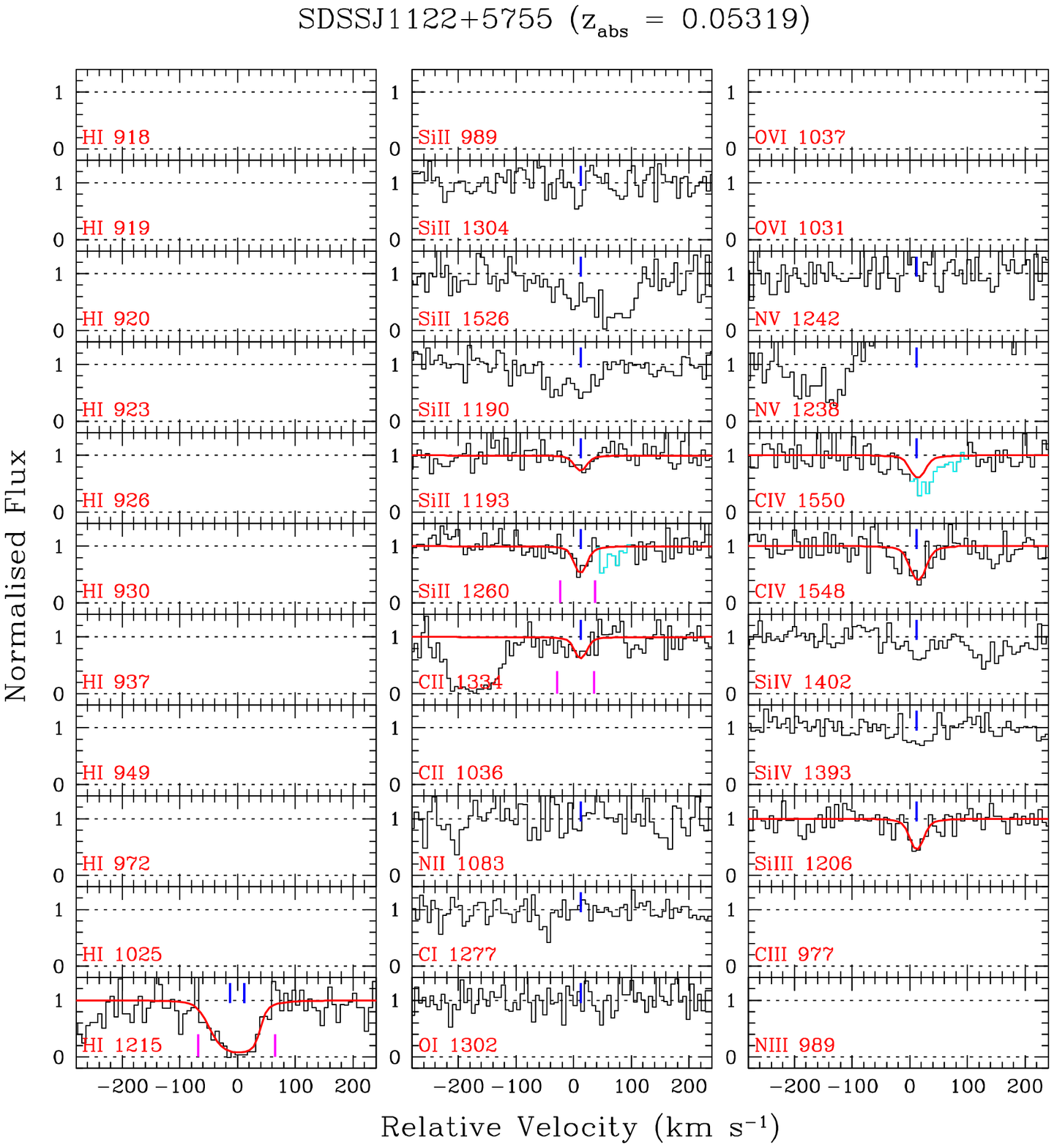} 
\vskip-0.8cm  
\caption{Similar to Fig.~\ref{PG0003_158_0.16512} but for the \zabs~$=0.05319$ system towards SDSSJ1122+5755. The absorption seen in the \SiII$\lambda1190$ and $\lambda1526$ panels actually arise from the Galactic \SII$\lambda1253$ and \FeII$\lambda1608$ lines, respectively. The wavelength rage of \NV$\lambda1238$ is affected by the geo-coronal \OI+\SiII\ emission lines. The single component, weak metal lines are fitted simultaneously keeping $z$ and $b$-parameters tied. Besides the weak low-ionization lines, strong \CIV\ and weaker \SiIV\ lines are detected. The \lya\ absorption shows an asymmetric profile indicating the presence of more than one component. We used two components to fit the \lya\ profile. {\sc vpfit} returns a large error ($\sim1.5$ dex) in $N(\HI)$. We assigned $Q=1$ for this system.}    
\label{SDSSJ1122_5755_0.05319} 
\end{figure*}

\begin{figure*} 
\includegraphics[width=0.98\textwidth]{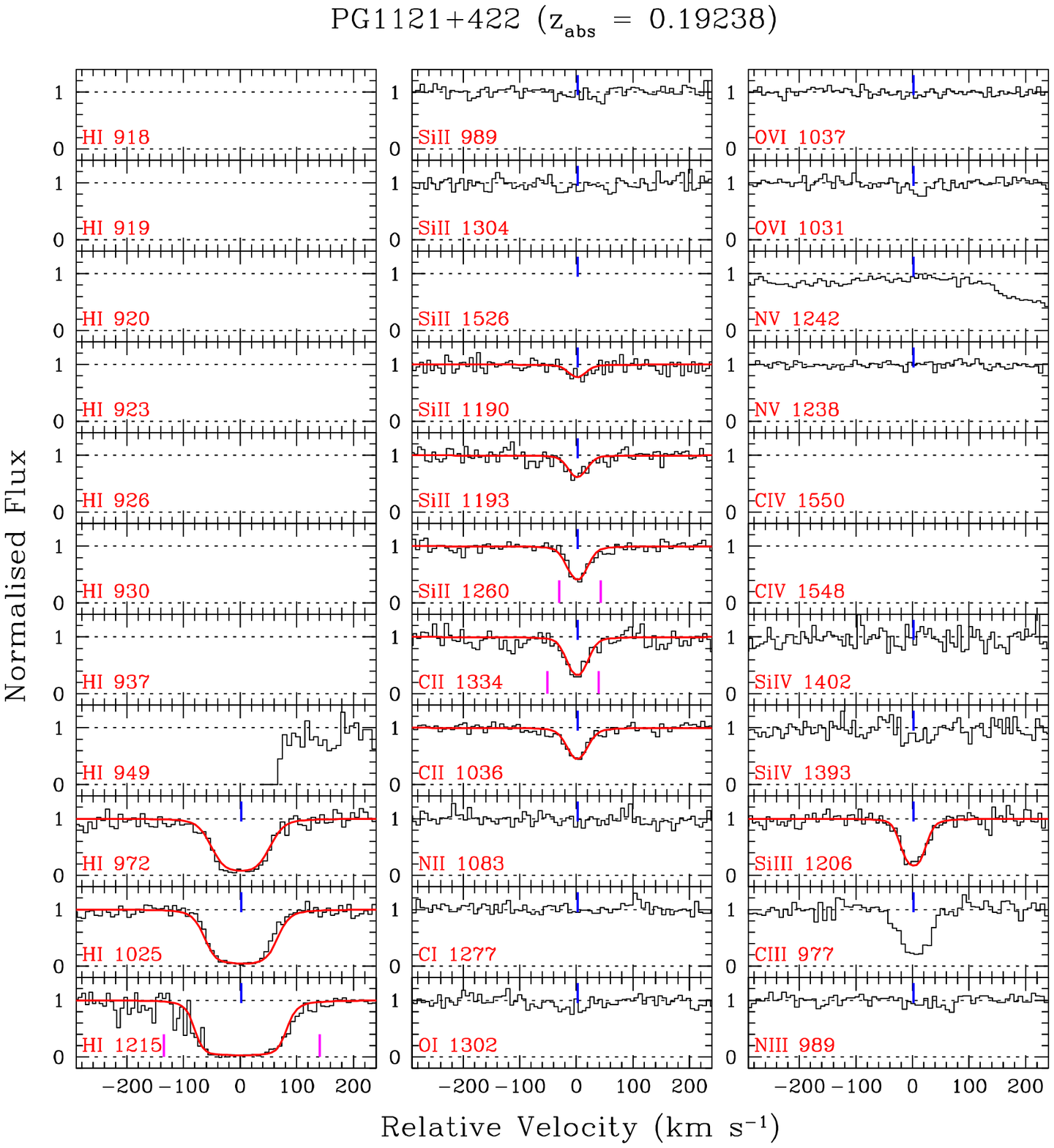} 
\vskip-0.8cm  
\caption{Similar to Fig.~\ref{PG0003_158_0.16512} but for the \zabs~$=0.19238$ system towards PG1121+422. Single-component metal lines are fitted simultaneously keeping $z$ and $b$-parameters tied with each other. The \SiIII\ line is mildly saturated. We note that the \CII$\lambda1036$ and \lyb\ lines require a velocity shift of $\sim-11$~\kms\ in order to be aligned with the other metal lines. All three Lyman series lines are heavily saturated. We obtained a solution with $\log N(\HI)=15.64\pm0.05$ and $b(\HI)=37\pm1$ \kms\ using a simultaneous, single-component fit. Since all three Lyman series lines are saturated we assign $Q=2$ for this system.}     
\label{PG1121_422_0.19238} 
\end{figure*}

\begin{figure*} 
\includegraphics[width=0.98\textwidth]{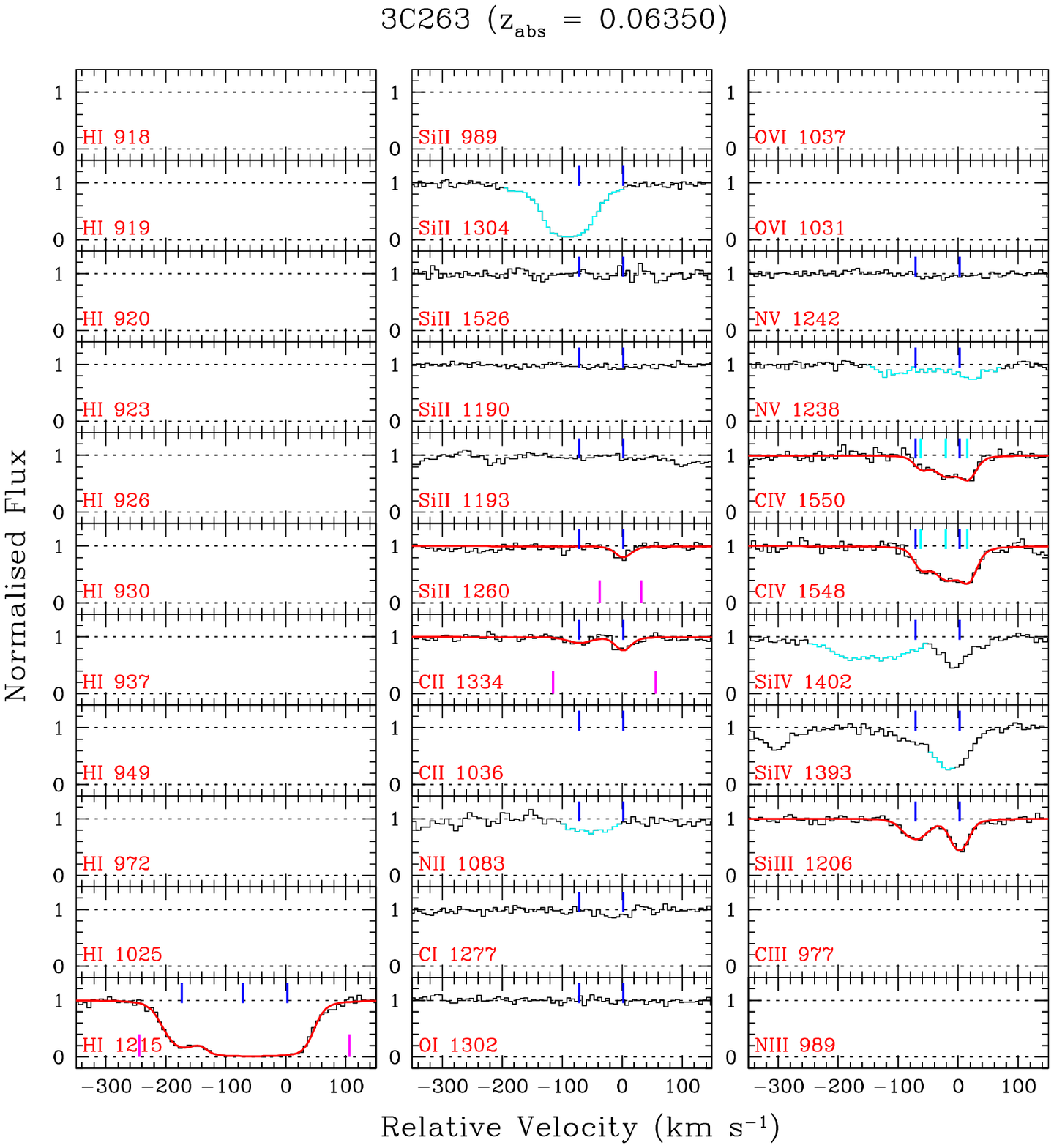} 
\vskip-0.8cm  
\caption{Similar to Fig.~\ref{PG0003_158_0.16512} but for the \zabs~$=0.06350$ system towards 3C263. Both the \SiIII\ and \CII$\lambda1334$ lines clearly show a two-component structure. The component at $\sim-70$~\kms, however, is not detected in the \SiII$\lambda1260$ line. Strong \CIV$\lambda\lambda1548,1550$ lines are present along with \SiIV$\lambda\lambda1393,1402$. The \SiIV$\lambda1393$ line is possibly blended with the \lya\ of the \zabs~$=0.21926$ system. The blue wing of the \SiIV$\lambda1402$ line is blended with the \lyb\ of the \zabs~$=0.45374$ system. Metal lines are fitted simultaneously. The \lya\ line is fitted using three components, with two of them locked at the redshifts of the metal line components. An extra \HI\ component is required to fit the weak absorption seen at $\sim-180$~\kms. However, we adopted the total $\log N(\HI)=15.40\pm0.12$ as measured by the \citet{Savage12} from the unsaturated \lyg\ profile using FUSE data. We thus assign a quality factor of $Q=4$ for this system.}    
\label{3C263_0.06350} 
\end{figure*}

\begin{figure*} 
\includegraphics[width=0.98\textwidth]{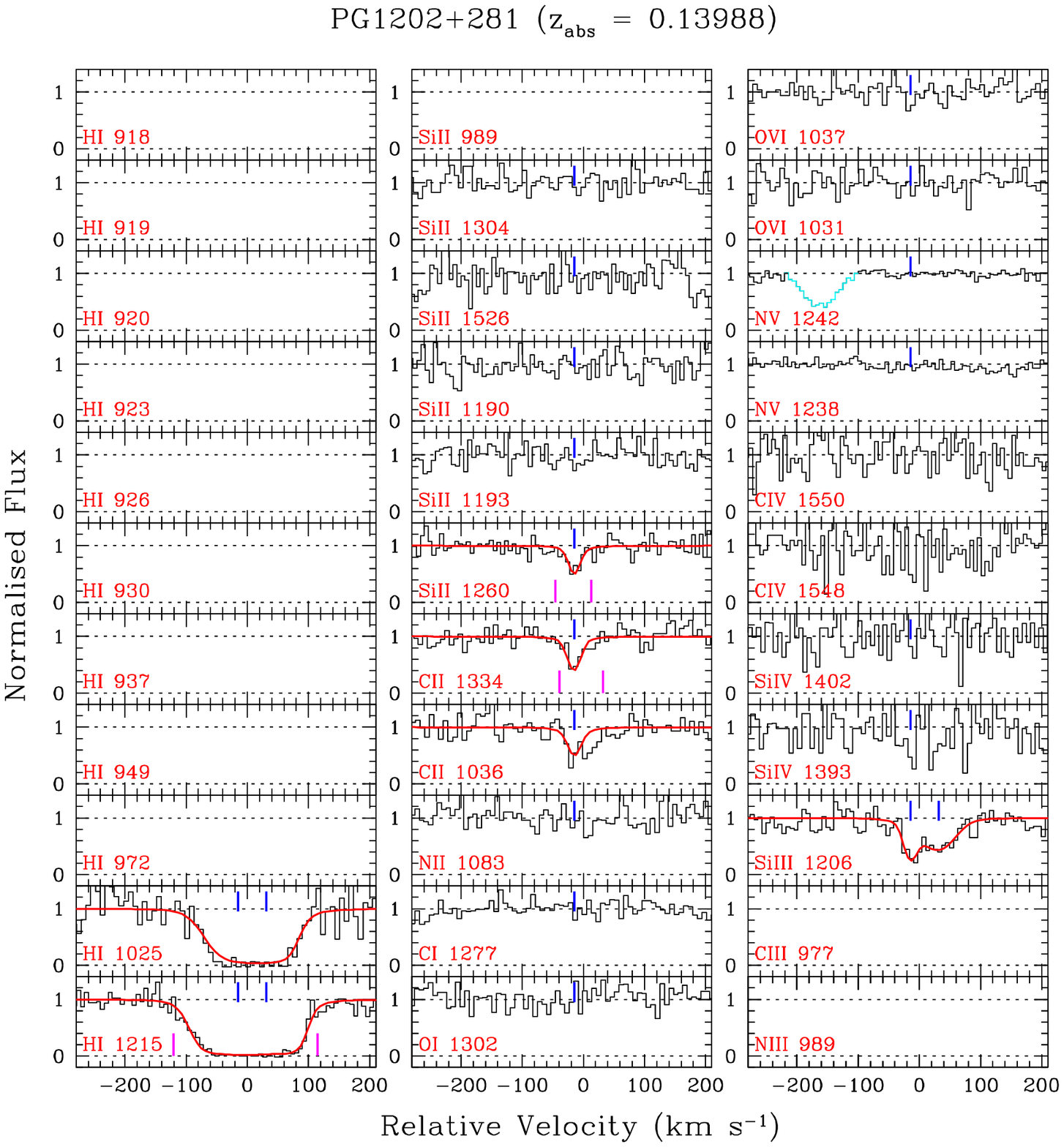} 
\vskip-0.8cm  
\caption{Similar to Fig.~\ref{PG0003_158_0.16512} but for the \zabs~$=0.13988$ system towards PG~1202+281. None of the high ionization lines are detected. The \SiIII\ line shows a two component absorption kinematics. The component at $+40$~\kms\ is not detected in the \SiII$\lambda1260$, \CII$\lambda1334$, and \CII$1036$ lines. We note that the \CII$\lambda1334$ line is offset by $\sim +10$~\kms\ with respect to the \lya/\SiIII\ line whereas the \lyb\ and \CII$\lambda1036$ lines are offset by $\sim+20$~\kms. The metal lines are fitted simultaneously keeping $z$ and $b$ tied for the component present in all of the transitions. A simultaneous fit is also performed for the \lya\ and \lyb\ lines using two Voigt profile components. The redshifts of the components are locked at the values obtained from the \SiIII\ fit. Since both the lines are heavily saturated we assign a quality factor of $Q=1$.}    
\label{PG1202_281_0.13988} 
\end{figure*}

\begin{figure*} 
\includegraphics[width=0.98\textwidth]{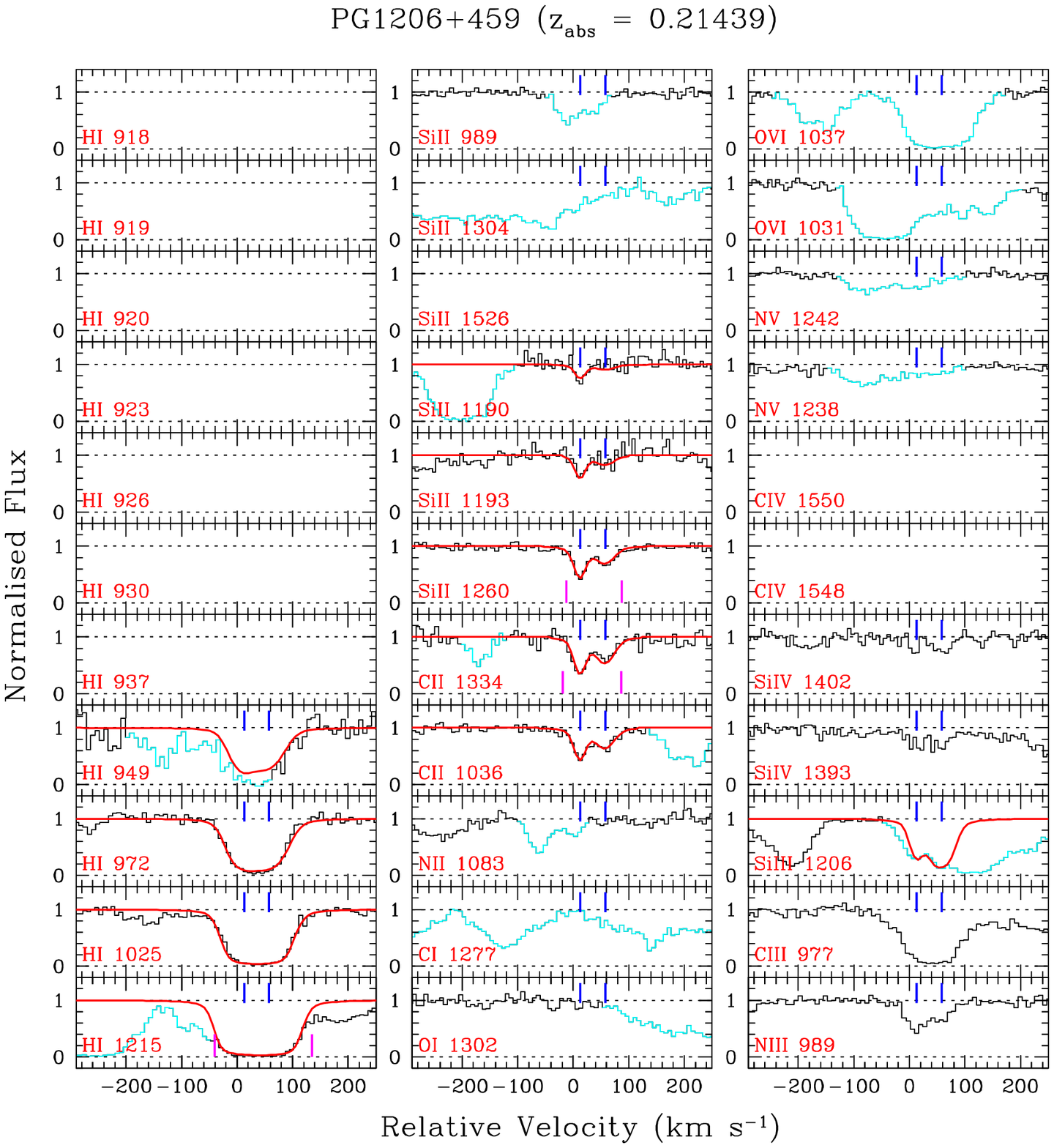} 
\vskip-0.8cm  
\caption{Similar to Fig.~\ref{PG0003_158_0.16512} but for the \zabs~$=0.21439$ system towards PG~1206+459. The narrow absorption seen in the blue wing of the \lya\ line is the \SII$\lambda765$ from the \zabs~$=0.9276$ system. The absorption seen in the \NV$\lambda1238$ panel is actually the \NeVIII~$\lambda780$ line from the same system \citep[see][for details]{Tripp11}. The \SiIII$\lambda1206$ line is severely blended with the \lya\ from the \zabs~$=0.20575$ system. The absorption in the \SiII$\lambda1304$ panel is actually the \NeVIII~$\lambda780$ line from a known intrinsic absorber at \zabs~$1.028$ studied by \citet{Muzahid13}. The \OVI~$\lambda1031$ line is blended with the Galactic \SII$\lambda1253$ line and the \lyb\ of the \zabs~$=0.22156$ system. The absorption in the \OVI~$\lambda1037$ panel is the Galactic \SiII~$\lambda1260$ line. So we could not confirm the presence/absence of \OVI\ absorption in this system. This spectrum suffers from a complex velocity alignment problem. However the clean two component structure makes it possible to gauge the uncertainty. The two components seen in the weak metal lines are fitted simultaneously. Since the \SiIII\ line is heavily blended, we have estimated the maximum column density allowed by the profile by keeping the $z$ and $b$-parameters locked at the values we obtained from the fitting of \SiII\ and \CII\ lines. The Lyman series lines are also fitted simultaneously using two components with the redshifts locked at the values we obtained for the metal lines. The \HI~$\lambda949$ line is partially blended by an unknown contaminant. However, we believe we have an adequate $N(\HI)$ measurement for this system with $Q=3$.}    
\label{PG1206_459_0.21439} 
\end{figure*}

\clearpage  
\begin{figure*} 
\includegraphics[width=0.98\textwidth]{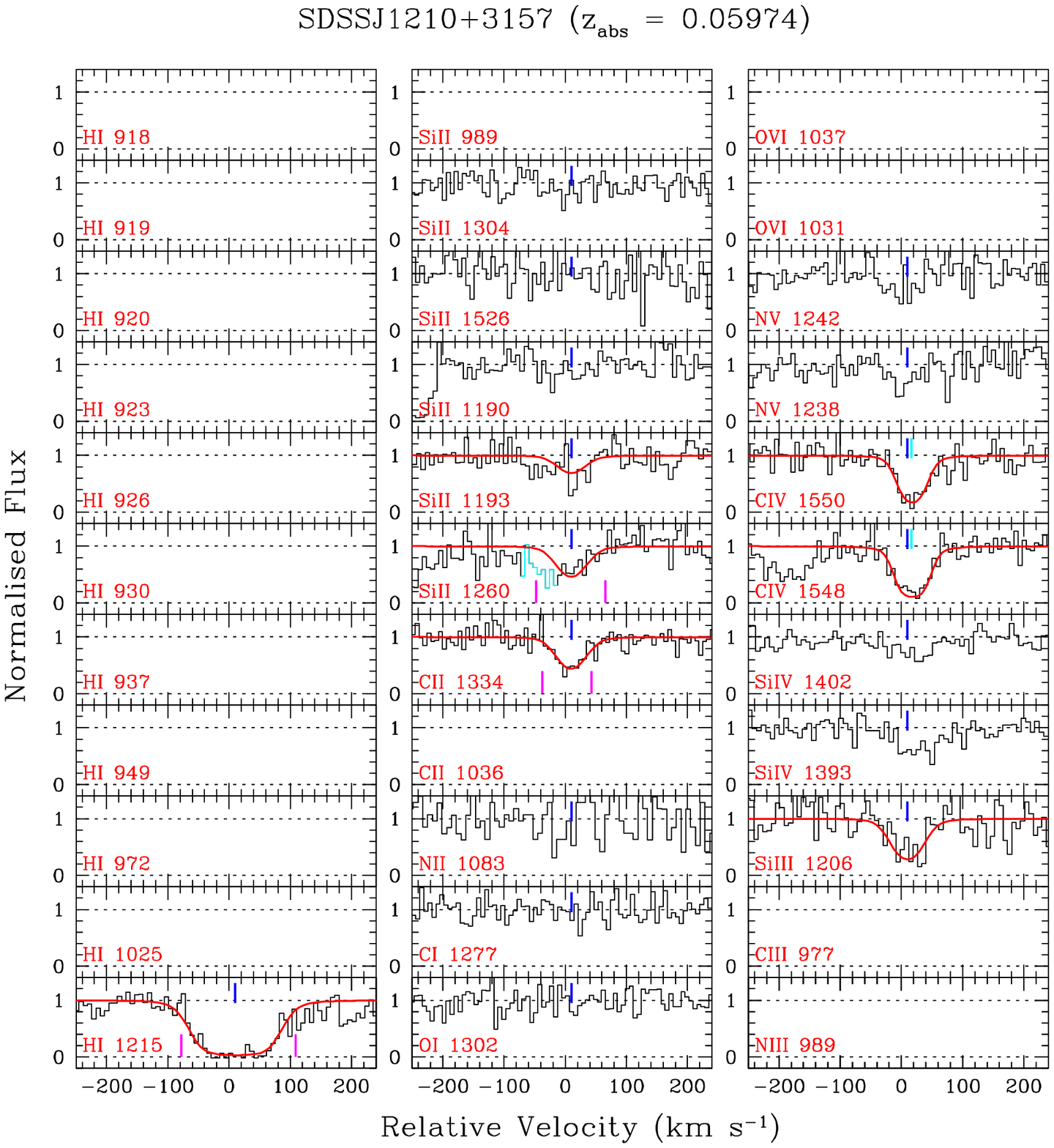} 
\vskip-0.8cm  
\caption{Similar to Fig.~\ref{PG0003_158_0.16512} but for the \zabs~$=0.05974$ system towards SDSSJ1210+3157. In addition to the low-ionization lines, strong \CIV, and weaker \SiIV, and \NV\ lines are detected. The blue wing of the \SiII$\lambda1260$ is blended with the Galactic \CII*$\lambda1335$ absorption. The equivalent width of the \SiII\ line is, thus, calculated from the model profile. The spectrum is very noisy at the position of the \SiII$\lambda1193$ line. A single component, simultaneous fit is performed for all the metal line transitions. The \lya\ line is fitted with a single component with redshift locked with the metal line component. Due to the lack of higher order unsaturated Lyman series lines, we assign $Q=1$ for this system.}    
\label{SDSSJ1210_3157_0.05974} 
\end{figure*}

\begin{figure*} 
\includegraphics[width=0.98\textwidth]{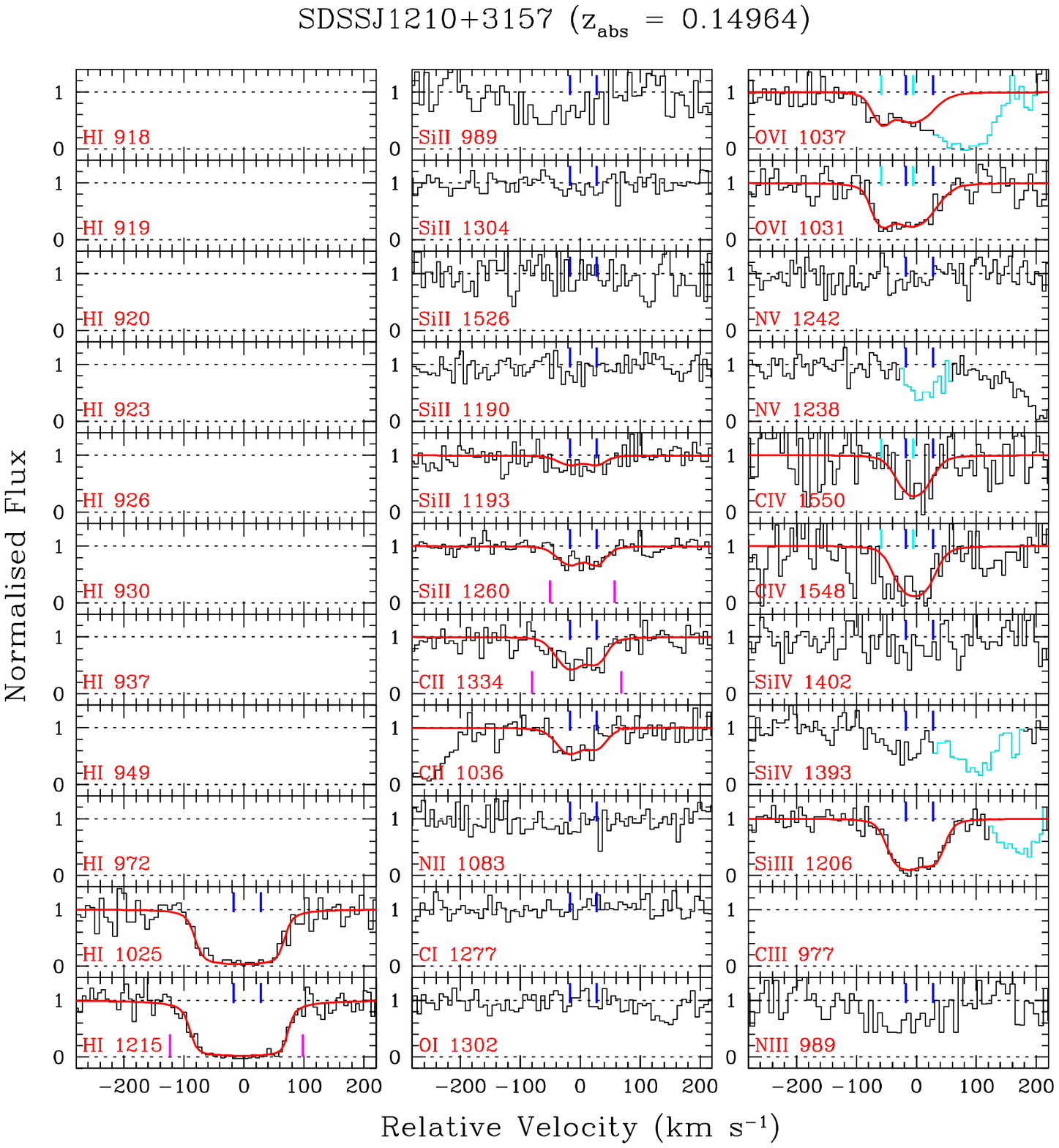} 
\caption{Similar to Fig.~\ref{PG0003_158_0.16512} but for the \zabs~$=0.14964$ system towards SDSSJ1210+3157. Besides several low-ionization lines, strong \OVI\ and \CIV\ lines are detected. The \OVI$\lambda1037$ line is blended with the Galactic \SiII$\lambda1193$ absorption. The absorption seen in the \NV$\lambda1238$ panel is the \lyb\ from the \zabs~$=0.38854$ associated system. Though the spectrum is noisy, a two component absorption kinematics is clearly seen in the \SiIII\ line. Note that the \SiIII\ line is strongly saturated. Thus the $N(\SiIII)$ we obtained from {\sc vpfit} should be considered as lower limit. Both the Lyman series lines are heavily saturated. We performed simultaneous fitting keeping both $z$ and $b$-parameters locked at the values we obtained from metal line fit. Since both the lines are heavily saturated we assign $Q=1$ for this system.}    
\label{SDSSJ1210_3157_0.14964} 
\end{figure*}

\begin{figure*} 
\includegraphics[width=0.98\textwidth]{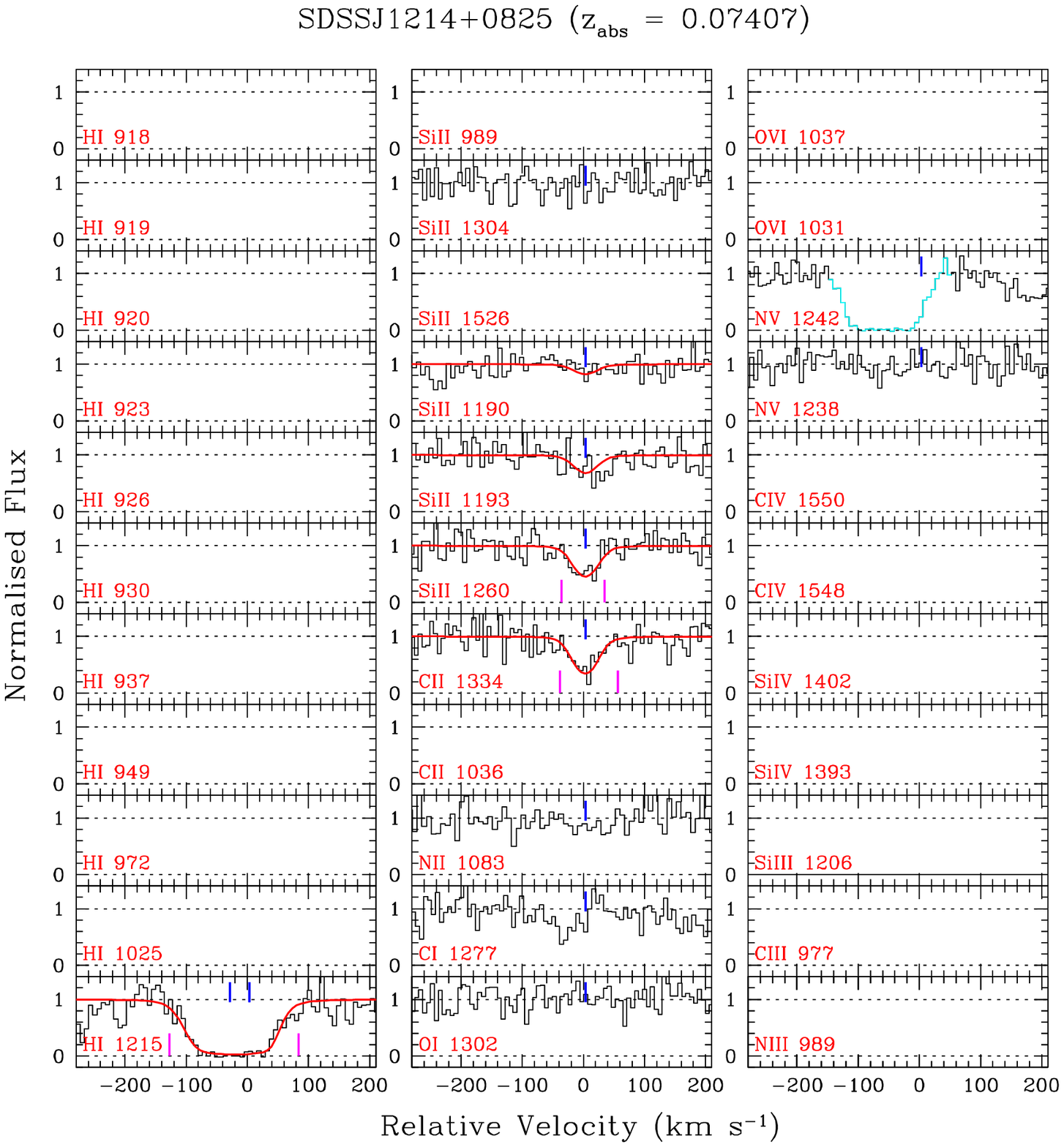} 
\vskip-0.8cm  
\caption{Similar to Fig.~\ref{PG0003_158_0.16512} but for the \zabs~$=0.07407$ system towards SDSSJ1214+0825. Only G130M data are available. \SiIII\ falls in the spectral gap. \lya\ falls on top of galactic \OI+\SiII\ emission lines. We combined the ``night-only" data separately for the \lya\ absorption. The \CII\ and \SiII\ lines can be fitted simultaneously using only one component.  The \lya\ absorption require one more component at $\sim-28$~\kms. Note that for the component at $0$~\kms\ we determine the maximum column density allowed by the profile to be $\log N(\HI)=15.0$, using a Doppler-parameter that we obtained for the metal lines ($b=22$~\kms). We assign a quality factor of $Q=1$ for the $N(\HI)$ upper limit. Because of the lack of \SiIII\ coverage, we do not consider this system for PI modeling.}        
\label{SDSSJ1214_0825_0.07407} 
\end{figure*}

\begin{figure*} 
\includegraphics[width=0.98\textwidth]{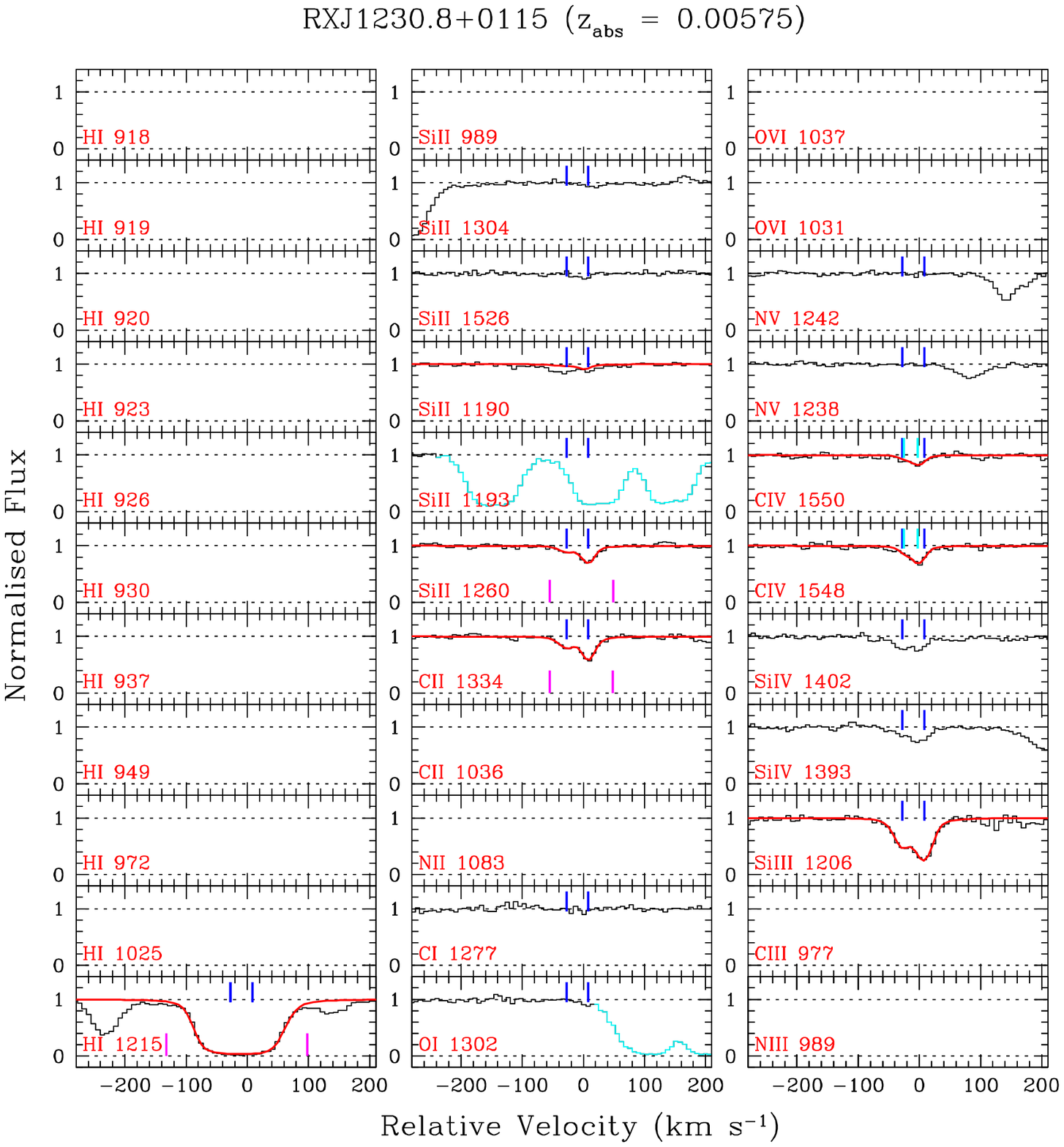} 
\vskip-0.8cm  
\caption{Similar to Fig.~\ref{PG0003_158_0.16512} but for the \zabs~$= 0.00575$ system towards RXJ~1230.8+0115. Besides low-ions, high-ionization metal lines (\CIV\ and \SiIV) are also detected. The \SiII$\lambda1260$ and \CII$\lambda1334$ lines are $\sim+6$ and $+4$~\kms\ offset, respectively, with respect to the \SiIII\ line. The \SiII$\lambda1193$ line is lost due to the strong Galactic \NI$\lambda1200$ line. The strong line seen in the \OI\ panel is actually \lya\ absorption from the \zabs~$=0.07770$ system. A clear two-component kinematics is seen in the low-ionization metal lines. The metal lines are fitted simultaneously. The \lya\ line is fitted using two components with redshifts locked at the values obtained from the metal line fitting. Due to the lack of unsaturated, higher order Lyman series lines, the estimated $N(\HI)$ is uncertain, and we assign a quality factor of $Q=1$.}      
\label{RXJ1230_0115_0.00575} 
\end{figure*}

\begin{figure*} 
\includegraphics[width=0.98\textwidth]{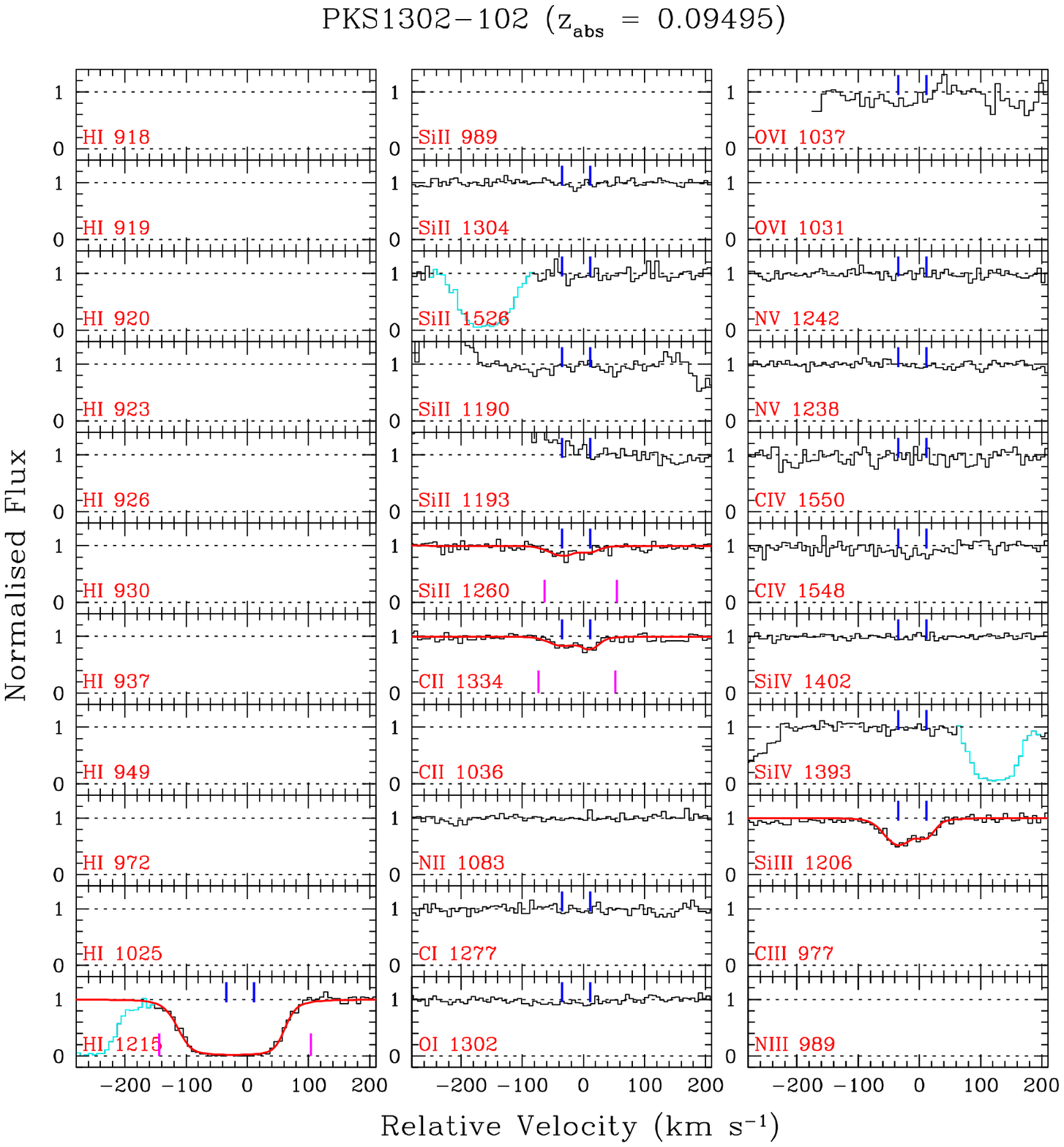} 
\vskip-0.8cm  
\caption{Similar to Fig.~\ref{PG0003_158_0.16512} but for the \zabs~$=0.09495$ system towards PKS~1302--102. The presence of high-ionization lines (e.g. \CIV/\OVI) is not obvious. A minimum of two components is required to fit the low-ionization lines. The \CII$\lambda1334$ line shows a velocity offset of $\sim-7$~\kms\ with respect to the other lines. The metal lines (\CII/\SiII) are fitted simultaneously. We use two Voigt profile components to fit the \lya\ absorption with line centroids locked at the values we obtained from the metal line fit. This is a partial-Lyman-limit system with $\log N(\HI) = 16.88\pm0.03$ as measured by \citet{Cooksey08} from the COG analysis of the full Lyman series lines using FUSE data. We adopted their $N(\HI)$ value and assign the highest quality factor of $Q=4$.}    
\label{PKS1302_102_0.09495} 
\end{figure*}

\begin{figure*} 
\includegraphics[width=0.98\textwidth]{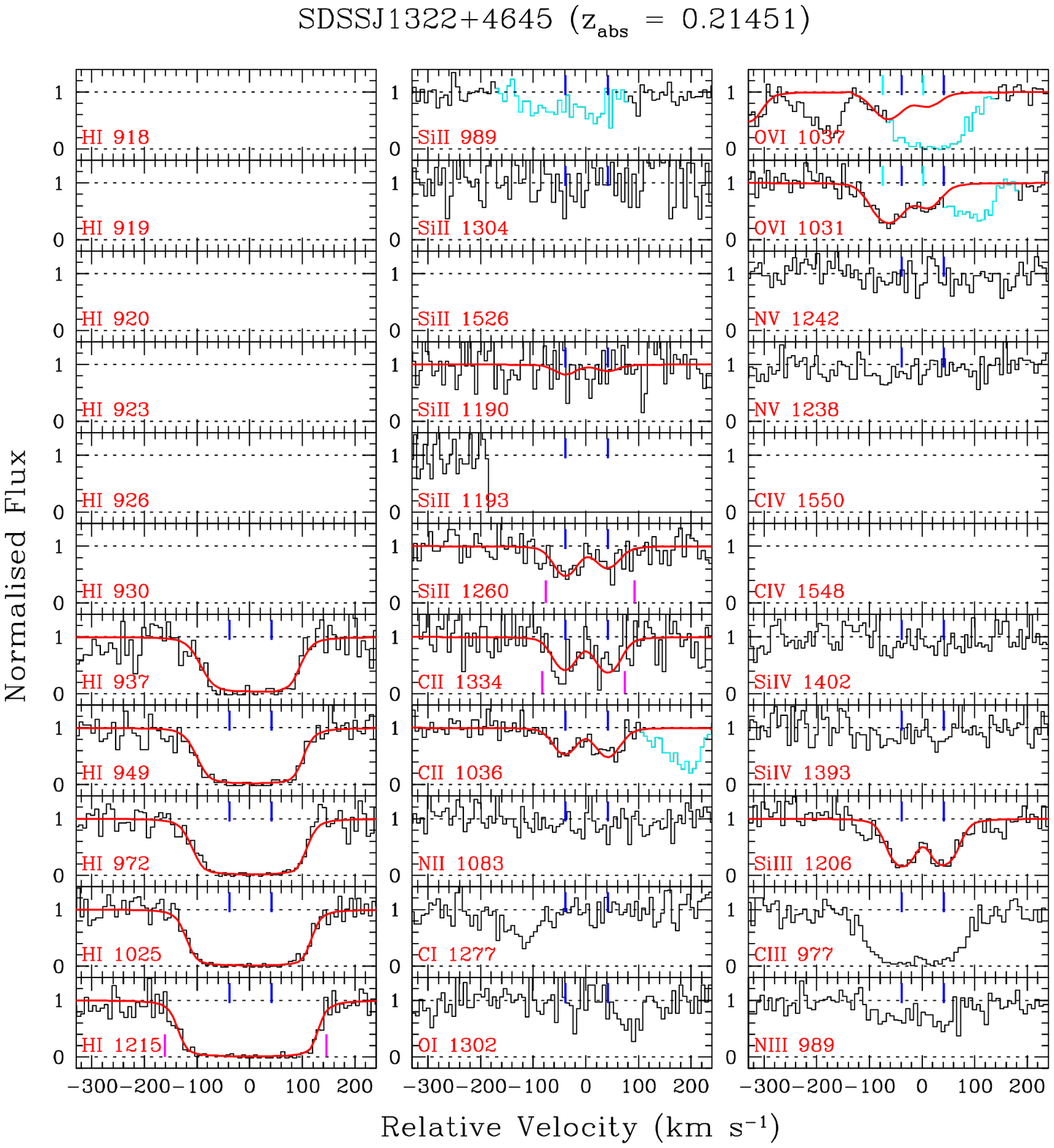} 
\vskip-0.8cm  
\caption{Similar to Fig.~\ref{PG0003_158_0.16512} but for the \zabs~$=0.21451$ system towards SDSSJ1322+4645. The \SiII$\lambda1193$ line falls in the spectral gap. The \OVI$\lambda1031$ line is blended with the Galactic \SII$\lambda1253$ line, whereas, the \OVI$\lambda1037$ line is lost in the strong Galactic \SiII$\lambda1260$ absorption. The spectrum is noisy. We used two components to fit the metal lines simultaneously. Both the \CII$\lambda1334$ and $\lambda1036$ lines are shifted by $\sim-10$~\kms\ with respect to the \SiIII/\SiII\ lines. We consider $N(\SiIII)$ as upper limit since the \SiIII\ line is strongly saturated. The \HI$\lambda937$ line falls at the extreme blue end of the spectrum. The line require a velocity shift of $\sim+12$~\kms. The Lyman series lines are also fitted simultaneously using two components by keeping redshifts locked but allowing $b$-parameters to vary. We, however, adopt the $N(\HI)$ value obtained from the Lyman limit break by \citet{Prochaska17}. We assign $Q=4$ for this system.}    
\label{SDSSJ1322_4645_0.21451} 
\end{figure*}

\begin{figure*} 
\includegraphics[width=0.98\textwidth]{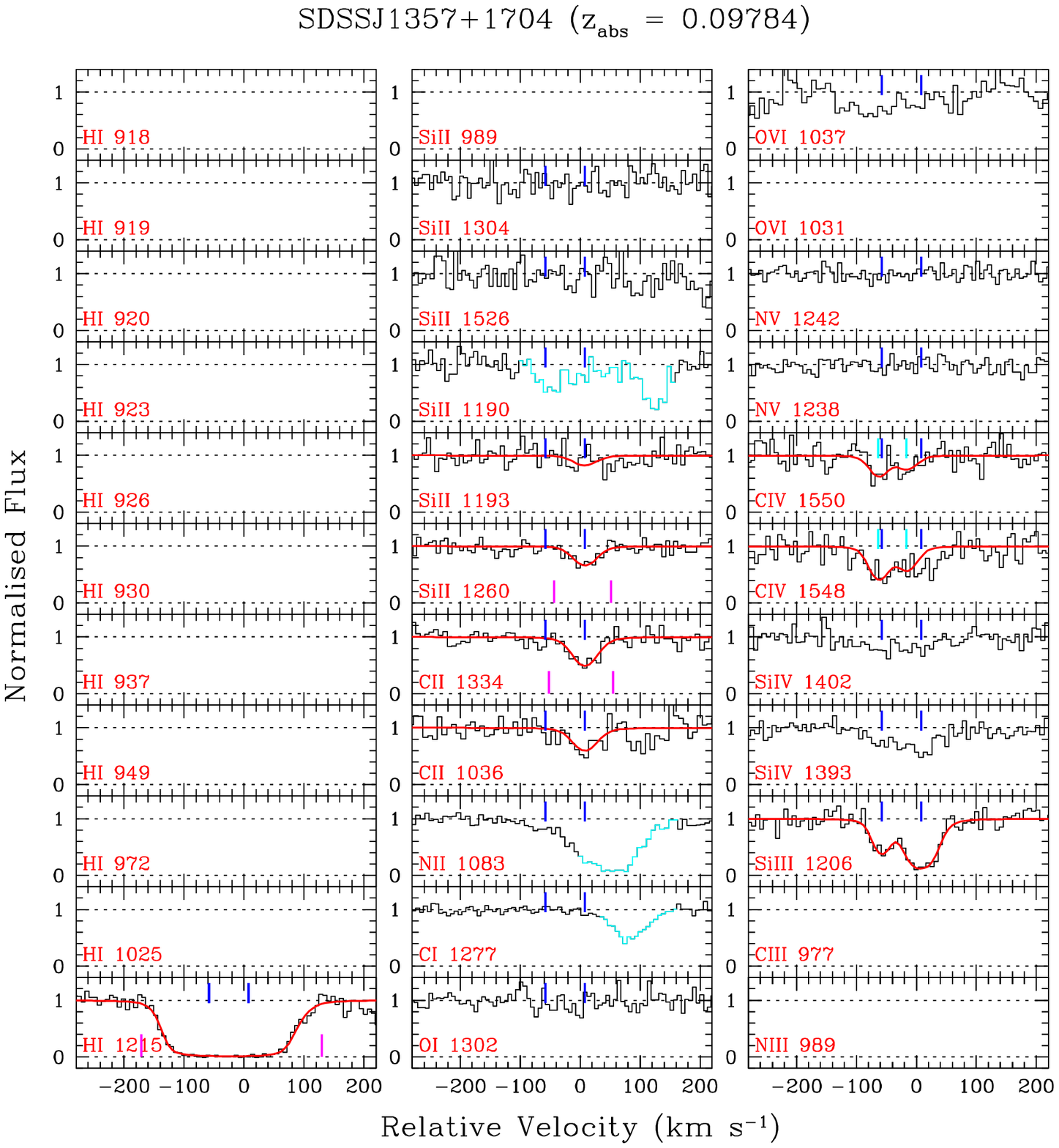} 
\vskip-0.8cm  
\caption{Similar to Fig.~\ref{PG0003_158_0.16512} but for the \zabs~$=0.09784$ system towards SDSSJ1357+1704. A two-component absorption kinematics, as seen in the \SiIII\ line, is also present in the \SiIV\ and \CIV\ doublets. However the \NV\ doublet is not detected. The \OVI$\lambda1031$ is not covered. Both \CII$\lambda1036$ and \OVI$\lambda1037$ fall at the extreme blue end of the spectrum. We note that the \CII$\lambda1036$ (and hence the \OVI$\lambda1037$) line require a velocity shift of $\sim+15$~\kms. Although some absorption is seen in the \OVI$\lambda1037$ panel, we could not confirm its association with this system. The \NII$\lambda1083$ line is blended with the Galactic \SiII$\lambda1190$ line. Note that the \SiIII\ component at $\sim0$~\kms\ is strongly saturated. The other component is not detected in the singly ionized carbon and silicon lines. The component at $\sim0$~\kms\ is fitted simultaneously keeping $z$ and $b$-parameter tied. Due to saturation, we consider $N(\SiIII)$ as a lower limit. The \lya\ line is fitted with two components as seen in the \SiIII\ line. We note that the \lya\ line is contaminated with Galactic \CII$\lambda1334$. Therefore, the $N(\HI)$ in this system is highly uncertain. We thus assign $Q=1$ for this system.}    
\label{SDSSJ1357_1704_0.09784} 
\end{figure*}

\begin{figure*} 
\includegraphics[width=0.98\textwidth]{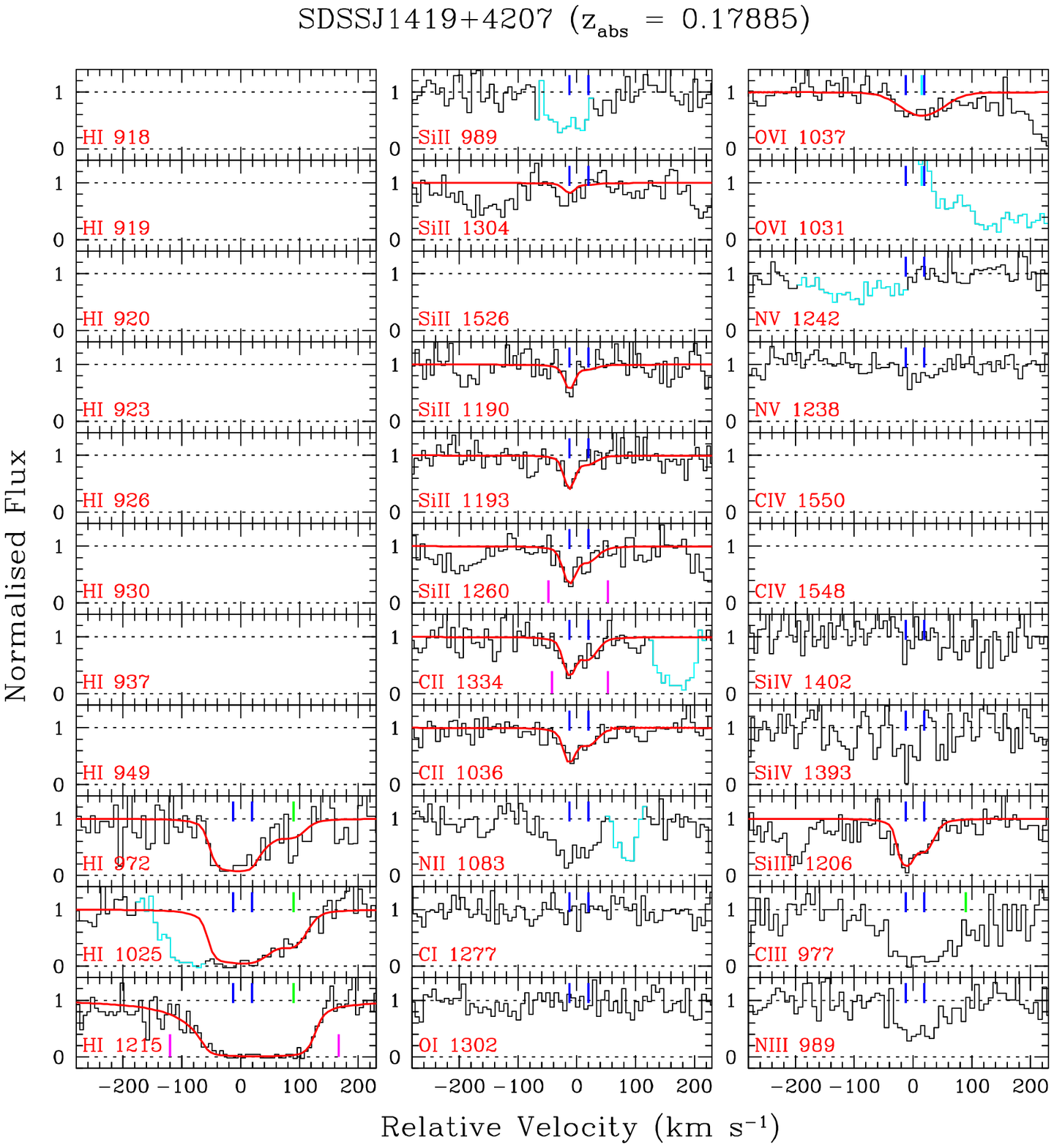} 
\vskip-0.8cm  
\caption{Similar to Fig.~\ref{PG0003_158_0.16512} but for the \zabs~$=0.17885$ system towards SDSSJ1419+4207. The \SiIII\ and \lya\ lines fall at the red edge of the G130M grating data. Both these lines require a velocity shift of $\sim-11$~\kms. The \SiIII, \CII$\lambda1036$ and $\lambda1334$ lines show one narrow and one broad component. The presence of the broad component is not obvious in the \SiII\ lines, particularly in the weaker lines. Metal lines are fitted simultaneously using two Voigt profile components. Note that we did not use the \SiII$\lambda1304$ and $\lambda989$ lines for fitting, because of blends. The \SiII$\lambda989$ is blended with the \NIII$\lambda989$ line. The blue wing of the \lyb\ line is blended with the \HI$\lambda937$ line of another intervening system at \zabs~$=0.28901$. The narrow component is saturated in the \CII\ and \SiIII\ lines. Thus, both $N(\CII)$ and $N(\SiIII)$ are considered as lower limits. Besides the two components seen in the metal lines, the Lyman series lines require an extra component at $\sim+80$~\kms, as indicated by the green tick. This additional component may be present in the \CIII\ line. Recently, \citet{Prochaska17} have published new $N(\HI)$ measurement for this system using the Lyman limit break seen in the G140L spectrum. We adopt their $N(\HI)$ value of $10^{16.63^{+0.09}_{-0.30}}$~\sqcm\ and assign $Q=4$.}    
\label{SDSSJ1419_4207_0.17885} 
\end{figure*}

\begin{figure*} 
\includegraphics[width=0.98\textwidth]{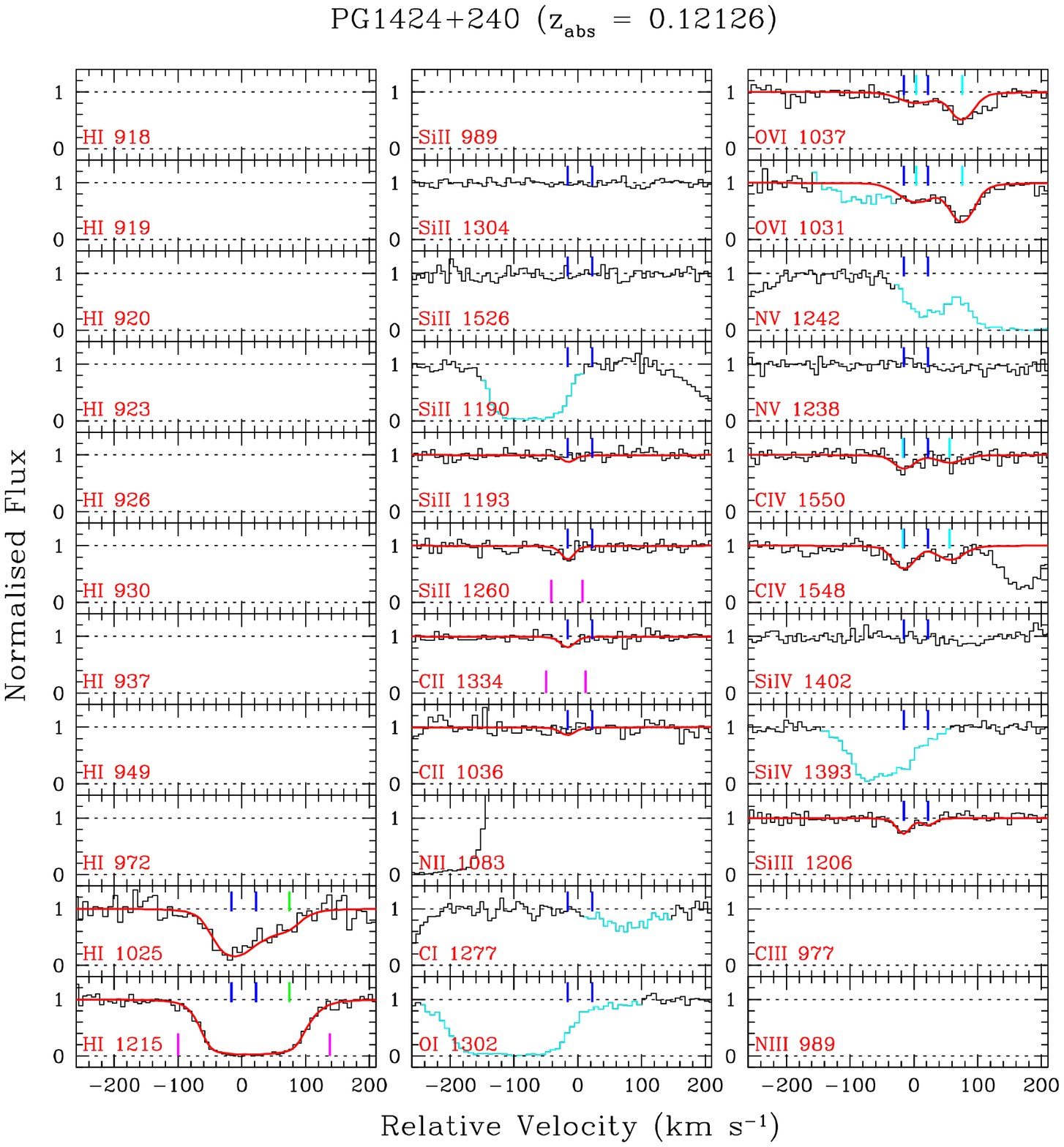} 
\vskip-0.8cm  
\caption{Similar to Fig.~\ref{PG0003_158_0.16512} but for the \zabs~$=0.12126$ system towards PG~1424+240. Two very weak components are detected in the \SiIII\ absorption. The weaker component at $\sim +20$~\kms, however, is not present in the \CII/\SiII\ absorption. We note that the \CII$\lambda1334$ line requires a $\sim-10$~\kms\ velocity shift. Metal lines are fitted simultaneously keeping $z$ and $b$-parameters tied for the component seen in all three transitions. Besides the weak low-ionization metal lines, \CIV\ and \OVI\ are also detected in this system, but \SiIV\ is not detected. The absorption seen in the \SiIV$\lambda1393$ panel is the \lya\ of the \zabs~$=0.28523$ system. The \NII$\lambda1083$ line is lost due to geo-coronal \lya\ emission. In addition to the two low-ionization metal line components, one more component, shown by the green tick, is required to fit the \lya\ and \lyb\ lines simultaneously. This additional component is also evident in the \OVI\ absorption. Since the \lyb\ line is only mildly saturated we assign $Q=3$ for this system.}    
\label{PG1424_240_0.12126} 
\end{figure*}

\begin{figure*} 
\includegraphics[width=0.98\textwidth]{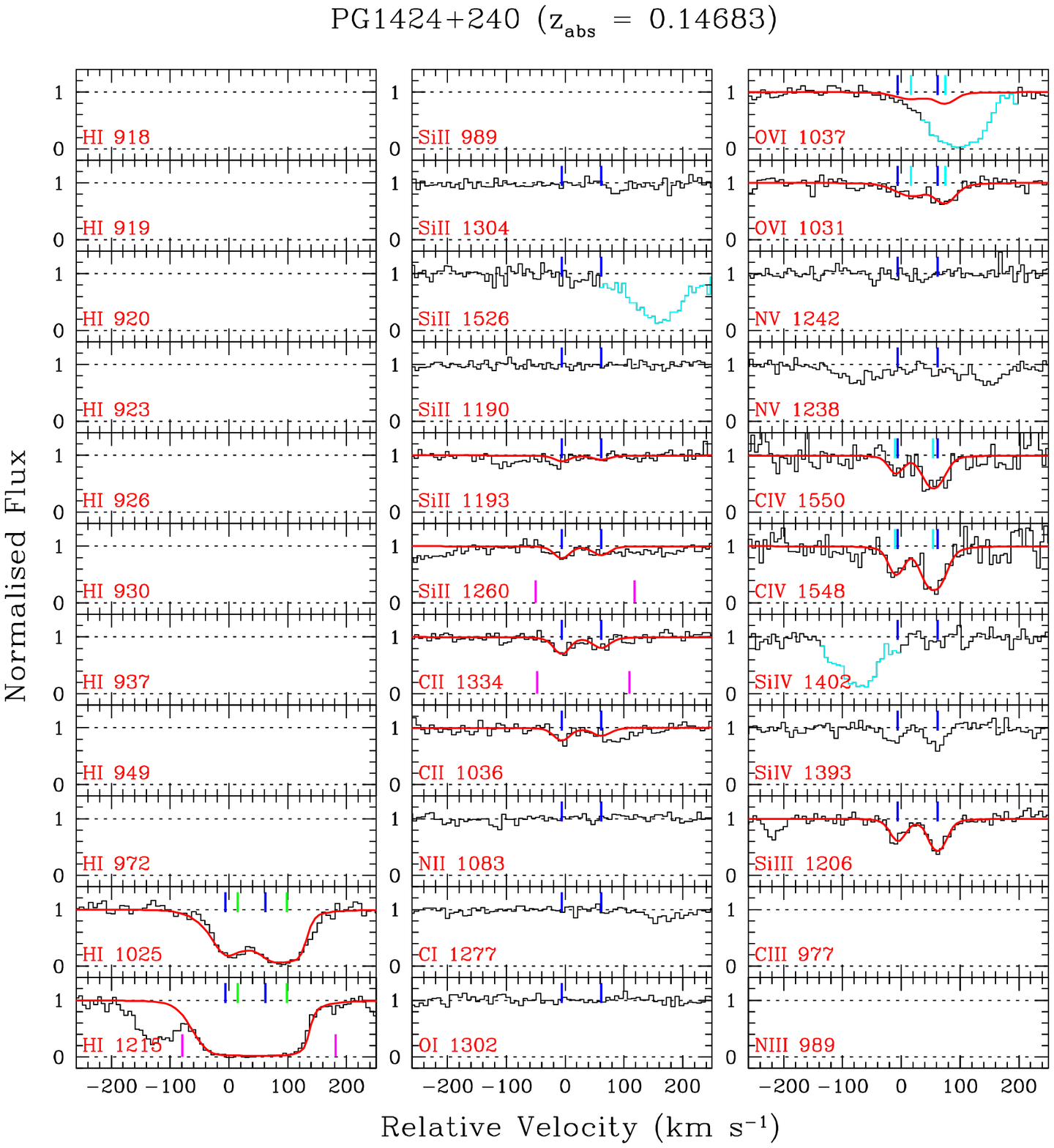} 
\vskip-0.8cm  
\caption{Similar to Fig.~\ref{PG0003_158_0.16512} but for the \zabs~$=0.14683$ system towards PG~1424+240. This is a very weak metal line system with high-ionization lines (\SiIV, \CIV, and \OVI) detected. The \OVI$\lambda1031$ line is blended with the Galactic \SiII$\lambda1190$ line. Two-component absorption kinematics is evident from the \SiIII\  line. Metal lines are fitted simultaneously by keeping $z$ and $b$-parameters tied. Interestingly, the centroids of the metal line components are not centered around the \lya\ and \lyb\ lines. In fact, they are shifted towards the blue wings. This is not due to the wavelength calibration uncertainty since the \CII$\lambda1036$ (close to the \lyb\ wavelength) and the \SiIII\ (close to the \lya\ wavelength) lines are properly aligned. Consequently, the \lya/\lyb\ lines require two additional components, as marked by the green ticks, for an adequate fit. Since there are not enough Lyman series lines to constrain our four-component fit, we assign $Q=1$ for this system.}    
\label{PG1424_240_0.14683} 
\end{figure*}

\begin{figure*} 
\includegraphics[width=0.98\textwidth]{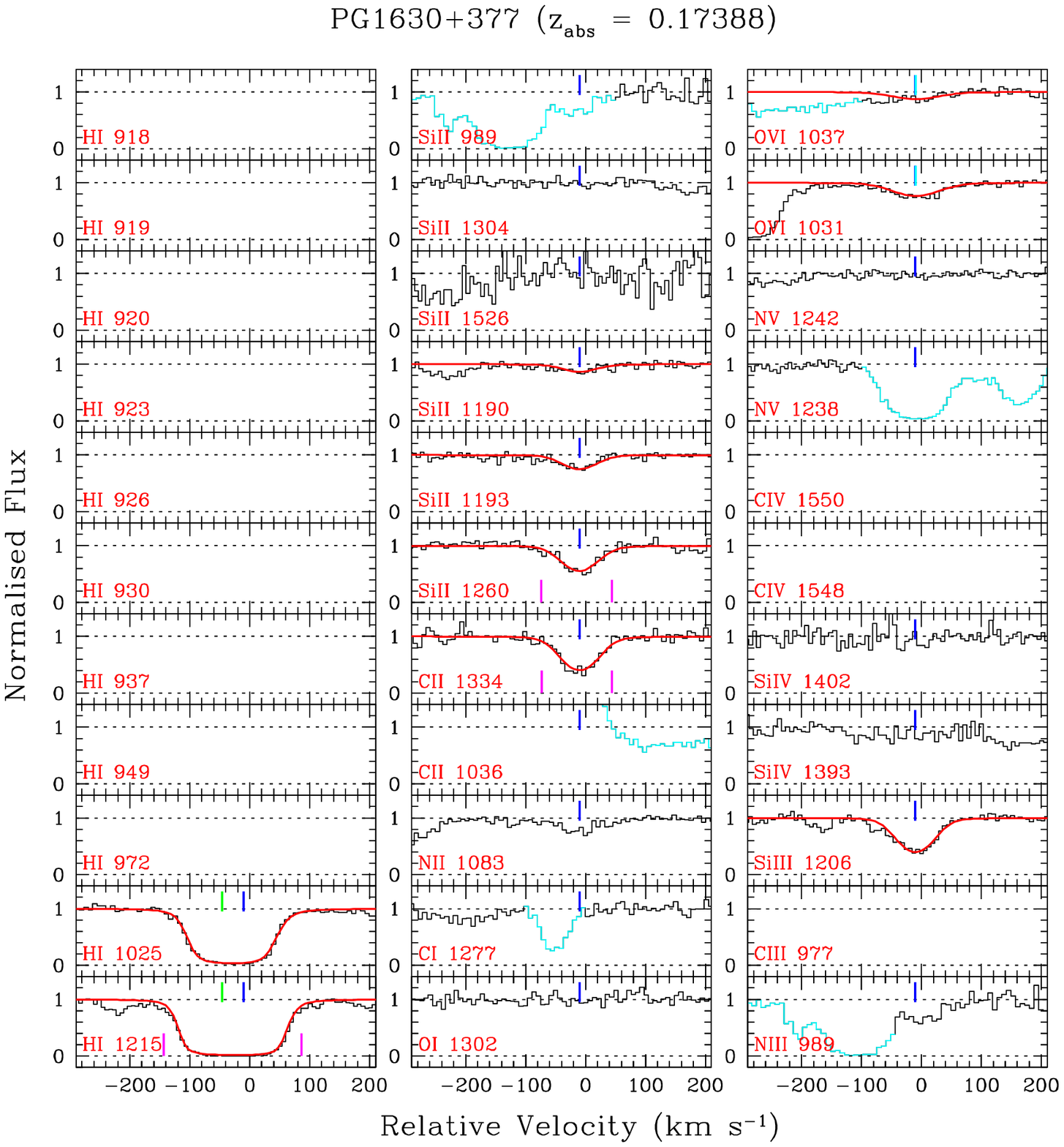} 
\vskip-0.8cm 
\caption{Similar to Fig.~\ref{PG0003_158_0.16512} but for the \zabs~$=0.17388$ system towards PG~1630+377. In addition to low-ionization metal lines, \OVI\ is detected. The \OVI$\lambda1031$ (and \CII$\lambda1036$) line is affected by Galactic \lya\ absorption. The strong absorption seen in the \NV$\lambda1238$ panel is the \lyb\ of the \zabs~$=0.41778$ system. We note that the \lyb\ and \OVI\ lines require a $\sim-20$~\kms\ velocity shift for properly alignment with the \lya\ line. The metal lines are fitted simultaneously. The Doppler parameter of $35$~\kms\ for our adopted single-component fit probably indicates the presence of multiple unresolved components. Two components are required to fit the \lya\ and \lyb\ absorption simultaneously. One of them is locked at the redshift corresponding to the low-ionization metal line component. The other component is possibly arising from the \OVI\ bearing gas phase as indicated by the green vertical tick. Since both the Lyman series lines are heavily saturated we assign $Q=1$ for this system.}    
\label{PG1630_377_0.17388} 
\end{figure*}

\begin{figure*} 
\includegraphics[width=0.98\textwidth]{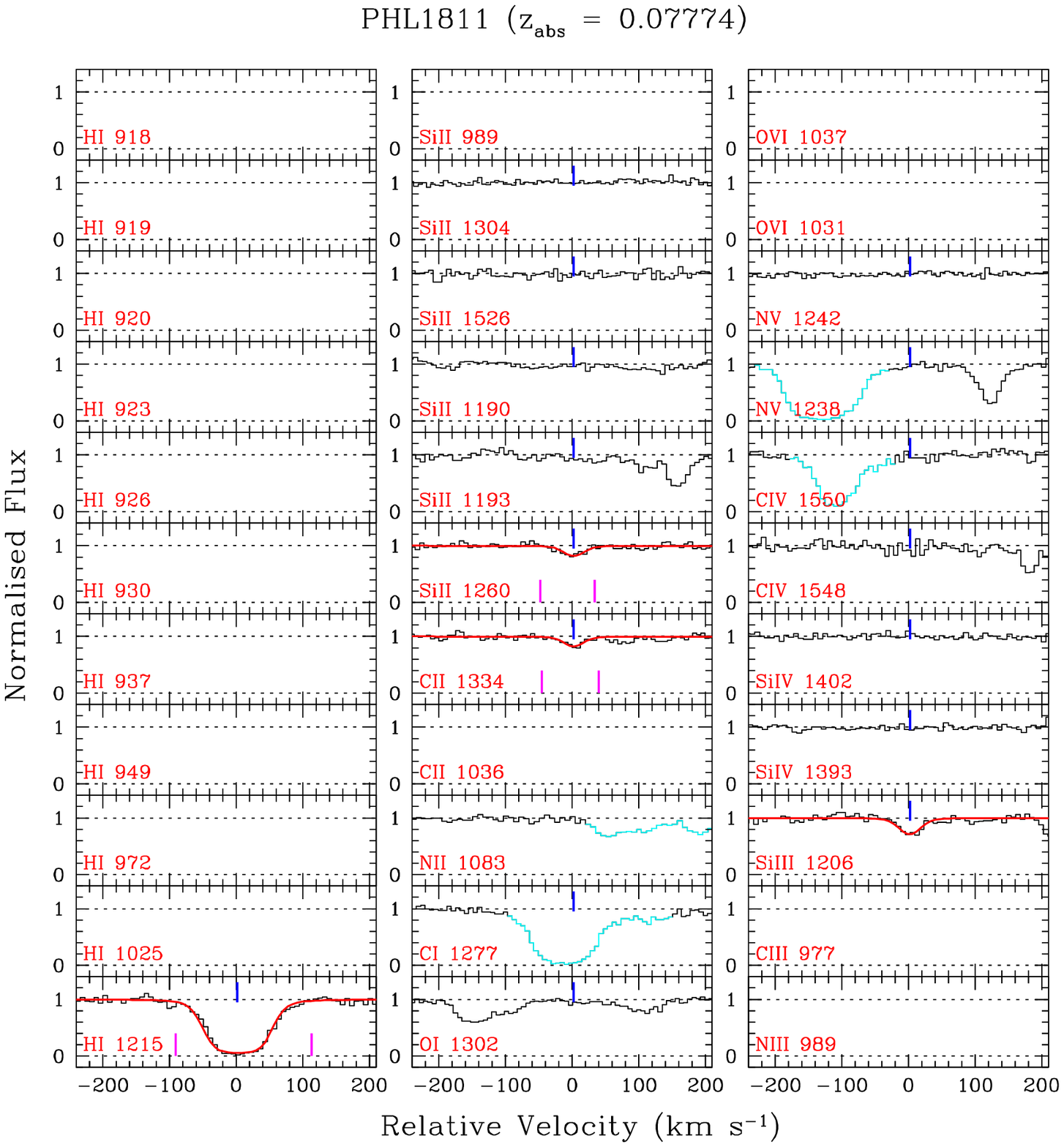} 
\vskip-0.8cm 
\caption{Similar to Fig.~\ref{PG0003_158_0.16512} but for the \zabs~$=0.07774$ system towards PHL1811. The low ionization metal lines are very weak. No high ionization metal lines are detected. Using unsaturated higher-order Lyman series lines (up to \HI$\lambda923$) covered by FUSE data and \lya\ from STIS data, \citet{Lacki10} have derived $\log N(\HI) =16.00\pm0.05$. Here we adopt their value and assign $Q=5$ for this system.}    
\label{PHL1811_0.07774} 
\end{figure*}

\begin{figure*} 
\includegraphics[width=0.98\textwidth]{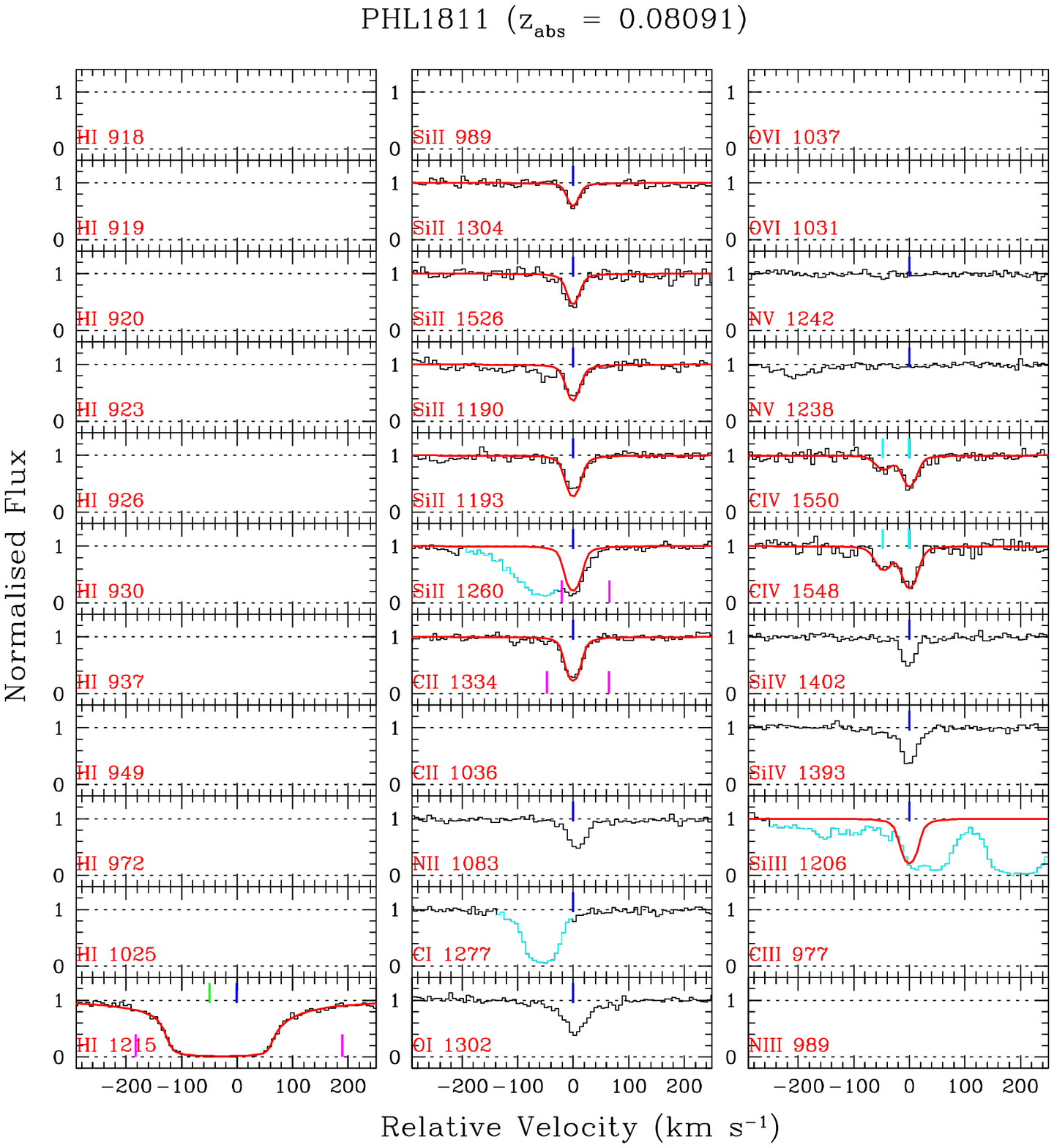} 
\vskip-0.8cm 
\caption{Similar to Fig.~\ref{PG0003_158_0.16512} but for the \zabs~$=0.08091$ system towards PHL1811. The \SiIII\ line is blended with Galactic \SiII$\lambda1304$. The \SiII$\lambda1260$ line is blended with the \lya\ of the \zabs~$=0.12051$ system. The single component \CII\ and \SiII\ lines are fitted simultaneously. We did not use the \SiII$\lambda1260$ line for fitting because of the blend; but the model profile corresponding to the best fitting parameters is shown. The \CII$\lambda1034$ line is heavily saturated. The \SiII$\lambda1193$ also shows an unresolved saturation effect. However the weaker, unsaturated \SiII$\lambda1304$ line gives an accurate $N(\SiII)$ estimate. Due to a severe blend in the \SiIII\ line we use the best fitting metal line parameters and estimate the maximum $N(\SiIII)$ allowed by the observed profile. In addition to the component seen in the low ionization metal lines, \lya\ requires another component at $\sim-50$~\kms. This component is seen in the high-ionization \CIV\ lines and is indicated by the green ticks. A strong Lyman limit break corresponding to this system is seen in a FUSE spectrum \citep[see Fig.~1 of][]{Jenkins03}. Ignoring the blue component \citet{Jenkins05} have derived $\log N(\HI)=17.98\pm0.05$ for the red component at $0$~\kms. It is consistent with our measurement (i.e. $\log N(\HI)=17.94\pm0.07$) for that component. We therefore assign $Q=4$ for this system.}          
\label{PHL1811_0.08091} 
\end{figure*}